\documentclass[fleqn,usenatbib]{mnras}

\usepackage{newtxtext,newtxmath}

\usepackage[T1]{fontenc}

\DeclareRobustCommand{\VAN}[3]{#2}
\let\VANthebibliography\thebibliography
\def\thebibliography{\DeclareRobustCommand{\VAN}[3]{##3}\VANthebibliography}

\usepackage{graphicx}	
\usepackage{amsmath}	
\usepackage{multicol}        
\usepackage{bm}		
\usepackage{pdflscape}	
\usepackage{threeparttable}
\usepackage{fontawesome}

\newcommand{\pyaneti}{\href{https://github.com/oscaribv/pyaneti}{\texttt{pyaneti}\,\faGithub}}
\newcommand{\logr}{$\log R'_{\rm HK}$}
\newcommand{\lbe}{$\lambda_{\rm e}$}
\newcommand{\lbp}{$\lambda_{\rm p}$}
\newcommand{\pgp}{$P_{\rm GP}$}
\newcommand{\ms}{${\rm m\,s^{-1}}$}
\newcommand{\kms}{${\rm km\,s^{-1}}$}

\newcommand{\citlalicue}{\texttt{citlalicue}}
\newcommand{\citlalatonac}{\texttt{citlalatonac}}

\title[\texttt{pyaneti} II]{\texttt{pyaneti} II: A multidimensional Gaussian process approach to analysing spectroscopic time-series}

\author[Barragán et al.,]{
Oscar Barragán,$^{1}$\thanks{E-mail: oscar.barragan@physics.ox.ac.uk, \href{https://twitter.com/oscaribv}{\faTwitter\texttt{@oscaribv}}} 
Suzanne Aigrain,$^{1}$
Vinesh M. Rajpaul,$^{2}$ and
Norbert Zicher$^{1}$
\\
$^{1}$Sub-department of Astrophysics, Department of Physics, University of Oxford, Oxford, OX1 3RH, UK\\
$^{2}$Astrophysics Group, Cavendish Laboratory, University of Cambridge, J. J. Thomson Avenue, Cambridge, CB3 0HE, UK\\
}

\date{Accepted XXX. Received YYY; in original form ZZZ}

\pubyear{2020}

\begin{document}
\label{firstpage}
\pagerange{\pageref{firstpage}--\pageref{lastpage}}
\maketitle

\begin{abstract}
The two most successful methods for exoplanet detection rely on the detection of planetary signals in photometric and radial velocity time-series. 
This depends on numerical techniques that exploit the synergy between data and theory to estimate planetary, orbital, and/or stellar parameters.
In this work we present a new version of the exoplanet modelling code \pyaneti. 
This new release has a special emphasis on the modelling of stellar signals in radial velocity time-series.
The code has a built-in multidimensional Gaussian process approach to modelling radial velocity and activity indicator time-series with different underlying covariance functions. 
This new version of the code also allows multi-band and single transit modelling; it runs on \texttt{Python 3}, and features overall improvements in performance.
We describe the new implementation and provide tests to validate the new routines that have direct application to exoplanet detection and characterisation.
We have made the code public  and freely available at \url{https://github.com/oscaribv/pyaneti}\href{https://github.com/oscaribv/pyaneti}{\faGithub}.
We also present the codes \citlalicue\ and \citlalatonac\ that allow one to create synthetic photometric and spectroscopic time-series, respectively, with planetary and stellar-like signals. 
\end{abstract}




\section{Introduction}

We are now living in a fascinating era in which we know that ``other Suns'' are the centres of concentric systems of many worlds \citep[cf.][]{Bruno1584}.
Over the past few decades, astronomers have developed techniques that allow us to ``translate'' photons into worlds \citep[see e.g.][]{exoplanets2016,Struve1952}. 
More than 4800\footnote{As of Aug 11, 2021, \url{http://exoplanet.eu}.} exoplanets discovered to date have shown that worlds are abundant and diverse. 
Interestingly, most of these exoplanets have been discovered indirectly by observing planet-induced variations in their host stars. 
The transit \citep[][]{Charbonneau2000,Henry1999} and radial velocity \citep[RV,][]{Mayor1995}  methods are the most successful techniques today in terms of number of discovered exoplanets. 
However, the current number of discovered and characterised exoplanets is dwarfed by the estimated number of exoplanets in the galaxy \citep[e.g., ][]{Batalha2014,Petigura2013}. 
Present and future exoplanet search and characterisation instruments, such as \emph{TESS} \citep[][]{Ricker2015}, \emph{CHEOPS} \citep[][]{Broeg2013}, \emph{PLATO} \citep[][]{Rauer2014}, \emph{ESPRESSO} \citep[][]{Pepe2010}, and \emph{SPIRou} \citep[][]{Donati2017}, will provide us with a plethora of photometric and spectroscopic data with exoplanet induced signals waiting to be discovered.

Light curves and RVs are compared with models to infer planetary, orbital, and/or stellar parameters. 
These analyses are generally performed numerically using diverse models combined with computational techniques. 
{Fortunately, the wealth of exoplanetary data has spurred the development of a variety of exoplanet numerical tools. 
To our knowledge, the codes that allow one to analyse jointly} RV and transit data are: \texttt{allesfitter} \citep[][]{allesfitter}, \texttt{EXOFAST} \citep[][]{exofast,exofast2}, \texttt{exoplanet} \citep[][]{exoplanet}, \texttt{Exo-Stricker} \citep[][]{exostricker}, \texttt{juliet} \citep[][]{juliet}, \texttt{MCMCI} \citep[][]{mcmci}, \texttt{PASTIS} \citep[][]{pastis,pastis2}, \texttt{PlanetPAck} \citep[][]{planetpack}, \texttt{TLCM} \citep[][]{tlcm}, and \pyaneti\ \citep[][]{pyaneti}.
These software packages cover a wide range of programming languages and models, and have been used extensively in the literature.

One of the current challenges in exoplanet detection is related to stellar signals in our data. Particularly, RV variations caused by stellar activity jeopardise our ability to detect planet-induced Doppler signals \citep[e.g.,][]{Queloz2001,Rajpaul2016}.
Some methods try to remove stellar activity during the RV extraction to produce activity-free RV times-series \citep[e.g.,][]{CollierCameron2020,Cretignier2020,Rajpaul2020}.
Others attempt to filter the activity induced signals in the RV time-series \citep[e.g.,][]{Barragan2018b,Hatzes2010,Hatzes2011,Pepe2013}.
Gaussian Processes (GPs) have become a widely used tool to model activity induced RVs given their ability to describe stochastic variations \citep[e.g.,][]{radvel,Grunblatt2015,Haywood2014}.
However, the flexibility that GPs offer may also be their major drawback if not used carefully. 
\citet[][]{Rajpaul2015} proposed a method of using spectroscopic activity-indicators together with RVs in order to constrain the activity induced signal in the RV time-series. This can be done by extending the GP approach to a multidimensional GP that exploits the correlations between the different time-series.
This method has proven useful in disentangling planetary and stellar induced signals in RV data with different levels of activity \citep[e.g.,][]{Barragan2019,Mayo2019}.

In this work we present a new version of the multi-planet modelling code \pyaneti\footnote{From the Italian word \emph{pianeti}, which means \emph{planets}.}\ \citep[][]{pyaneti}. 
This updated version of the code focuses on multidimensional GP regression in order to model planetary and activity induced signals in spectroscopic times-series. This new version also allows one to model multi-band transits and single transit events, and features various performance improvements.
This new version of \pyaneti\ has already been used in recent exoplanet characterisation works \citep[e.g.,][]{Carleo2020,Eisner2020b,Eisner2020a,Georgieva2021}.

This manuscript is part of a series of papers under the project 
\emph{GPRV: Overcoming stellar activity in radial velocity planet searches} funded by the European Research Council (ERC, P.I.~S.~Aigrain).
The paper is organised as follows: for the sake of self-completeness, we provide a short recap on Gaussian processes in Section~\ref{sec:gps}. Section~\ref{sec:multigps} describes the multidimensional GPs, with a special emphasis on connecting the activity indicators and the RV time-series. We describe the new implementation of \pyaneti\ in Section~\ref{sec:newpyaneti}. Section~\ref{sec:tests} describes the tests used to validate the code and we conclude in Section~\ref{sec:conlusions}.

\section{A brief overview of Gaussian Processes}
\label{sec:gps}

In this manuscript we do not provide a detailed description of Gaussian processes. Instead, we will provide the basics of GPs needed in order to apply them in data analysis, specifically, in the context of RV and light curve modelling. For further details, we advise the reader to consult specialist literature \citep[e.g.,][]{Rasmussen2006,Roberts2013}.

A \emph{stochastic~process} is a system which evolves in continuous space (in our case time) while undergoing fluctuations.
It is possible to describe the system as a finite set of random variables that are related by a given mathematical entity \citep[][]{Coleman1974}. If the mathematical object that describes the relation between the random variables is a multi-variate normal distribution, then the stochastic process is a \emph{Gaussian~process}.
{
Following \citet[][]{Tracey2018},
a GP assumes that the marginal joint distribution of function values at any finite set of input locations, $\bm{t}={t_{i,(i=1,\cdots,N)}}$, is given by a multi-variate Gaussian distribution } 
\begin{equation}
P(\bm{t}) = \frac{1}{\sqrt{(2\pi)^N |\bm{K}|}}
\exp \left[ 
-\frac{1}{2} \left(\bm{t} - \bm{\mu}\right)^{\rm T} \bm{K}^{-1} (\bm{t} - \bm{\mu})
\right],
\label{eq:multivariate}
\end{equation}

\noindent
where $\bm{\mu}$ is a vector containing mean values, and $\bm{K}$ is a matrix containing the information about the correlation between the variables. 
The only condition about the matrix $\bm{K}$ is that it has to be symmetric and positive semi-definite. 
We note that for a given $\bm{\mu}$ and $\bm{K}$ we can draw an infinite number of curves as random samples of eq.~\eqref{eq:multivariate}. As these curves are not characterised by explicit sets of parameters, GPs are referred as non-parametric functions \citep[see e.g.,][ for more details]{Roberts2013}.

The important aspect is then to find a way to compute our $\bm{\mu}$ and $\bm{K}$ entities that describe a particular GP.
One advantage of GPs being defined over a continuous space is that the mean vector, $\bm{\mu}$, and covariance matrix, $\bm{K}$, can be computed from evaluations of continuous parametric functions at the positions $\bm{t}$. 
Equation~\eqref{eq:multivariate} depends only on $\bm{\mu}$ and $\bm{K}$; therefore, a GP can be fully described with a mean and a covariance kernel function \citep[see][for more details]{Rasmussen2006}.

\subsection{Mean and covariance functions}
\label{sec:kernels}

Mean functions are the deterministic part of a GP. It can be any function $\mu(t;\bm{\phi})$ that depends on a set of parameters, $\bm{\phi}$, and the variable describing the continuous space, $t$. For example, a mean function can be a straight line, a sinusoid, a Keplerian, or a transit model. 

A covariance (also called kernel) function $\gamma(t_i,t_j;\bm{\Phi})$ describes how two locations, $t_i$ and $t_j$, are related according to some parameters $\bm{\Phi}$. 
Such kernel functions can be tuned in order to describe physical/instrumental signals, such as noise, periodicity, long-term evolution, etc.
We describe below some examples of covariance kernel functions widely used in astronomical literature.

\begin{figure*}
    \centering
    \includegraphics[width=0.99\textwidth]{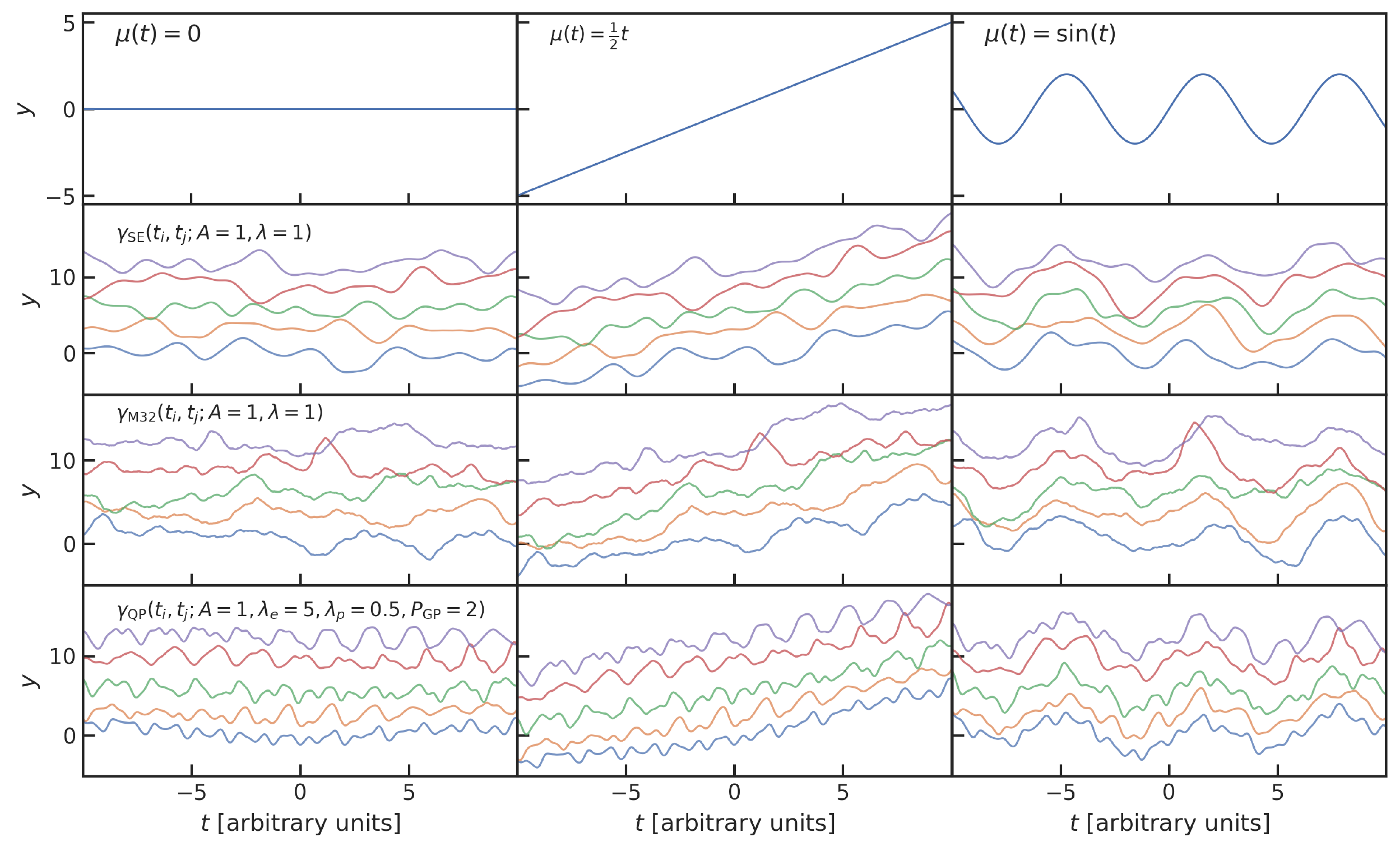}
    \caption{Example of functions generated by GPs with different mean and covariance functions. Left, middle, and right panels corresponds to GPs with mean functions $\mu=0$, $\mu = 1/2\, t$, and  $\mu=\sin(t)$, respectively. Top panel shows a plot with the respective mean function. 
    From top to bottom, the second, third, and fourth panels show five GPs samples from square exponential (with $A=1, \lambda=1$), Mat\'ern 3/2 (with $A=1, \lambda=1$), and Quasi-Periodic kernels (with $A=1, \lambda_e=5$,$\lambda_p=0.5$,$P_{\rm GP}=2$), respectively.  }
    \label{fig:gp_examples}
\end{figure*}

One of the simplest covariance matrix is computed with the white noise kernel
\begin{equation}
    \gamma_{\rm WN}(t_i,t_j) = 
    \sigma_i^2 
    \delta_{ij},
\end{equation}
\noindent
where $\sigma_i$ is the error associated with the datum $i$ and $\delta_{ij}$ is the Kronecker delta. This kernel creates a diagonal covariance matrix, and is used to take into account uncertainties in data.
Another widely used kernel is the squared exponential
\begin{equation}
    \gamma_{\rm SE}({t_i,t_j}) = A^2 \exp \left(
    - \frac{ \lvert t_i - t_j \rvert^2 }{2 \lambda^2}
    \right)
    ,
    \label{eq:se}
\end{equation}
\noindent
where $A$ is an amplitude that works as a scale factor that determines the typical deviation from the mean function, and $\lambda$ is the length scale, which can be interpreted as the characteristic distance for which two points are strongly correlated. 
This kernel generates smooth functions with a typical length scale $\lambda$. Figure~\ref{fig:gp_examples} shows some examples of functions drawn using the $\gamma_{\rm SE}$ kernel and different mean functions. 

Also widely used are the Mat\'ern family of kernels. They are based on the standard Gamma function and the modified Bessel function of second order \citep[see e.g.,][for more details]{Rasmussen2006}. 
Two examples of the Mat\'ern kernels are the Mat\'ern 3/2 Kernel
\begin{equation}
    \gamma_{\rm M32}({t_i,t_j})  = A^2 
    \left(
    1 +  t_{3/2}
     \right)
    \exp \left( -  t_{3/2}  \right)
    ,
    \label{eq:m32}
\end{equation}
\noindent
with $t_{3/2} \equiv \sqrt{3} \lvert t_i - t_j \rvert \lambda^{-1}$, and the Mat\'ern 5/2 Kernel
\begin{equation}
    \gamma_{\rm M52}({t_i,t_j}) = A^2 
    \left(
    1 +  t_{5/2}
     + \frac{ t_{5/2}^2}{3}
     \right)
    \exp \left( - t_{5/2}  \right)
    ,
\end{equation}
\noindent
with $t_{5/2} \equiv \sqrt{5} \lvert t_i - t_j \rvert \lambda^{-1}$. The parameters $A$ and $\lambda$ have the same role as for the Squared Exponential kernel, but in these cases the resulting functions are less smooth. 
Figure~\ref{fig:gp_examples} shows GP samples drawn using the $\gamma_{\rm M32}$ kernel. 

A widely used kernel in astronomy, especially in exoplanet  research, is the Quasi-Periodic (QP) kernel \citep[as defined by][]{Roberts2013}
\begin{equation}
    \gamma_{\rm QP}(t_i,t_j) = A^2 \exp 
    \left\{
    - \frac{\sin^2\left[\pi \left(t_i - t_j \right)/P_{\rm GP}\right]}{2 \lambda_{\rm p}^2}
    - \frac{\left(t_i - t_j\right)^2}{2\lambda_{\rm e}^2}
    \right\}
    \label{eq:qp}
    ,
\end{equation}
\noindent
where $A$ has the same meaning as for the Squared Exponential kernel, $P_{\rm GP}$ is the characteristic period of the GP, \lbp\ the inverse of the harmonic complexity (how complex variations are inside each period), and \lbe\ is the long term evolution timescale (similar to the $\lambda$ for the squared exponential kernel). 
We show some examples of functions created using the $\gamma_{\rm QP}$ kernel in Fig.~\ref{fig:gp_examples}.

Because the QP kernel generates stochastic periodic signal, this choice of covariance function is widely used to model stellar activity signals in both photometry and RVs \citep{Haywood2014,Rajpaul2015}.
In a general context the GP period, $P_{\rm GP}$, can be interpreted as the stellar rotation period, the long term evolution time-scale, \lbe, can be associated with the active region lifetime on the stellar surface; and the inverse harmonic complexity, \lbp, can be associated with the activity regions distribution on the stellar surface \citep[see e.g.,][]{Aigrain2015}.

We note that for the cases in which the GP has a relatively small evolution time scale (\lbe\,$\lesssim P_{\rm GP}$), the periodicity of the GP is practically irrelevant \citep[as pointed out by][]{Rajpaul2015}. Therefore, special care has to be taken when dealing with signals in which the evolution time scale is smaller than the expected periodicity. It is better to use a QP kernel only in cases when $P_{\rm GP}<\,$\lbe. If that is not case, the QP kernel might not be appropriate and some other kernels should be considered. In this work we ensure that when we use the QP kernel, \lbe\,$> P_{\rm GP}$ is satisfied.

Figure \ref{fig:gp_examples} also shows an example of the non-parametric behaviour of the GPs \citep[see][for more details]{Rasmussen2006,Roberts2013}. While in a parametric deterministic model a given set of parameters will give always the same curve, in the non-parametric case, a given set of parameters can give different curves with the condition that the random variables satisfy their intrinsic correlation. Given that the parameters of GPs do not have the same interpretation as for parametric functions, they are often called \emph{hyper-parameters}.

\subsection{Gaussian process Regression}
\label{sec:gpr}

We can perform regression using GPs if we assume that our data (a finite set of variables $\bm{y}$, taken at times $\bm{t}$) are samples of a GP.
The mean function $\mu$ can be a physically motivated parametric model (e.g., Keplerian or transit curves), and the covariance kernel function, $\gamma(t_i,t_j)$, can encompass any intrinsic correlation in our data set (e.g., stellar activity and/or instrumental systematics).

Given that a finite set of variables drawn from a GP is described by a multi-variate normal distribution, we can use this property to write a logarithmic Gaussian likelihood to marginalise over variables as 
\begin{equation}
\ln \mathcal{L}(\bm{\phi},\bm{\Phi}) = 
- \frac{1}{2} \left(
 N_{\rm obs} \ln 2\pi + \ln \lvert \bm{K} \rvert + \bm{r}^{\intercal} \bm{K}^{-1} \bm{r}
\right),
\label{eq:loglikelihood}
\end{equation}
\noindent
where $\bm{\phi}$ and $ \bm{\Phi}$ are the mean and covariance functions parameters, respectively; $\bm{r} = \bm{y - \mu}$ is the vector of residuals of the data $\bm{y}$ and the mean function $\bm{\mu}=\mu({\bm{t})}$ evaluated at the times $\bm{t}$, $\bm{K}$ is the covariance matrix for the observations $\bm{t}$ given a kernel function $\gamma(t_i,t_j)$, and $N_{\rm obs}$ is the number of observations.
Equation~\eqref{eq:loglikelihood} can be optimised or sampled with different numerical techniques in order to infer the parameters of the mean and covariance functions.

GP regression is the cornerstone of the updated version of \pyaneti. Mean functions can be constructed easily with Keplerian and transit models, while any other intrinsic correlation can be absorbed by the covariance function. If we use a physically motivated kernel function, we can also learn about the underlying mechanism that gives rise to the correlation (e.g., stellar activity).
Section~\ref{sec:newpyaneti} describes how GP regression is included inside \pyaneti.

\section{multidimensional Gaussian Processes}
\label{sec:multigps}

We have described how we can use a GP to describe non-parametric functions over a continuous space on $\mathbb{R}$ (in our case time). 
It is possible to extend this idea and to assume that the same GP can describe correlations in multiple continuous spaces that may be related to each other. 

multidimensional (also called multi-variate) GPs provide a solid and unified framework to make the prediction of GPs on $\mathbb{R}^{N}$, where $N$ is the number of dimensions \citep[for more details about the mathematical formalism of multidimensional GPs see e.g.,][]{Alvarez2011,Chen2020}.
An useful application of multidimensional GP is regression.

The general idea of multidimensional GP regression is to transform the multidimensional problem into a ``big'' one-dimensional GP regression.
This is done by vectorising the set of observations $[\bm{t},\bm{y}]_i$ (where $i = 1,\cdots, N$) as a big vector in $\mathbb{R}$. The residual vector $\bm{r}$ is created as a concatenation of residuals vectors $\bm{r}_i$, for each dimension $i$.
The covariance matrix $\bm{K}$ is constructed of small sub-matrices $\bm{k}^{l,m}$ that describe the correlations between the different dimensions $l$ and $m$.
This allows one to reformulate the multidimensional GP regression as a conventional GP regression (see Sect.~\ref{sec:gpr}).
In the remainder of this section we will describe how we can use multidimensional GPs to model activity induced signals in spectroscopic time-series.

\subsection{multidimensional GPs for spectroscopic time-series}
\label{sec:mgpforrv}

\citet[][]{Rajpaul2015} proposed a framework to model stellar activity in RV time-series simultaneously with the activity indicators using a multidimensional GP approach.
This approach assumes that the stellar induced signals in all observables can be described by the same latent GP and its time derivative.
\citet[][]{Jones2017} and \citet[][]{Gilbertson2020} expanded the work of \citet[][]{Rajpaul2015} by adding higher derivatives and generalising the work to a generic set of activity indicators. In this work we also generalise the work of \citet[][]{Rajpaul2015} to a generic set of activity indicators, but we maintain the approach of using only the first GP derivative. 

\subsubsection{Physical motivation}

It has been shown that there is an intrinsic relation between the area covered by active regions on the stellar surface and the stellar-induced RV variations \citep[e.g.,][]{Aigrain2012,Boisse2009}. Following this idea, we can assume that we can describe a function, $G(t)$, which is a latent unobserved variable that represents the projected area of the visible stellar disc that is covered by active regions as function of time. 
Such active regions affect photometric and spectroscopic observed parameters (e.g., RVs or activity indicators) in different ways. 
Some of them are only affected by the projected area that is covered by active regions, i.e. they can be described as $G(t)$ with some scale factor; others are also affected by how these regions evolve in time on the stellar surface, i.e., they are described by $G(t)$ and its time derivatives ($\dot{G}(t)$, $\ddot{G}(t)$, etc.).

RVs are affected by the position of the active regions on the stellar surface and how these regions evolve in time \citep[see e.g.,][]{Dumusque2014}. Therefore, in our assumption that $G(t)$ describes the area covered by active regions of the visible stellar disc, activity-induced RV data can be described by $G(t)$ (e.g., to account for the convective blue-shift) and its time derivatives (to account for the evolution of the spots on the stellar surface).
Some activity indicators, such as \logr (flux of the \ion{Calcium}{II} H~\&~K lines relative to the bolometric flux) or $S_{\rm HK}$ (flux of the \ion{Calcium}{II} H~\&~K lines relative to the local continuum), are only affected by the fraction of the area that is covered by the stellar surface \citep[see e.g.,][]{Isaacson2010,Thompson2017}, i.e., they can be described only by $G(t)$. 
Some other activity indicators, such as the bisector inverse slope (BIS), are also affected by how the active regions evolve on the stellar surface \citep[see e.g.,][]{Dumusque2014}, requiring higher time derivatives of $G(t)$ in order to describe them. 

A set of contemporaneous time-series that contain activity-induced signals of a given star (RVs, \logr, etc.) can then be modelled simultaneously assuming they are described by the same underlying function $G(t)$ and its derivatives. 
We can assume that our function $G(t)$ is generated by a GP, given its flexibility to model stochastic signals and the property that any affine operator (including linear and/or derivative) applied to a GP yields another GP \citep[see][and references therein]{Rajpaul2015}.
We also note that, if we assume that our underlying function comes from a GP, we can use a multidimensional GP approach to exploit the correlation between observations in the different time-series, assuming each one is a different continuous space described by the same underlying GP-drawn function.
The advantage of this approach is that RV time-series contain stellar activity and planet-induced variations, while activity indicators are only sensitive to activity. Therefore, activity indicators can help to constrain the activity induced signal in RV time-series, thus allowing one to disentangle the planetary signals from the activity signals \citep[][]{Rajpaul2015}.

\subsubsection{Theoretical approach}
\label{sec:mgpr}

We follow the approach described by \citet[][]{Rajpaul2015} and we will assume that we have a set of $N$ time-series, $\mathcal{A}_{i=1,...,N}$, each one with $M$ points. 
Although this is not necessary for the multidimensional GP approach, we make this assumption given that for spectroscopic time-series the RVs and activity indicators are computed from the same spectra.
A set of $N$ time-series that are characterised by the same GP-drawn function $G(t)$ and its derivative $\dot{G}(t)$ can be described as
\begin{equation}
\begin{matrix}
 \mathcal{A}_1 =  A_{1} G(t) + B_{1} \dot{G}(t) \\
\vdots \\
\mathcal{A}_N =  A_{N} G(t) + B_{N} \dot{G}(t), \\
\end{matrix}
\label{eq:gps}
\end{equation}
\noindent
where the variables $A_{1}$, $B_{1}$, $\cdots$, $A_{N}$, $B_{N}$, are free parameters which relate the individual time-series to $G(t)$ and $\dot{G}(t)$. 
We note that each $\mathcal{A}_i$ is a GP by itself, given the property that any derivative operator applied to a GP and the sum of two GPs is also a GP.
We can assume that the GP has zero mean because in the GP regression (see Sect.~\ref{sec:gpr}) we use the residuals vector to evaluate the likelihood, i.e., we remove the mean function from the observations. 

To create the covariance matrix, we need to define how points are correlated between all time-series $\mathcal{A}_i$ and $\mathcal{A}_j$.
Following \citet[][]{Rajpaul2015}, the covariance between two observations at times $t_i$ and $t_j$ between the time-series $\mathcal{A}_l$ and $\mathcal{A}_m$ is given by
\begin{equation}
\begin{aligned}
k^{l,m}({i,j})  = & \, A_l A_m \gamma^{G,G}({i,j}) + B_l B_m  \gamma^{dG,dG}({i,j}) \\
 & + A_l B_m \gamma^{G,dG}({i,j}) + A_m B_l \gamma^{dG,G}({i,j}),
 \label{eq:smallks}
\end{aligned}
\end{equation}
\noindent
where $\gamma^{G,G}({i,j})$ denotes the covariance between (non-derivative) observations of $G$ at times $t_i$ and $t_j$; $\gamma^{G,dG}({i,j})$ refers to the covariance between an observation of $G$ at time $t_i$ and an observation of $\dot{G}$ at time $t_j$; $\gamma^{dG,G}({i,j})$ refers to the covariance between an observation of $\dot{G}$ at time $t_i$ and an observation of ${G}$ at time $t_j$; and $\gamma^{dG,dG}({i,j)}$ denotes the covariance between two observations of $\dot{G}$ at times $t_i$ and $t_j$. 
In appendix~\ref{ap:derivatives} we show the gamma terms ($\gamma^{G,G}$, $\gamma^{dG,G}$, $\gamma^{G,dG}$, and $\gamma^{dG,dG}$) for the  squared exponential, Mat\'ern 5/2, and QP kernels.

The ``big'' covariance matrix $\bm{K}_{\rm big}$ that describes the covariance between all the $N$ time-series is 
\begin{equation}
    \bm{K}_{\rm big} =
    \begin{pmatrix}
    \bm{k}^{1,1} & \bm{k}^{1,2} & \cdots & \bm{k}^{1,N}  \\
    \bm{k}^{2,1} & \bm{k}^{2,2} & \cdots & \bm{k}^{2,N}  \\
    \vdots & \vdots & \ddots & \vdots  \\
    \bm{k}^{N,1} & \bm{k}^{N,2} & \cdots & \bm{k}^{N,N}  \\
    \end{pmatrix}
    ,
    \label{eq:bigk}
\end{equation}
\noindent
where each $\bm{k}^{l,m}$ is computed using eq.~\eqref{eq:smallks} for any covariance function $\gamma$.
If $\gamma$ is a valid kernel function, the matrix $\bm{K}$ is a valid covariance matrix that we can use in GP regression. This matrix also has the property of being symmetric and positive definite, therefore, we need to compute only the upper triangle part of the matrix, while the lower panel matrices can be computed as $\bm{k}^{j,i} = (\bm{k}^{i,j})^{^{\intercal}}$.

At this point we have all the mathematical framework needed to perform multidimensional GP regression for RVs and activity indicators using the GP regression described in Sect.~\ref{sec:gpr}. We can create a residual vector $\bm{r}_i$ for each dimension $i$, e.g., the residuals corresponding to the  RVs can be computed by subtracting Keplerian models, while the residuals of an activity indicator can be computed by subtracting constant offsets. 
We then create the residual vector $\bm{r}$ by concatenating the $\bm{r}_i$ for each dimension.
We can create the ``big'' covariance matrix $\bm{K}_{\rm big}$ using eq.~\eqref{eq:bigk} with any valid kernel (e.g. the ones in Sect.~\ref{sec:kernels}). Once we have $\bm{r}$ and $\bm{K}_{\rm big}$ we can use the likelihood given by eq.~\eqref{eq:loglikelihood} to infer the parameters of our multidimensional GP using our preferred numerical method. Section~\ref{sec:newpyaneti} describes how this framework is included inside \pyaneti.

\subsection{Comparison between multidimensional GP and other approaches}
\label{sec:gpcomparisons}

{
A common approach in the literature to describe stellar signals in RV time-series consists of modelling ancillary time-series (these can be light curves or activity indicators) with a GP to infer the hyper-parameters for a given kernel. The hyper-parameters inferred from one or more ancillary time series are then used to inform hyper-parameters priors for the GP modelling of the RV data using the same kernel \citep[see e.g.,][]{Grunblatt2015,Haywood2014}. 
The process of retrieving kernel hyper-parameters for a GP modelling is known as \emph{training} a GP, and we therefore refer to this approach as the "training-GP" approach. Training the GP in this way ensures that the stellar activity signal in the RVs is modelled with the same characteristic features --such as periodicity, degree of smoothness, characteristic evolution timescale-- as the ancillary time-series. In this approach, the functions describing the ancillary time-series and the RVs share the similar covariance properties, but they are otherwise independent of each other, and their shapes are entirely unrelated. 

A slight variation on the training-GP approach involves  modelling the ancillary time-series and the RVs simultaneously, using independent GPs sharing the same covariance function \citep[see e.g.,][]{Osborn2021,Suarez2020}. In this case, both the RVs and the ancillary time-series are used to constrain the GP hyper-parameters, and the functions describing them share exactly the same covariance properties, but they are still independent of each other and have unrelated shapes. 

These approaches make weaker assumptions about the relationship between the ancillary time-series and the RVs than those made in the multidimensional model developed by \citet{Rajpaul2015} and implemented in \pyaneti. They result in more flexible models for the RV time-series, with the associated risk of over-fitting (where potential planetary signals can be absorbed or modified by the activity model). On the other hand, our assumptions are based on a fairly simplistic toy model, whose limitations will no doubt become apparent once the framework is applied to a large enough sample of high-precision datasets. Such failures should however be easy to diagnose, as they would lead to a poor fit to the data.
}
In Section~\ref{sec:toy2} we present a comparison of planetary signal recovering between the ``training-GP'' method described in this section and the multi-GP approach.

It is also important to note that, in our multidimensional GP model, the functions used to describe the activity signal in the RVs have different covariance properties from those used to model the activity indicators. The fact that the RVs and activity indicators are modelled as different linear combinations of the underlying GP and its time-derivative results in markedly different harmonic complexity, for example (see Sect.~\ref{sec:gpderivatives} for a more detailed discussion). 
While we are not aware of examples in the literature, the "training-GP" approach could be generalised to take into account the derivatives of a GP to describe time-series. In Sect.~\ref{sec:run2} we show an example on how to train a GP to use the hyper-parameters to model RVs taking into account the derivative of the chosen kernel.
A more complete quantitative comparison between different methods to model stellar signals is given by \citet{Ahrer2021}.

\subsection{On the usefulness of the GP derivatives}
\label{sec:gpderivatives}

In this section we describe the importance of taking into account the derivatives of the GP to model RV data when assuming that our GP generates a function that describes the surface covered by active regions on the stellar surface. 
We base our discussion on the QP kernel, but the conclusions can be extended to other kernels.

Let us suppose we have a 2-dimensional GP to describe two time-series, $S_1$ and $S_2$, that behave as 
\begin{equation}
\begin{aligned}
S_1 = &  G(t),  \\
S_2 = & \dot{G}(t). \\
\end{aligned}
\label{eq:sgps}
\end{equation}

\begin{figure*}
    \centering
    \includegraphics[width=0.99\textwidth]{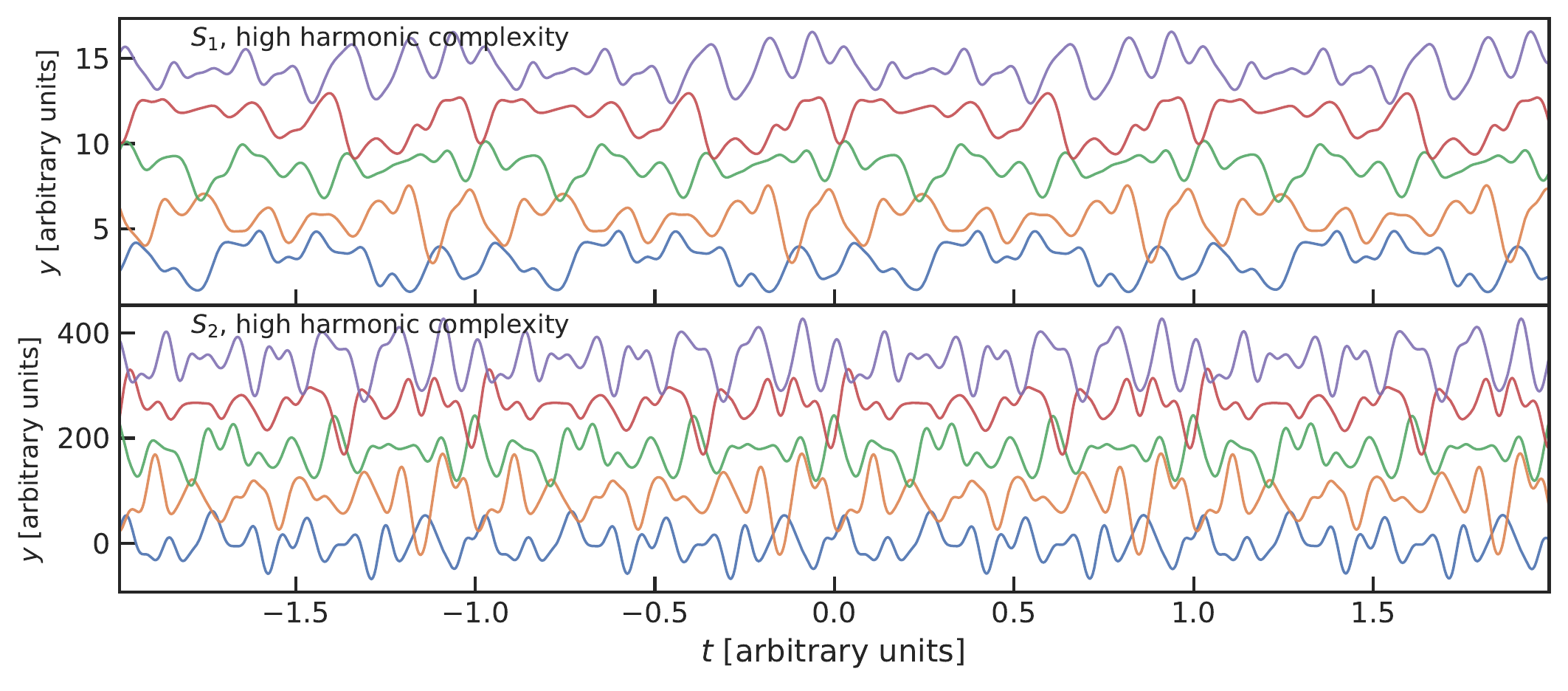}\\
    \includegraphics[width=0.99\textwidth]{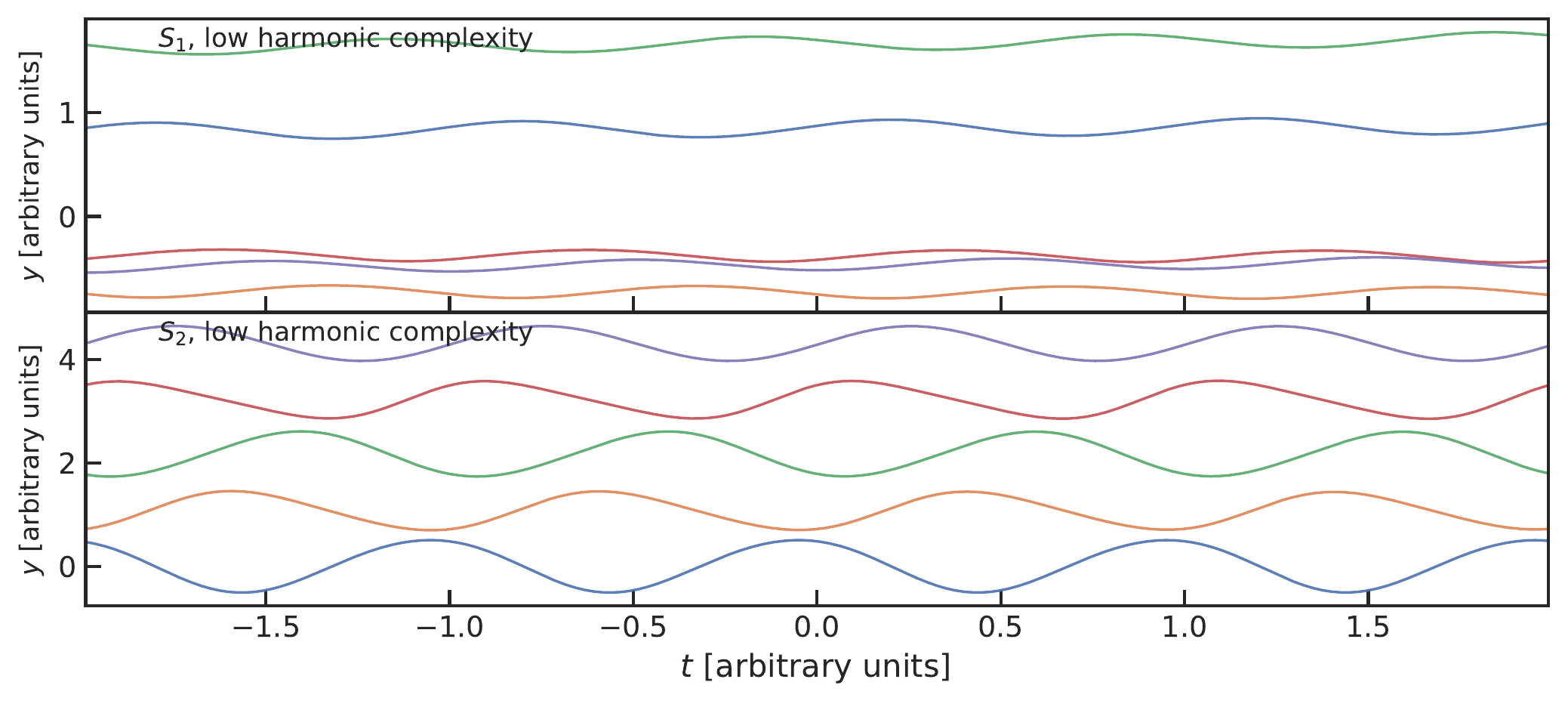}\\
    \caption{Example of samples from a multidimensional GP (two dimensions in this case) created with a QP kernel with $P_{\rm GP} = 1$, \lbe$=10$ and high (\lbp$=0.1$, top pannel) and low \lbp$=10$, lower pannel) levels of harmonic complexity. Each panel shows two subplots with samples of $S_1$ and $S_2$ time-series (See eq.~\ref{eq:sgps}). Corresponding $S_1$ and $S_2$ draws from the same sample are shown with the same colour.}
    \label{fig:example_multigp}
\end{figure*}

Equation~\eqref{eq:sgps} was computed from equation~\eqref{eq:gps} with $A_1=B_2=1$, and $A_2=B_1 = 0$. 
Figure~\ref{fig:example_multigp} shows some samples of $S_1$ and $S_2$ time-series that were created using a QP kernel with $P_{\rm GP} = 1$, \lbe\,$= 10$, and different values of \lbp.

From the examples in the top panel of  Fig.~\ref{fig:example_multigp} we can see that if the $S_1$ signal has a high harmonic complexity, then the contemporaneous $S_2$ would have an apparently higher harmonic complexity. 
We can see that this behaviour is expected from the derivatives of the QP kernel (see Appendix~\ref{ap:derivatives}). From equations \eqref{eq:qpgdg} and \eqref{eq:qpdgdg} we can see that when \lbp\,$\lesssim 1$ there are some ``$\tau$ terms'' (terms that include the $\tau$ parameter) that add extra ``wiggles'' to the behaviour of the $S_2$ curve.
This has a direct implication when training GPs with ancillary observations (that may behave as $S_1$) to model RVs (that may behave as $S_2$). Setting a prior on \lbp\ for the $S_2$ signal based on our $S_1$ signal may lead to biased results. 

The previous discussion has special implications when training GPs to model RVs using light curves. In general, light curves and RV data are not taken simultaneously, so the active regions on the stellar disc might not be the same between the two data sets \citep[see e.g.,][]{Aigrain2012,Barragan2021}. But even if they were, the harmonic complexity extracted from a light curve may not be the same as for the activity induced RV signal if modelled only with $G(t)$, i.e., without accounting for the GP derivatives. 
There are examples of the importance of using the GP derivative when modelling RVs with high harmonic complexity \citep[e.g.,][]{Barragan2019,Barragan2021b}.

In the other extreme of low harmonic complexity (\lbp$\gg 1$) we see that $S_1$ and $S_2$ behave as quasi-sinusoidal signals (lower panel of Fig.~\ref{fig:example_multigp}). If we consult equations \eqref{eq:qpgdg} and \eqref{eq:qpdgdg}, we can see that in the limit when \lbp\,$\gg 1$, all $\tau$ terms are irrelevant and the behaviour of the GP and derivatives is quasi-sinusoidal with a long-term evolution regulated by \lbe.
In this case using the derivative of the GP to model the RVs may not be crucial for modelling the activity induced signal. There are some examples in the literature of activity induced RV signals that behave similar to the activity indicators in the low harmonic complexity regime (e.g., Serrano et al., submitted.).

We also note that the fact that RVs depends on how active regions evolve in the stellar surface may cause an apparent asynchrony with the activity indicators.
This phenomenon of RV and activity indicators being out of phase has been reported in the literature \citep[e.g,][]{CollierCameron2019}.
If the stellar signal induced in the RVs depends on a combination of the kind $A G(t) + B \dot{G}$, this can generate curves that may seem similar to the one of the activity indicator, that may vary as $A G(t)$, but with an apparent phase shift. 
This may be most applicable to the low harmonic complexity case in which signals tend to look more sinusoidal.
For example, in \cite{Georgieva2021} the stellar signal in the RV and activity indicators time-series are similar, but the RV curve seems to have a different phase (seemingly ahead) of the contemporary activity indicator time-series. In the multidimensional GP approach the time derivatives of $G(t)$ allow this behaviour to be accounted for.

We have discussed the importance of the GP derivatives based on the QP kernel. 
However, any other kernel that has strong variations on short time scales may suffer from the same effect in which the derivatives of the GP are relevant to model RV time-series.
We note that this topic needs to be explored further (e.g., Nicholson et al., in prep.), but a detailed description of this issue is beyond the scope of this paper. 
We mention it however, to apprise the reader of the usefulness of including the derivatives of the GP when modelling RV time-series with \pyaneti.

\subsection{Limitations of the multidimensional GP approach}

We warn the reader that the multidimensional GP approach should not be taken as a magic recipe that will always solve our stellar induced RV variation problems. 

The first limitation is that this framework assumes a relation between all time-series via a function $G(t)$ and its time derivatives. This may work in many cases, but we warn that this is still only a first order approach to the problem. Stellar activity signals may be more complicated than the underlying model assumed for this framework. Therefore the multidimensional GP approach may lead to activity signals being fit imperfectly. 

Problems may also arise if the data sampling is not optimal.
The main idea behind GP regression is to exploit the correlation between our observations.
This implies that if observations have time separations larger than the GP time-scales that we want to characterise, then the GP model may be poorly constrained. Therefore, special care has to be taken when using GPs to modelling data with sparse sampling. 
In Sect.~\ref{sec:citlalatonac} we show and describe \citlalatonac, a code that simulates spectroscopic-like time-series that can be helpful to plan RV observation campaigns for active stars.

It is also worth mentioning that a downside of GP regression is the computational cost of matrix inversion.
This is particularly relevant to the multidimensional GP approach where the dimension of the matrix to invert increases with the number of time-series to model. However, in the case of spectroscopic time-series, there are rarely more than several hundred observations per star. Therefore, the multidimensional GP regression to model RVs and activity indicators remains computationally treatable.  

\section{The new \texttt{PYANETI}} 
\label{sec:newpyaneti}

\citet[][]{pyaneti} describes the data analysis approach taken by \pyaneti. Briefly, \pyaneti\ uses a implementation of a Markov chain Monte Carlo (MCMC) sampler based on \texttt{emcee} \citep[][]{emcee} to create marginalised posterior distributions for exoplanet RV and transit parameters. The code uses a built-in Gaussian likelihood together with user-input priors for each sampled parameter. 
The demanding computational routines are written in \texttt{FORTRAN} and the subroutines are wrapped into python using the \texttt{f2py} tool included in the \texttt{numpy} \citep[][]{numpy} package.

The new version of \pyaneti\ is an extension of the original package presented in \citet[][]{pyaneti}. 
The biggest update is the generalisation of the Gaussian likelihood that includes correlation using a GPR following eq.~\eqref{eq:loglikelihood}.
The code can also perform multi-band transit  and single transit modelling.
The rest of this section describes in more detail the new additions to the code.

\subsection{Gaussian Processes}

\citet[][]{pyaneti} describes the functions used to model multi-planet signals in both RV and transit data sets. These equations are used as mean functions of the GP to compute the residual vector $\bm{r}$, allowing \pyaneti\ to perform multi-planet fits with RV and/or transit data together with GP regression using eq.~\eqref{eq:loglikelihood} and user input priors. 
The elements of the covariance matrix inside \pyaneti\ are created as
\begin{equation}
    K(t_i,t_j) = \gamma \left( t_i, t_j \right) + 
    \left(\sigma_i^2 + \sigma_{{\rm jitter}}^2\right) 
    \delta_{ij},
    \label{eq:covarianceposterior}
\end{equation}
\noindent
where $\gamma(t_i,t_j)$ is any valid kernel function, $\delta_{ij}$ is the Kronecker delta, $\sigma_i$ the white noise associated with the datum $i$, and $\sigma_{{\rm jitter}}$ is a jitter term.
We note that the same jitter term can be shared by a collection of points with the same underlying systematics, e.g., a jitter term associated to a common instrument.
We note that in the case where no correlation is assumed in the data, i.e. $\gamma(t_i,t_j) = 0$, eq.~\eqref{eq:loglikelihood} reduces to the white noise Gaussian likelihood implemented in the previous version of \pyaneti\ \citep{pyaneti}.
We have implemented in \pyaneti\ all kernels described in Sect.~\ref{sec:kernels}, but more can be added easily if needed. 

We also incorporated the multidimensional GP approach described in Sect.~\ref{sec:multigps} into \pyaneti\ within the RV modelling routines. 
The new version of \pyaneti\ can reproduce the original approach given by \citet[][]{Rajpaul2015}, but also allows one to combine arbitrarily many time-series to use multiple activity indicators.
The residual vector for the RV data is computed subtracting Keplerian signals, and for the activity indicators with a constant offset. We note that \pyaneti\ can deal with different instrumental offsets for the RV and activity indicators.

To create the ``big'' covariance matrix, eq.~\eqref{eq:bigk}, we define the sub-matrices $k^{l,m}_{\rm wn}({i,j})$ to account for white noise as
\begin{equation}
    k^{l,m}_{\rm wn} (i,j) = k^{l,m}(i,j) +    \left(\sigma_{i,l}^2 + \sigma_{{\rm jitter},l}^2\right)
    \delta_{ij} \delta_{lm},
    \label{eq:smallkswn}
\end{equation}
\noindent
where $k^{l,m}(i,j)$ is computed using eq.~\eqref{eq:smallks} and a valid covariance function, $\sigma_{i,l}$ and $\sigma_{{\rm jitter},l}$ are the white noise and jitter term associated to the dimension $l$; and $\delta_{ij}$ and $\delta_{lm}$ are Kronecker deltas. The $\delta_{lm}$ between two different dimensions, $l$ and $m$, ensures that the nominal errors are only added in the diagonal of the ``big'' matrix.
\pyaneti\ creates the $\bm{K}_{\rm big}$ covariance matrix using the sub-matrices computed  with eq.~\eqref{eq:smallkswn}. The code has built-in routines to perform multidimensional GP regression using the squared exponential (eq.~\ref{eq:se}), Mat\'ern 5/2 (eq.~{\ref{eq:m32}}), and QP (eq.~\ref{eq:qp}) Kernels. Appendix~\ref{ap:derivatives} show the derivatives included into \pyaneti\ to compute eq.~\eqref{eq:smallks} for these covariance functions. 

\subsubsection{Dimensionality problem}

We note that \pyaneti\ suffers from the high computational cost of GP regression, which in general entails matrix inversion. 
The computational time of a matrix inversion scales as $\mathcal{O}(N^3)$, where $N$ is the number of data points. 
This is generally not a problem for RV data sets, which include relative few observations (usually less than 1000). However, it becomes a problem when modelling light curve time-series with thousands of observations. 

Fortunately, in the case of light curves, the time-scales of the planetary transits are small compared with those associated with stellar variability.
This makes it relatively easy to remove such trends from light curves. This \emph{detrending} is a common approach in the literature when one is only interested on modelling transit signals in flattened light curves \citep[see e.g.,][]{Hippke2019}. 
A reason for fitting GPs to full light curves with transits would be to argue that the light curve might provide information on GP hyper-parameters that can also be used simultaneously with RV data. 
Yet this may not be optimal because usually light curves and RVs are not observed simultaneously, and there is evidence that light curves and RVs may not be constrained by the same time scales, as discussed in Sect.~\ref{sec:gpderivatives}. 
Since the GP implementation of \pyaneti\ is mainly focused on the RV analysis, we did not explore GP computation acceleration for light curve analyses. 
We note that \pyaneti\ could still be used to model binned light curves in order to try to estimate hyper-parameters of the light curve, but a GP modelling of the light curve including transits is not advisable with \pyaneti.

{
We note that progress has been made in fast matrix inversion for GP regression. We considered implementing the GP regression operations included in \pyaneti\ using the \texttt{george} package \citep{george}; this would require explicitly coding the derivative kernels we use to model each time-series in \texttt{george}, a feasible but non-trivial endeavour. It has not been necessary to do it so far, given the typical size of the datasets we are modelling, but it is something to be considered for a future implementation of \pyaneti. Another, even faster option for GP regression on large datasets is the \texttt{celerite} package \citep{celerite}, but that is not suitable for the present work as it is restricted to 1-dimensional datasets.}

\subsection{Multi-band fit}

Multi-band photometric follow-up of transiting planets has become common. This is because the number of ground and space-based instruments has increased, and more transiting planets are found around relatively bright stars.
For this reason, we have added multi-band transit modelling into \pyaneti. The code solves for the same orbital parameters for all bands, but independently samples the wavelength dependent parameters, i.e., limb darkening coefficients, cadence and integration time, and scaled planet radius.

As in the previous version, the code uses the \citet[][]{Mandel2002} equations to model transits by assuming the star limb darkening can be modelled as a quadratic law. 
The code samples for two limb darkening coefficients for each band following the $q_1$ and $q_2$ parametrization described in \citet[][]{Kipping2013}. 
The code also allows for a different cadence and integration time for each band \citep[see][]{Kipping2010}. 
Finally, the code also allows modelling for the same scaled planet radius $R_{\rm p}/R_\star$ for all bands, as well as an independent $R_{{\rm p},i}/R_\star$ for each band $i$. The latter can be useful to test false-positive scenarios \citep[e.g.,][]{Parviainen2019}, or to fit transit depths at different wavelengths as used in transmission spectroscopy \citep[e.g.,~][]{Charbonneau2002}.

\subsection{Single transit fit}

Single transit events can be caused by transiting planets with periods longer than the observational window. Fortunately, they can be detected by methods that do not rely on periodicity of transit-like events \citep[see e.g.][]{Eisner2020b,Osborn2016}.

The problem when dealing with mono-transits is that the period and the semi-major axis cannot be determined. 
These parameters are important because they determine the velocity at which the planet moves during the transit. 
In order to solve this, in the new version of \pyaneti\, we fix a period to a dummy value larger than our observing window, and we sample for a dummy scaled semi-major axis $a_{\rm dummy}$. While $a_{\rm dummy}$ does not have a physical sense, but it ensures that the transit shape is sampled.

Therefore, in order to model a single transit, \pyaneti\ samples the time of mid-transit $T_0$, impact parameter $b$, scaled planet radius $R_{\rm p}$, $a_{\rm dummy}$, and limb darkening coefficients $q_1$ and $q_2$ \citep[][]{Kipping2013}. 
With these parameters we can estimate the transit duration, and if we assume that the planetary orbit is circular, we can estimate the orbital period albeit with relatively large uncertainty \citep[see][]{Osborn2016}.




\subsection{\texttt{citlalicue} and \texttt{citlalatonac}}
\label{ap:citla}

We have created codes to create synthetic stellar photometric and spectroscopic time-series, called \citlalicue\ and \citlalatonac, respectively\footnote{In Aztec mythology, Citlalicue (goddess) and Citlalatonac (god) are the creators of the stars. The words root, \emph{Citlali}, is the Nahuatl word for star.}.

\subsubsection{\citlalicue: the light curves creator}
\label{sec:citlalicue}

\citlalicue\ is a \texttt{Python} module that allows one to create synthetic light curves. It is totally independent of \pyaneti. It can be easily installed using \texttt{pip install citlalicue}.
The module has a class called \texttt{citlali} that contains all the attributes and methods needed to create a synthetic stellar light curves with transits, periodic modulation, and white noise.
The current version of the code uses a QP kernel (eq.~\ref{eq:qp}) to simulate stellar variability, and it allows one to create transits for any number of planets using \texttt{pytransit} \citep{pytransit}. 
An example of how to create a light curve using \citlalicue\ is given \href{https://github.com/oscaribv/citlalicue/blob/master/example_light_curves.ipynb}{here \faGithub}.

\citlalicue\ also has a class called \texttt{detrend} that allows one to detrend light curves using GPs. \texttt{detrend} takes a plain-text file with light curve data containing time and flux (and errors as input if available). 
The code allows  one to mask out the transits from the data as well as to fit simultaneously for the transits and GP. The former is recommended given that it is faster.
The code uses \texttt{george} \citep[][]{george} to perform a fast GP regression that enables the modelling of the variability in the light curve.
The code allows for an iterative optimisation with a sigma clipping algorithm, where the threshold can be tuned by the user (the default is 5). Once the optimal model is found by the code, the inferred trend is removed from the light curve, creating a flattened signal with transits.
An example of how to detrend a light curve using \citlalicue\ is available \href{https://github.com/oscaribv/citlalicue/blob/master/example_detrending.ipynb}{here \faGithub}.
Examples of the detrending capability of \citlalicue\ can be found in e.g., \citet[][]{Barragan2021} and \citet{Georgieva2021}.

\subsubsection{\citlalatonac: the spectroscopic time-series creator}
\label{sec:citlalatonac}

The \texttt{Python} package \citlalatonac\  uses \pyaneti\ in order to create synthetic spectroscopic (RVs and activity indicators-like) time-series. 
This package comes together with \pyaneti\ when the latter is cloned directly from its GitHub repository.
The code creates samples of a multidimensional GP following eq.~\eqref{eq:gps}. This simulates spectroscopic-like signals (RVs and activity indicators) of an active star, assuming they all are generated by the same underlying GP.

The main class of the package is named \texttt{citlali}. When \texttt{citlali} is called in \texttt{Python}, the user needs to specify the time range in which the synthetic data will be created (e.g., this range can be an observing season), the number of time-series to create, the amplitudes of the signals, following eq.~\eqref{eq:gps}, the kernel to use, and the kernel parameters.  
By default the first time-series is called \texttt{rv} and it is always treated as RV-like, i.e., the planet-induced signals are added to this time-series only.
The class includes methods that allow one to add as many planets as needed, as well as white and red noise.

The package also includes the \texttt{create\_real\_times} function that allows one to create realistic sampling of targets at a given observatory using \texttt{astropy} \citep[][]{astropy1,astropy2}. 
This utility can be useful for estimating the number of data points needed in order to measure the Doppler semi-amplitude of a given target, even if the star is active. Such numbers can be valuable while writing a telescope proposal.
A practical example of how to use \citlalatonac\ to create synthetic time-series of a target observed at a given observatory can be found \href{https://github.com/oscaribv/pyaneti/blob/master/pyaneti_extras/synthetic_k2100.ipynb}{here \faGithub}.

\section{Tests}
\label{sec:tests}

\subsection{Recovering multi-GP hyper-parameters}
 \label{sec:toy1}

\begin{figure*}
    \centering
    \includegraphics[width=0.99\textwidth]{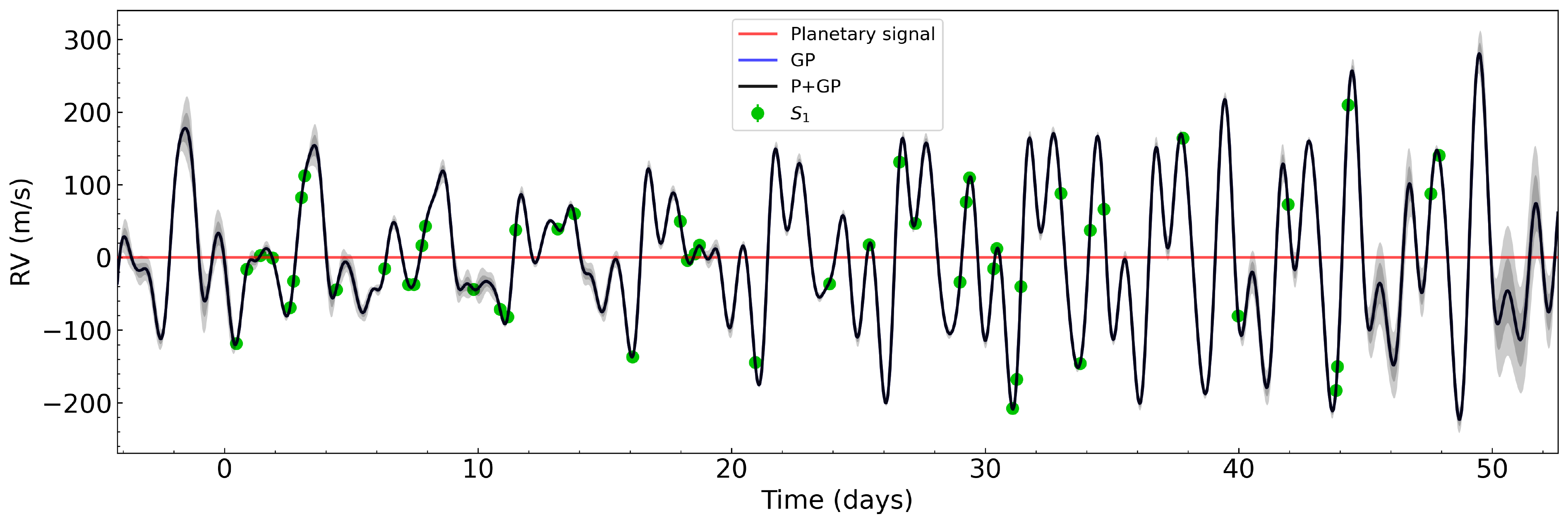}
    \includegraphics[width=0.99\textwidth]{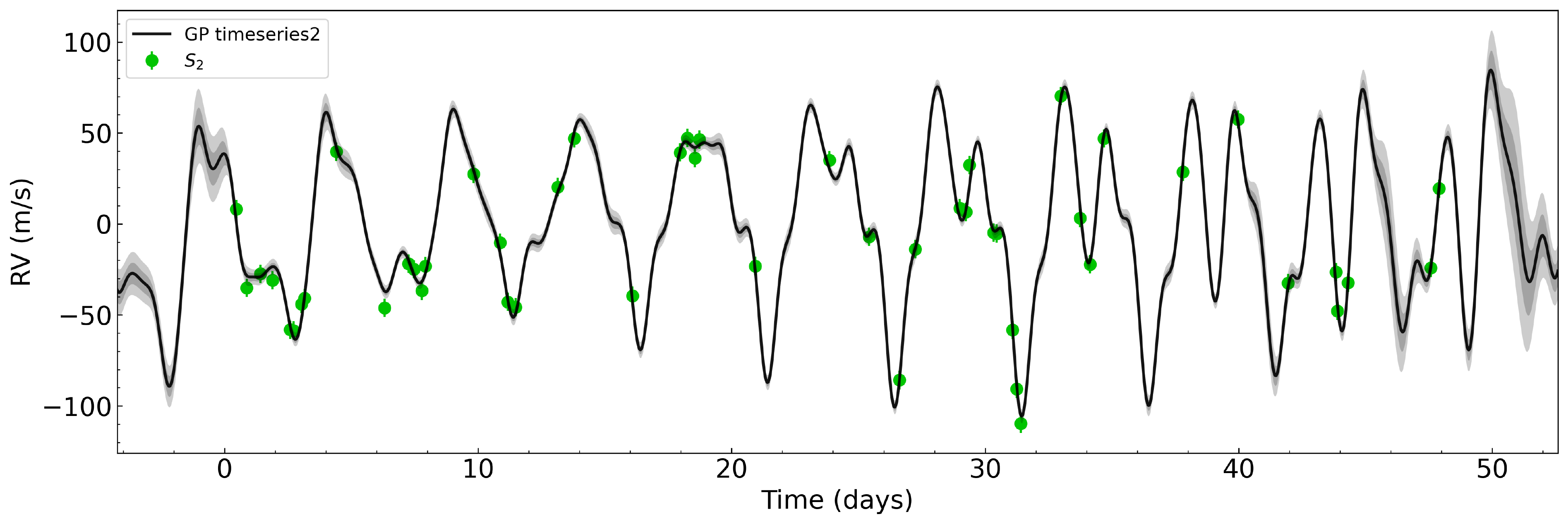}
    \includegraphics[width=0.99\textwidth]{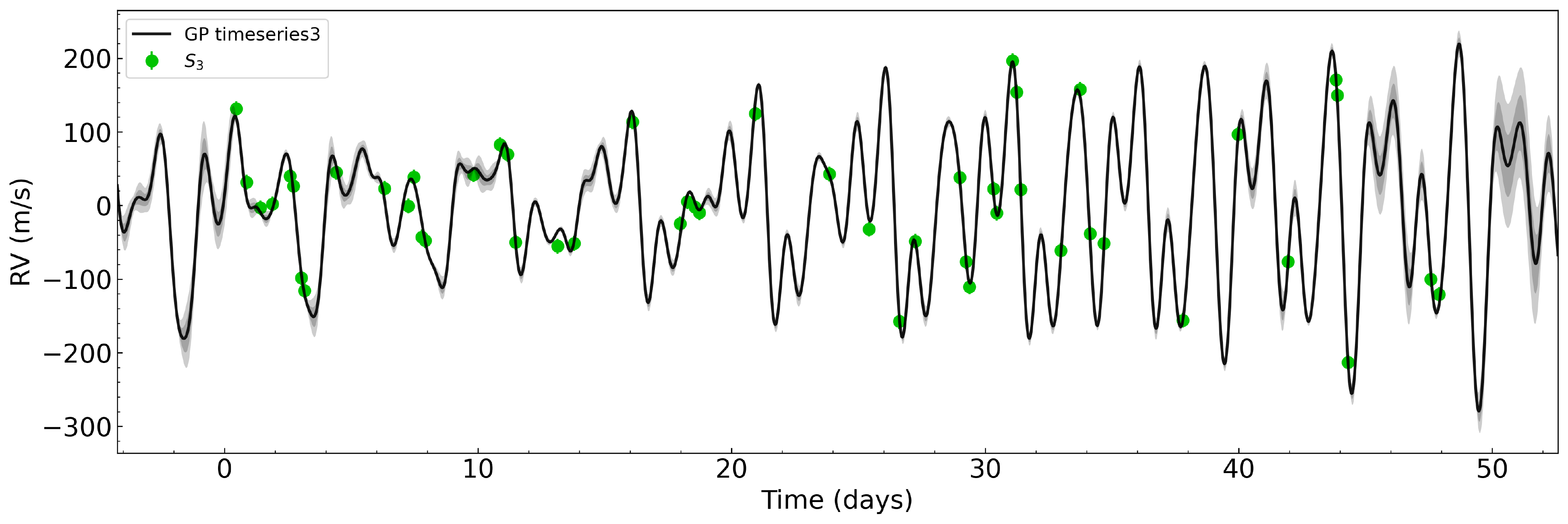}
    \caption{
    $S_1$, $S_2$, and $S_3$ time-series described in Sect.~\ref{sec:toy1}.
    The green markers in each panel represent the synthetic observations with inferred offsets extracted. 
    The solid dark lines shows the inferred mean of the predictive distribution of our multidimensional GP, with dark and light shaded areas showing the one and two sigma credible intervals of the corresponding GP model, respectively. These plots are shown as they are provided by \pyaneti. The code assumes that the first time-series is RV-like, and it may contain Keplerian signals that are represented with a red curve, in this case no Keplerian signals are included and the curve appears as a horizontal red line at 0.
    }
    \label{fig:toymodel}
\end{figure*}

We created a set of synthetic spectroscopic-like time-series using \citlalatonac\ (see Appendix~\ref{ap:citla}) in order to test the ability of \pyaneti\ to recover parameters using a multidimensional GP.
We assume that we have 3 time-series that are described by the same underlying function, $G(t)$, as
\begin{equation}
    \begin{aligned}
    S_1 & = & A_1 G(t) & + B_1 \dot{G}(t), \\
    S_2 & = & A_2 G(t), \\
    S_3 & = & A_3 G(t) & + B_3 \dot{G}(t).
    \label{eq:toymodel}
\end{aligned}
\end{equation}
\noindent
We compute eq.~\eqref{eq:toymodel} using eq.~\eqref{eq:gps} with 3 time-series. We set the values for the amplitudes $A_1 = 0.005$\,\kms, $B_1= 0.05$\,\kms\,d, $A_2=0.02$\,\kms, $B_2 = 0$\,\kms\,d, $A_3=0.02$\,\kms, $B_3=-0.05$\,\kms\,d. 
We assume a mean function of zero for all three time series, and we use a QP covariance function with hyper-parameters \lbe$ = 30$ d, \lbp$= 0.3$ and $P_{\rm GP} = 5$ d.
We created 50 simultaneous observations taken randomly in a window of 50 d.
We added white noise with standard deviation of $0.001$\,\kms\ for $S_1$, 0.005\,\kms\ for $S_2$ and 0.010\,\kms\ for $S_3$. 
The synthetic time-series data are shown in Fig.~\ref{fig:toymodel}. The \texttt{Jupyter} notebook used to create the synthetic data is available \href{https://github.com/oscaribv/pyaneti/blob/master/inpy/example_toyp1/toy_model1.ipynb}{here \faGithub}.

We performed multidimensional GP modelling of the data set using \pyaneti. Priors and parameters used are defined in Table~\ref{tab:parameterstoy}. 
We perform an MCMC analysis with 100 Markov chains. We use the last 5000 iterations of converged chains, with a thin factor of 10, to create the posterior distributions.
{
We assume chains have converged when their \citet[][]{Gelman2004} criterion $\hat{R}$ is smaller than $1.02$ for all the sampled parameters \citep[for more details see][]{pyaneti,Gelman2004}.
}
The inferred parameters are shown in Table~\ref{tab:parameterstoy}, and the inferred models, together with the data, are shown in Fig.~\ref{fig:toymodel}.
We have made available the data and input file in \pyaneti\ for this example;
it can be run as \texttt{./pyaneti.py example\_toyp1} from the main \pyaneti\ directory.

From Table~\ref{tab:parameterstoy} we can see that the code is able to recover the injected amplitudes and kernel parameters within the error bars. Something to note is that the code is able to recover the value of $B_2 = 0$\,\kms\,d. This is important because it means that the analysis is able to differentiate between the pure $G(t)$ curves from those that depends on $\dot{G}(t)$. This has a practical application to understand the behaviour of the activity indicators that we use in our modelling.

\newcommand{\toffsetone}[1][${\rm km\,s^{-1}}$]   {$0.0007 _{ - 0.0043 } ^ { + 0.0047 }$~#1} 
\newcommand{\toffsettwo}[1][${\rm km\,s^{-1}}$]   {$-0.007 _{ - 0.022 } ^ { + 0.027 }$~#1} 
\newcommand{\toffsetthree}[1][${\rm km\,s^{-1}}$]   {$-0.0016 _{ - 0.003 } ^ { + 0.0029 }$~#1} 
\newcommand{\tAzero}[1][]   {$0.0103 _{ - 0.0042 } ^ { + 0.0053 }$~#1} 
\newcommand{\tAone}[1][]   {$0.063 _{ - 0.011 } ^ { + 0.014 }$~#1} 
\newcommand{\tAtwo}[1][]   {$0.064 _{ - 0.011 } ^ { + 0.014 }$~#1} 
\newcommand{\tAthree}[1][]   {$-0.00041 _{ - 0.00062 } ^ { + 0.00055 }$~#1} 
\newcommand{\tAfour}[1][]   {$0.0031 _{ - 0.0047 } ^ { + 0.0043 }$~#1} 
\newcommand{\tAfive}[1][]   {$-0.063 _{ - 0.014 } ^ { + 0.011 }$~#1} 
\newcommand{\tlambdae}[1][]   {$20.01 _{ - 1.34 } ^ { + 1.34 }$~#1} 
\newcommand{\tlambdap}[1][]   {$0.331 _{ - 0.021 } ^ { + 0.02 }$~#1} 
\newcommand{\tPGP}[1][]   {$4.9949 _{ - 0.0054 } ^ { + 0.0059 }$~#1} 

\begin{table}
  \caption{System parameters for toy model. \label{tab:parameterstoy}}
  \begin{tabular}{lccc}
  \hline
  Parameter & Real value & Prior$^{(\mathrm{a})}$ & Inferred value \\
  \hline
  \noalign{\smallskip}
     $A_1$ (\kms) & 0.005 &  $\mathcal{U}[0,0.5]$ & \tAzero \\
     $B_1$ (\kms\,d) & 0.05 &  $\mathcal{U}[0,0.5]$ & \tAone \\
     $A_2$ (\kms) & 0.05 &  $\mathcal{U}[-0.5,0.5]$ &  \tAtwo \\
     $B_2$ (\kms\,d) & 0 &  $\mathcal{U}[-0.5,0.5]$ & \tAthree \\
     $A_3$ (d) & 0.005 &  $\mathcal{U}[-0.5,0.5]$ &  \tAfour \\
     $B_3$ (\kms\,d) & -0.05 &  $\mathcal{U}[-0.5,0.5]$ & \tAfive \\
    \lbe\ (d) & 20 & $\mathcal{U}[1,80]$ & \tlambdae \\
    \lbp & 0.3 & $\mathcal{U}[0.01,5]$ & \tlambdap \\
    \pgp\, (d) & 5 & $\mathcal{U}[4,6]$ & \tPGP \\
    Offset $S_1$ & 0 & $\mathcal{U}[-0.7,0.7]$  & \toffsetone[] \\
    Offset $S_2$ & 0 & $\mathcal{U}[-0.7,0.7]$  & \toffsettwo[] \\
    Offset $S_3$ & 0 & $\mathcal{U}[-0.7,0.7]$  & \toffsetthree[] \\
    \hline
   \noalign{\smallskip}
  \end{tabular}
  \begin{tablenotes}\footnotesize
  \item \emph{Note} -- $^{(\mathrm{a})}$ $\mathcal{U}[a,b]$ refers to uniform priors between $a$ and $b$, $\mathcal{N}[a,b]$ to Gaussian priors with mean $a$ and standard deviation $b$.
  $^{(\mathrm{b})}$  Inferred parameters and errors are defined as the median and 68.3\% credible interval of the posterior distribution. 
\end{tablenotes}
\end{table}

\subsection{Recovering Keplerian signals with multi-instrument data}
\label{sec:toy2}

We perform another test similar to the one described in Sect.~\ref{sec:toy1}, but this time we added some more challenges to the test.
In this case we assume that we have contemporaneous observations of two time-series that behave as
\begin{equation}
    \begin{aligned}
    S_1 & = & A_1 G(t) & + B_1 \dot{G}(t), \\
    S_2 & = & A_2 G(t), \\
    \label{eq:toymodel2}
\end{aligned}
\end{equation}
\noindent
with $A_1 = 0.005$~\kms, $B_1= 0.05$~\kms\,d, $A_2=0.02$~\kms, and $B_2 = 0$~\kms\,d.
We use a QP covariance function with hyper-parameters \lbe$ = 20$\,d, \lbp$= 0.5$ and \pgp$ = 5$\,d. 
We assume that the spectroscopic data comes from two different instruments, $I_1$ and $I_2$, with an offset of zero for each time-series. 
For instrument $I_1$ we created 20 random observations between 0 and 60 days, each datum with an error bar of 0.003~\kms. 
For instrument $I_2$ we created 30 random observations in the same range, each one with an error bar of 0.005~\kms.
We included two Keplerian signals in the RV-like time-series ($S_1$). One signal is associated with a circular orbit \citep[$\sqrt{e} \sin \omega = 0$, and $\sqrt{e} \cos \omega = 0$, following][parametrisation]{Anderson2011} and the other one with an eccentricity of $0.3$ and angle of periastron of $\pi/3$ ($\sqrt{e} \sin \omega = 0.47$, and $\sqrt{e} \cos \omega = 0.27$). The amplitude of the Keplerian signals is significantly smaller than the amplitudes of the activity-like signal.
The parameters used to create both signals are listed in Table~\ref{tab:parameterstoy2}.
The \texttt{Jupyter} notebook used to create the synthetic data is available \href{https://github.com/oscaribv/pyaneti/blob/master/inpy/example_toyp2/toy_model2.ipynb}{here \faGithub}. We show the synthetic time-series in Fig.~\ref{fig:toymodel2}.
We perform four different analyses to the data using different techniques. For all the cases we assume that we know the ephemeris of the Keplerian signals and set Gaussian priors on the time of minimum conjunction, $T_0$, and period, $P$, for the two signals.

\subsubsection{Run 1}
\label{sec:run1}

We first perform a 1-dimensional GP modelling with \pyaneti\ modelling only the RVs with a QP kernel.  
We assume that the only information that we have of the GP hyper-parameters are the ranges where the true values lie. 
Table~\ref{tab:parameterstoy2} shows the sampled parameters and priors we use for this run that we name as \emph{Run 1}.
We perform an MCMC sampling with 100 Markov chains. 
We create the posterior distributions with the last 5000 iterations of converged chains with a thin factor of 10. This generates distributions with 50\,000 independent points per each sampled parameter.

The inferred parameters are shown in Table~\ref{tab:parameterstoy2}.
The first thing we note is that for this case the recovered \lbe\ and \pgp\ are consistent with the true values, but the value of \lbp\ is smaller than the true vale used to create the time-series. This is expected given that we are modelling the data without taking into account the GP derivative (see discussion in Sect.~\ref{sec:gpderivatives}). However, the parameter true values used to create the Keplerian signals are recovered within the confidence interval.

\subsubsection{Run 2}
\label{sec:run2}

We then perform a \emph{Run 2} in which we train our GP based on our activity-indicator-like signal ($S_2$). As we mentioned in Sect.~\ref{sec:gpcomparisons}, this is a common approach in the literature. 
We first do a 1-dimensional GP modelling of the $S_2$ signal using a QP kernel in order to obtain posterior distributions for the hyper-parameters \lbe, \lbp, and \pgp. 
We then model the RV data following the normal approach in the literature \citep[e.g.,][]{Grunblatt2015}, i.e., we  use a QP kernel with Gaussian priors on \lbe, \lbp, and \pgp\ based on our $S_2$ analysis.
The MCMC details are identical to the ones described in Sect.~\ref{sec:run1}.
Table~\ref{tab:parameterstoy2} shows priors and inferred values for all the sampled parameters.

From Table~\ref{tab:parameterstoy2} we can see that the results for this run are similar to the ones from the \emph{Run 1}. We note that despite the Gaussian prior that we set on \lbp\, the recovered value for this parameter is significantly smaller than the Gaussian prior mean.
This is again  expected given that we are modelling the RV data only with a QP kernel, without accounting for the time derivative (See Sect.~\ref{sec:gpderivatives}). 
Nonetheless, the recovered values of the Doppler semi-amplitudes of the of the coherent signals are recovered with a significance similar to the ones in \emph{Run 1}.

\subsubsection{Run 3}

We explored another possibility on the modelling of the RV time-series training the GP. But this time for the RV modelling we are including the first time derivative of the GP. The GP training comes from the same analysis of the $S_2$ time-series described in Sect.~\ref{sec:run2}. 
But, for the RV 1-dimensional GP regression we construct our covariance matrix with the kernel
\begin{equation}
    \gamma(t_i,t_j) = A_1^2 \gamma_{\rm QP}^{G,G} + B_1^2 \gamma_{\rm QP}^{dG,dG},
    \label{eq:newgammaqp}
\end{equation}
\noindent
that includes the derivative term of the QP kernel. Appendix~\ref{sec:gpderivatives} shows the full form of the $\gamma_{\rm QP}^{G,G}$ and $\gamma_{\rm QP}^{dG,dG}$ terms. We perform a 1-dimensional GP modelling of the RV data following the priors described in Table~\ref{tab:parameterstoy2} as Run 3.
The MCMC configuration follows the same parameters as in Sect.~\ref{sec:run1}.

Table~\ref{tab:parameterstoy2} show the recovered parameters for this run. 
We note that for this case, the recovered value for \lbp\ is consistent with the true value, as expected now that we are including the derivative of the GP. 
However, we note that even if we include the derivative, the recovered values of the Doppler semi-amplitudes are not more precise that the values recovered in Runs 1 and 2.
This can be explained because even if we are using a better model (that we know agrees with the model used to create the synthetic data), this approach still does not constrain better the shape of the underlying function describing the stellar signal in the RV data.

\subsubsection{Run 4}

We then perform a two-dimensional GP modelling with \pyaneti\ following the approach described in Sect.~\ref{sec:mgpforrv}. Table~\ref{tab:parameterstoy2} shows the sampled parameters, priors we use, and derived parameters.
again, the MCMC configuration is the same as the one described in Sect.~\ref{sec:run1}. This example can be reproduced by running \texttt{./pyaneti.py example\_toyp2} in the main \pyaneti\ directory.

\begin{figure*}
    \centering
    \includegraphics[width=0.99\textwidth]{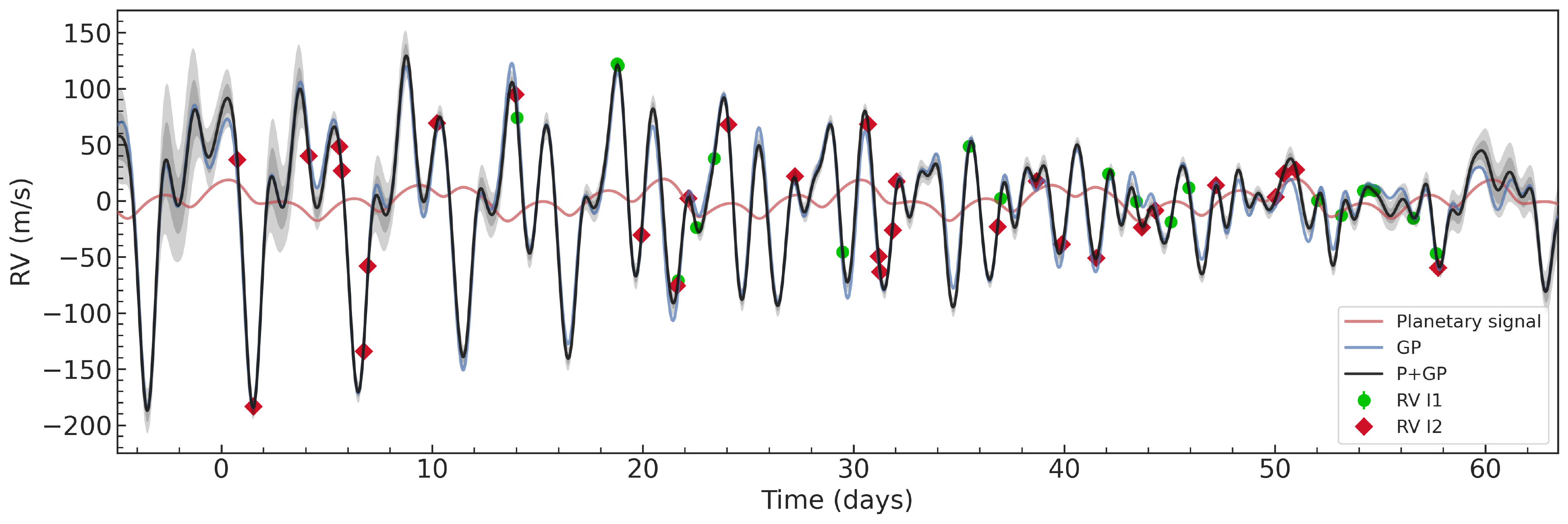} \\
    \includegraphics[width=0.99\textwidth]{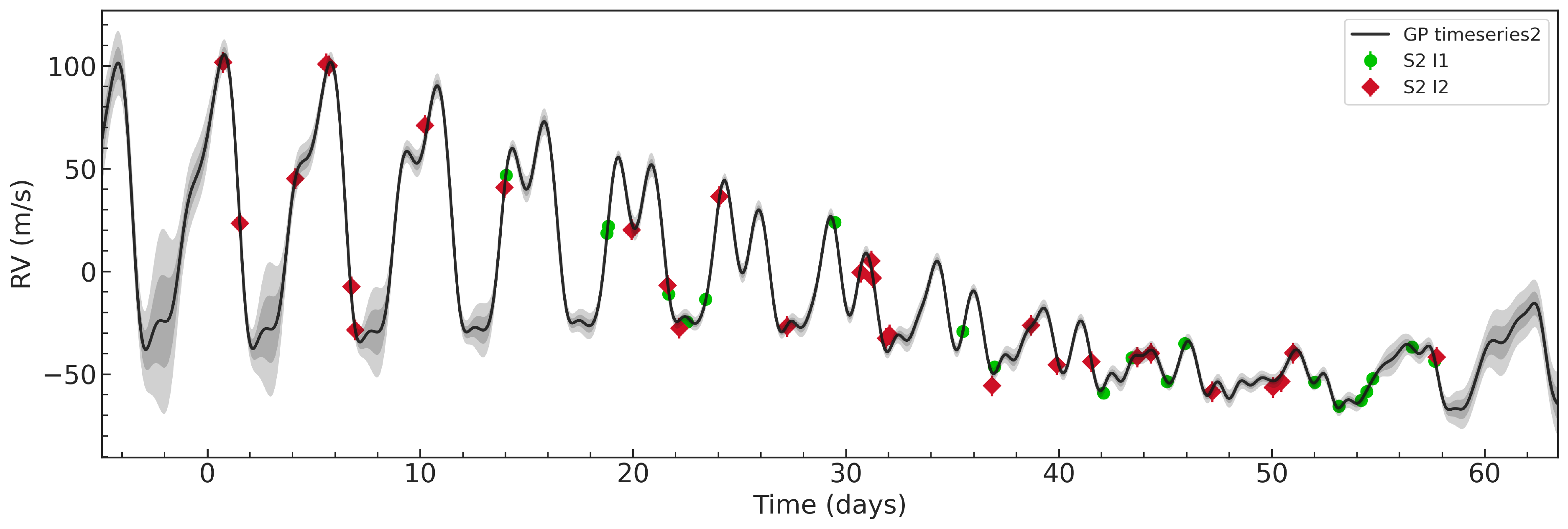} \\
    \includegraphics[width=0.495\textwidth]{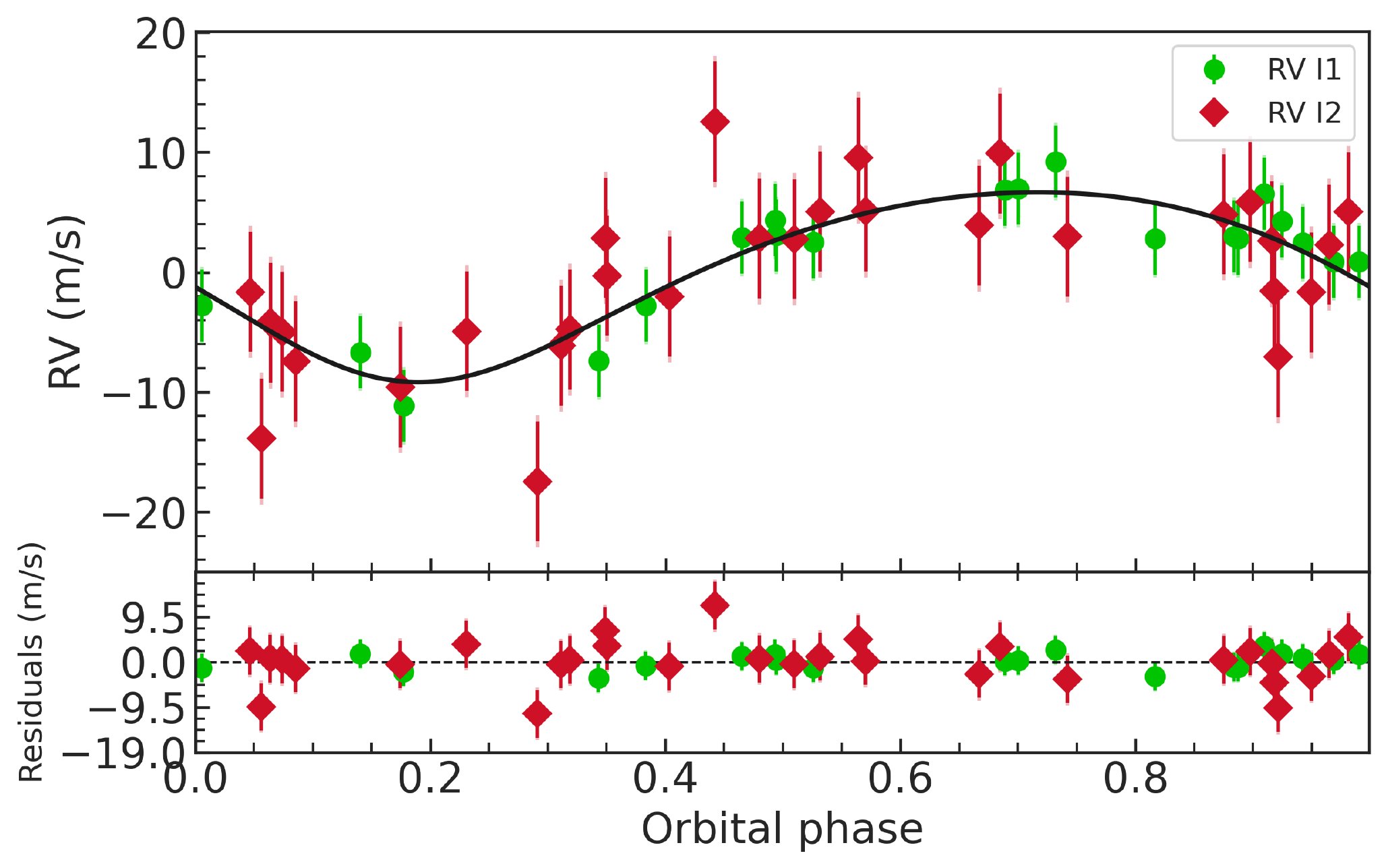}
    \includegraphics[width=0.495\textwidth]{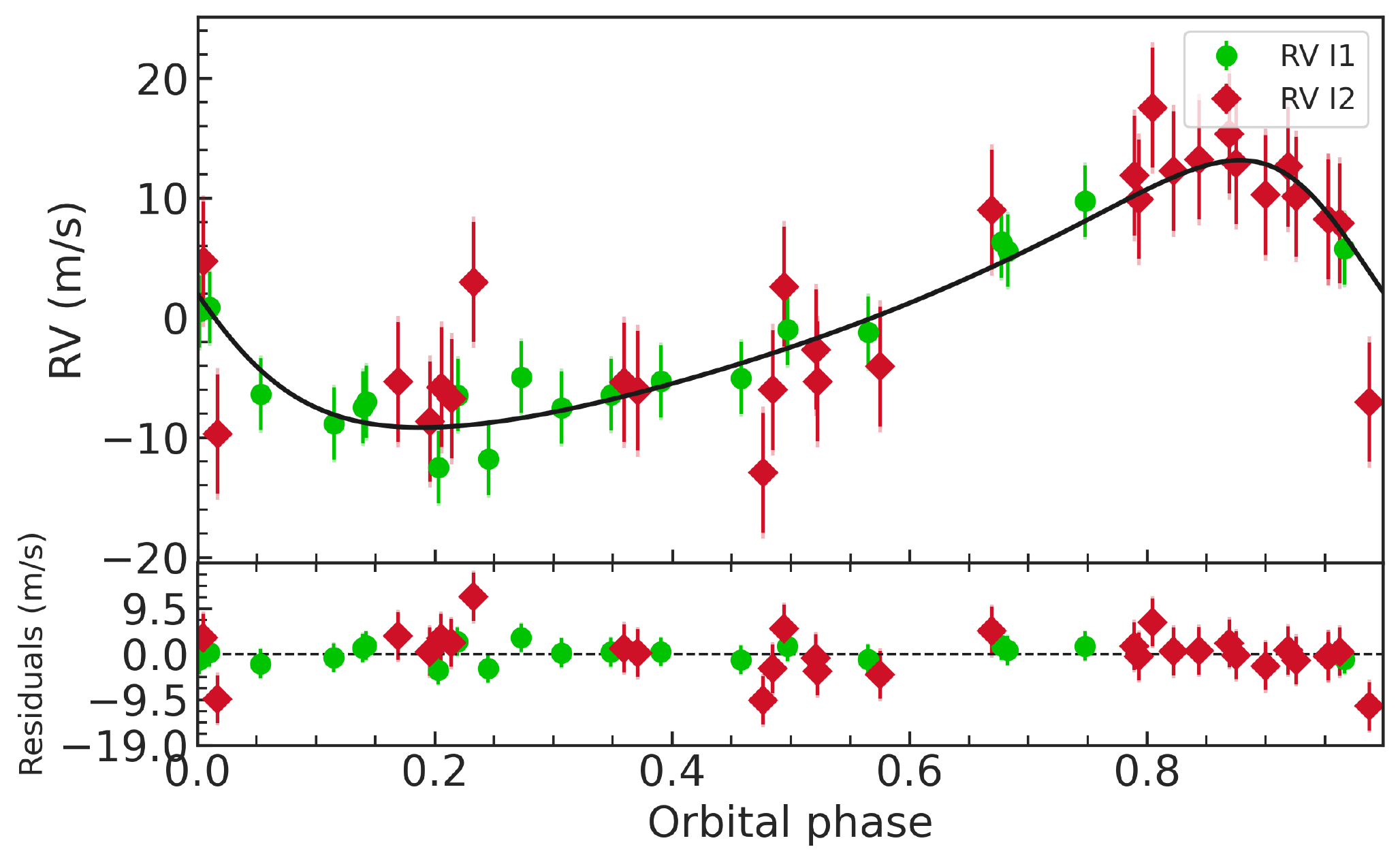}
    \caption{   $S_1$, and $S_2$ time-series described in Sect.~\ref{sec:toy2}.
    Top panel: The green (instrument $I_1$) and red markers (instrument $I_2$) in each panel represent the synthetic measurements with inferred offsets extracted. Solid dark line and shadow regions are as in Figure \ref{fig:toymodel}. The RV-like time-series also included the RV model for the two planets (red line).
    Bottom panel: Synthetic RV-like data folded on the orbital period of each injected planet following the subtraction of the systemic velocities, GP signal, and the other planet. The plots also show the inferred RV model for each planet (solid black line).
    These plots were generated automatically by \pyaneti.}
    \label{fig:toymodel2}
\end{figure*}

Figure~\ref{fig:toymodel2} shows the derived time-series and phase-folded models for this case.
We can see in Table~\ref{tab:parameterstoy2} that all derived parameters agree with the true values within the error bars. Specially the value of \lbp\ agrees with the true value as in the \emph{Run 3}, as expected given that we are using the GP derivative in this case.
We note that in this case, the derived semi-amplitudes for both Keplerian signals are recovered with a relative higher precision than in the previous runs.

\newcommand{\rTzerob}[1][days]   {$1. \pm 0.001 $~#1} 
\newcommand{\rPb}[1][days]   {$3.  \pm 0.001 $~#1} 
\newcommand{\resinb}[1][ ]   {$0.19 _{ - 0.48 } ^ { + 0.39 }$~#1} 
\newcommand{\recosb}[1][ ]   {$-0.1 _{ - 0.3 } ^ { + 0.36 }$~#1} 
\newcommand{\rkb}[1][${\rm m\,s^{-1}}$]   {$7.41 _{ - 2.6 } ^ { + 3.0 }$~#1} 
\newcommand{\rmpb}[1][$M_{\oplus}$]   {$15.45 _{ - 5.67 } ^ { + 5.91 }$~#1} 
\newcommand{\rTperib}[1][days]   {$1.04 _{ - 0.61 } ^ { + 0.65 }$~#1} 
\newcommand{\reb}[1][ ]   {$0.25 _{ - 0.17 } ^ { + 0.27 }$~#1} 
\newcommand{\rwb}[1][deg]   {$70.1 _{ - 177.8 } ^ { + 62.9 }$~#1} 
\newcommand{\rTzeroc}[1][days]   {$2. \pm 0.001 $~#1} 
\newcommand{\rPc}[1][days]   {$10. \pm 0.001$ ~#1} 
\newcommand{\resinc}[1][ ]   {$0.34 _{ - 0.57 } ^ { + 0.34 }$~#1} 
\newcommand{\recosc}[1][ ]   {$0.25 _{ - 0.21 } ^ { + 0.2 }$~#1} 
\newcommand{\rkc}[1][${\rm m\,s^{-1}}$]   {$11.49 _{ - 2.87 } ^ { + 3.67 }$~#1} 
\newcommand{\rmpc}[1][$M_{\oplus}$]   {$35.86 _{ - 8.86 } ^ { + 9.51 }$~#1} 
\newcommand{\rTperic}[1][days]   {$1.54 _{ - 2.47 } ^ { + 0.4 }$~#1} 
\newcommand{\rec}[1][ ]   {$0.33 _{ - 0.22 } ^ { + 0.23 }$~#1} 
\newcommand{\rwc}[1][deg]   {$47.0 _{ - 96.9 } ^ { + 31.6 }$~#1} 
\newcommand{\rRVone}[1][${\rm km\,s^{-1}}$]   {$0.005 _{ - 0.029 } ^ { + 0.031 }$~#1} 
\newcommand{\rRVtwo}[1][${\rm km\,s^{-1}}$]   {$0.001 _{ - 0.029 } ^ { + 0.031 }$~#1} 
\newcommand{\rAzero}[1][]   {$0.075 _{ - 0.015 } ^ { + 0.026 }$~#1} 
\newcommand{\rAone}[1][]   {$0.0 _{ - 0.0 } ^ { + 0.0 }$~#1} 
\newcommand{\rlambdae}[1][]   {$26.59 _{ - 4.84 } ^ { + 5.94 }$~#1} 
\newcommand{\rlambdap}[1][]   {$0.273 _{ - 0.035 } ^ { + 0.037 }$~#1} 
\newcommand{\rPGP}[1][]   {$5.032 _{ - 0.014 } ^ { + 0.012 }$~#1} 


\newcommand{\sTzerob}[1][days]   {$ 1. \pm 0.001 $~#1} 
\newcommand{\sPb}[1][days]   {$ 3. \pm 0.001 $~#1} 
\newcommand{\sesinb}[1][ ]   {$0.16 _{ - 0.48 } ^ { + 0.4 }$~#1} 
\newcommand{\secosb}[1][ ]   {$-0.14 _{ - 0.26 } ^ { + 0.35 }$~#1} 
\newcommand{\skb}[1][${\rm m\,s^{-1}}$]   {$7.27 _{ - 2.66 } ^ { + 3.02 }$~#1} 
\newcommand{\smpb}[1][$M_{\oplus}$]   {$15.19 _{ - 5.89 } ^ { + 6.18 }$~#1} 
\newcommand{\sTperib}[1][days]   {$1.09 _{ - 0.46 } ^ { + 0.77 }$~#1} 
\newcommand{\seb}[1][ ]   {$0.23 _{ - 0.15 } ^ { + 0.26 }$~#1} 
\newcommand{\swb}[1][deg]   {$64.8 _{ - 190.8 } ^ { + 69.0 }$~#1} 
\newcommand{\sTzeroc}[1][days]   {$ 2. \pm 0.001 $~#1} 
\newcommand{\sPc}[1][days]   {$ 10. \pm 0.001 $~#1} 
\newcommand{\sesinc}[1][ ]   {$0.34 _{ - 0.59 } ^ { + 0.33 }$~#1} 
\newcommand{\secosc}[1][ ]   {$0.26 _{ - 0.2 } ^ { + 0.2 }$~#1} 
\newcommand{\skc}[1][${\rm m\,s^{-1}}$]   {$11.47 _{ - 2.77 } ^ { + 3.26 }$~#1} 
\newcommand{\smpc}[1][$M_{\oplus}$]   {$35.75 _{ - 8.6 } ^ { + 8.76 }$~#1} 
\newcommand{\sTperic}[1][days]   {$1.5 _{ - 2.51 } ^ { + 0.43 }$~#1} 
\newcommand{\swc}[1][deg]   {$45.4 _{ - 97.0 } ^ { + 31.7 }$~#1} 
\newcommand{\sRVone}[1][${\rm km\,s^{-1}}$]   {$0.003 _{ - 0.03 } ^ { + 0.033 }$~#1} 
\newcommand{\sRVtwo}[1][${\rm km\,s^{-1}}$]   {$-0.001 _{ - 0.03 } ^ { + 0.033 }$~#1} 
\newcommand{\sAzero}[1][]   {$0.081 _{ - 0.016 } ^ { + 0.022 }$~#1} 
\newcommand{\slambdae}[1][]   {$24.16 _{ - 3.4 } ^ { + 3.62 }$~#1} 
\newcommand{\slambdap}[1][]   {$0.308 _{ - 0.037 } ^ { + 0.045 }$~#1} 
\newcommand{\sPGP}[1][]   {$5.031 _{ - 0.013 } ^ { + 0.012 }$~#1} 

\newcommand{\rrTzerob}[1][days]   {$1. \pm 0.001$~#1}
\newcommand{\rrPb}[1][days]   {$3. \pm 0.001 $~#1}
\newcommand{\rresinb}[1][ ]   {$0.05 _{ - 0.46 } ^ { + 0.43 }$~#1}
\newcommand{\rrecosb}[1][ ]   {$-0.15 _{ - 0.25 } ^ { + 0.38 }$~#1}
\newcommand{\rrkb}[1][${\rm m\,s^{-1}}$]   {$6.82 _{ - 2.7 } ^ { + 2.87 }$~#1}
\newcommand{\rrmpb}[1][$M_{\oplus}$]   {$14.25 _{ - 5.89 } ^ { + 5.94 }$~#1}
\newcommand{\rrTperib}[1][days]   {$1.15 _{ - 0.67 } ^ { + 0.8 }$~#1}
\newcommand{\rreb}[1][ ]   {$0.22 _{ - 0.15 } ^ { + 0.26 }$~#1}
\newcommand{\rrwb}[1][deg]   {$30.0 _{ - 162.0 } ^ { + 106.0 }$~#1}
\newcommand{\rrTzeroc}[1][days]   {$2. \pm 0.001$~#1}
\newcommand{\rrPc}[1][days]   {$10. \pm 0.001$~#1}
\newcommand{\rresinc}[1][ ]   {$0.3 _{ - 0.59 } ^ { + 0.34 }$~#1}
\newcommand{\rrecosc}[1][ ]   {$0.28 _{ - 0.2 } ^ { + 0.19 }$~#1}
\newcommand{\rrkc}[1][${\rm m\,s^{-1}}$]   {$11.61 _{ - 2.59 } ^ { + 3.06 }$~#1}
\newcommand{\rrmpc}[1][$M_{\oplus}$]   {$36.35 _{ - 8.11 } ^ { + 8.29 }$~#1}
\newcommand{\rrTperic}[1][days]   {$1.33 _{ - 2.51 } ^ { + 0.56 }$~#1}
\newcommand{\rrec}[1][ ]   {$0.32 _{ - 0.19 } ^ { + 0.21 }$~#1}
\newcommand{\rrwc}[1][deg]   {$39.0 _{ - 91.7 } ^ { + 34.0 }$~#1}
\newcommand{\rrRVone}[1][${\rm km\,s^{-1}}$]   {$-0.0019 _{ - 0.0063 } ^ { + 0.0088 }$~#1}
\newcommand{\rrRVtwo}[1][${\rm km\,s^{-1}}$]   {$-0.0052 _{ - 0.0066 } ^ { + 0.009 }$~#1}
\newcommand{\rrAzero}[1][]   {$0.0109 _{ - 0.0083 } ^ { + 0.0209 }$~#1}
\newcommand{\rrAone}[1][]   {$0.051 _{ - 0.013 } ^ { + 0.018 }$~#1}
\newcommand{\rrlambdae}[1][]   {$22.64 _{ - 3.46 } ^ { + 3.8 }$~#1}
\newcommand{\rrlambdap}[1][]   {$0.48 _{ - 0.055 } ^ { + 0.06 }$~#1}
\newcommand{\rrPGP}[1][]   {$5.03 _{ - 0.015 } ^ { + 0.014 }$~#1}

\newcommand{\smass}[1][$M_{\odot}$]{ $ 1.0000000 _{- 0.1000000}^{ + 0.1000000} $ #1} 
\newcommand{\sradius}[1][$R_{\odot}$]{ $1.0000000 _{ - 0.1000000}^{ + 0.1000000} $ #1}
\newcommand{\stemp}[1][$\mathrm{K}$]{ $ 5772.0000000 _{- 100.0000000}^{ + 100.0000000} $ #1 }
\newcommand{\Tzerob}[1][d]   {$1. \pm 0.001 $~#1} 
\newcommand{\Pb}[1][d]   {$3. \pm 0.001 $~#1} 
\newcommand{\esinb}[1][ ]   {$0.08 _{ - 0.31 } ^ { + 0.27 }$~#1} 
\newcommand{\ecosb}[1][ ]   {$0.15 _{ - 0.23 } ^ { + 0.17 }$~#1} 
\newcommand{\kb}[1][${\rm m\,s^{-1}}$]   {$4.89 _{ - 0.88 } ^ { + 1.01 }$~#1} 
\newcommand{\mpb}[1][$M_{\oplus}$]   {$10.84 _{ - 2.04 } ^ { + 2.39 }$~#1} 
\newcommand{\Tperib}[1][d]   {$0.71 _{ - 0.7 } ^ { + 0.57 }$~#1} 
\newcommand{\eb}[1][ ]   {$0.112 _{ - 0.075 } ^ { + 0.115 }$~#1} 
\newcommand{\wb}[1][deg]   {$24.1 _{ - 92.6 } ^ { + 63.9 }$~#1} 
\newcommand{\Tzeroc}[1][d]   {$2. \pm 0.001$~#1} 
\newcommand{\Pc}[1][d]   {$10. \pm 0.001$~#1} 
\newcommand{\esinc}[1][ ]   {$0.552 _{ - 0.136 } ^ { + 0.088 }$~#1} 
\newcommand{\ecosc}[1][ ]   {$0.21 _{ - 0.14 } ^ { + 0.15 }$~#1} 
\newcommand{\kc}[1][${\rm m\,s^{-1}}$]   {$9.97 _{ - 0.89 } ^ { + 1.0 }$~#1} 
\newcommand{\mpc}[1][$M_{\oplus}$]   {$31.2 _{ - 3.18 } ^ { + 3.38 }$~#1} 
\newcommand{\Tperic}[1][d]   {$1.76 _{ - 0.31 } ^ { + 0.18 }$~#1} 
\newcommand{\ec}[1][ ]   {$0.362 _{ - 0.089 } ^ { + 0.082 }$~#1} 
\newcommand{\wc}[1][deg]   {$69.4 _{ - 18.3 } ^ { + 14.1 }$~#1} 
\newcommand{\RVone}[1][${\rm km\,s^{-1}}$]   {$-0.0006 _{ - 0.0028 } ^ { + 0.0027 }$~#1} 
\newcommand{\RVtwo}[1][${\rm km\,s^{-1}}$]   {$-0.0013 _{ - 0.0027 } ^ { + 0.0028 }$~#1} 
\newcommand{\Actone}[1][${\rm km\,s^{-1}}$]   {$-0.008 _{ - 0.024 } ^ { + 0.025 }$~#1} 
\newcommand{\Acttwo}[1][${\rm km\,s^{-1}}$]   {$-0.009 _{ - 0.024 } ^ { + 0.025 }$~#1} 
\newcommand{\Azero}[1][]   {$0.006 _{ - 0.0018 } ^ { + 0.0024 }$~#1} 
\newcommand{\Aone}[1][]   {$0.056 _{ - 0.011 } ^ { + 0.015 }$~#1} 
\newcommand{\Atwo}[1][]   {$0.054 _{ - 0.011 } ^ { + 0.015 }$~#1} 
\newcommand{\Athree}[1][]   {$0.15 _{ - 0.47 } ^ { + 0.49 } 10^{-3}$~#1} 
\newcommand{\lambdae}[1][]   {$20.56 _{ - 1.53 } ^ { + 1.6 }$~#1} 
\newcommand{\lambdap}[1][]   {$0.487 _{ - 0.033 } ^ { + 0.041 }$~#1} 
\newcommand{\PGP}[1][]   {$5.012 _{ - 0.011 } ^ { + 0.01 }$~#1} 

\begin{table*}
  \caption{System parameters for toy model in Sect.~\ref{sec:toy2}. \label{tab:parameterstoy2}}
  \begin{tabular}{lccccccccc}
  \hline
   & Real & \multicolumn{2}{c}{Run 1}
   & \multicolumn{2}{c}{Run 2}  & \multicolumn{2}{c}{Run 3} & \multicolumn{2}{c}{Run 4} \\
  Parameter & Value &Prior$^{(\mathrm{a})}$ &
  value$^{(\mathrm{b})}$ & Prior$^{(\mathrm{a})}$ &  value$^{(\mathrm{b})}$ & Prior$^{(\mathrm{a})}$ &  value$^{(\mathrm{b})}$ & Prior$^{(\mathrm{a})}$ & value$^{(\mathrm{b})}$ \\
  \hline
  \noalign{\smallskip}
    $T_{0,b}$ (d) & $1$ & 
    $\mathcal{N}[1,10^{-3}]$ & \rTzerob[] &
    $\mathcal{N}[1,10^{-3}]$ & \sTzerob[] &
    $\mathcal{N}[1,10^{-3}]$ & \rrTzerob[] &
    $\mathcal{N}[1,10^{-3}]$ & \Tzerob[]  \\
    $P_{b}$ (d) & $3$ & 
    $\mathcal{N}[3,10^{-3}]$ & \rPb[] & 
    $\mathcal{N}[3,10^{-3}]$ & \sPb[]& 
    $\mathcal{N}[3,10^{-3}]$ & \rrPb[]& 
    $\mathcal{N}[3,10^{-3}]$ & \Pb[] \\
    $\sqrt{e}_{b}\sin\omega_b$ & $0$ & 
    $\mathcal{U}[-1,1]$ & \resinb[] & 
    $\mathcal{U}[-1,1]$ & \sesinb[] & 
    $\mathcal{U}[-1,1]$ & \rresinb[] & 
    $\mathcal{U}[-1,1]$ & \esinb[] \\
    $\sqrt{e}_{b}\cos\omega_b$ & $0$ & 
    $\mathcal{U}[-1,1]$ & \recosb[] & 
    $\mathcal{U}[-1,1]$ & \secosb[] & 
    $\mathcal{U}[-1,1]$ & \rrecosb[] & 
    $\mathcal{U}[-1,1]$ & \ecosb[] \\
    $K_{b}$ (\ms) & $5$ & 
    $\mathcal{U}[0,500]$ & \rkb[] &
    $\mathcal{U}[0,500]$ & \skb[] &
    $\mathcal{U}[0,500]$ & \rrkb[] &
    $\mathcal{U}[0,500]$ & \kb[] \\
    $T_{0,c}$ (d) & $2$ & 
    $\mathcal{N}[2,10^{-3}]$ & \rTzeroc[] & 
    $\mathcal{N}[2,10^{-3}]$ & \sTzeroc[] & 
    $\mathcal{N}[2,10^{-3}]$ & \rrTzeroc[] & 
    $\mathcal{N}[2,10^{-3}]$ & \Tzeroc[]  \\
    $P_{c}$ (d) & $10$ & 
    $\mathcal{N}[10,10^{-3}]$ & \rPc[] & 
    $\mathcal{N}[10,10^{-3}]$ & \sPc[] & 
    $\mathcal{N}[10,10^{-3}]$ & \rrPc[] & 
    $\mathcal{N}[10,10^{-3}]$ & \Pc[] \\
    $\sqrt{e}_{c}\sin\omega_c$ & $0.47$ &
    $\mathcal{U}[-1,1]$ &  \resinb[] &
    $\mathcal{U}[-1,1]$ &  \sesinb[] &
    $\mathcal{U}[-1,1]$ &  \rresinb[] &
    $\mathcal{U}[-1,1]$ &  \esinb[] \\
    $\sqrt{e}_{c}\cos\omega_c$ & $0.27$ & 
    $\mathcal{U}[-1,1]$ & \recosb[] & 
    $\mathcal{U}[-1,1]$ & \secosb[] & 
    $\mathcal{U}[-1,1]$ & \rrecosb[] & 
    $\mathcal{U}[-1,1]$ & \ecosb[]  \\
    $K_{c}$ (\ms) & $10$ & 
    $\mathcal{U}[0,500]$ & \rkc[] & 
    $\mathcal{U}[0,500]$ & \skc[] & 
    $\mathcal{U}[0,500]$ & \rrkc[] & 
    $\mathcal{U}[0,500]$ & \kc[] \\
    $A_1$ (\kms) & 0.005 &  
    $\mathcal{U}[0,0.5]$ & \rAzero[] &
    $\mathcal{U}[0,0.5]$ & \sAzero[] & 
    $\mathcal{U}[0,0.5]$ & \rrAzero[] & 
    $\mathcal{U}[0,0.5]$ & \Azero[]  \\
    $B_1$ (\kms d) & 0.05 &  
    $\cdots$ &  $\cdots$ & 
    $\cdots$ &  $\cdots$ & 
    $\mathcal{U}[0,0.5]$ &   \rrAone[] & 
    $\mathcal{U}[0,0.5]$ &  \Aone[] \\
    $A_2$ (\kms) & 0.05 &  
    $\cdots$ & $\cdots$ &  
    $\cdots$ & $\cdots$ & 
    $\cdots$ & $\cdots$ & 
    $\mathcal{U}[-0.5,0.5]$ & \Atwo[] \\
    $B_2$ (\kms d) & 0 &  
    $\cdots$ &  $\cdots$  &  
    $\cdots$ & $\cdots$ & 
    $\cdots$ & $\cdots$ & 
    $\mathcal{U}[-0.5,0.5]$ &  \Athree[] \\
    \lbe (d) & 20 & 
    $\mathcal{U}[1,80]$ & \rlambdae[] & 
    $\mathcal{N}[21,5]$ &\slambdae[] & 
    $\mathcal{N}[21,5]$ &\rrlambdae[] & 
    $\mathcal{U}[1,80]$ & \lambdae[] \\
    \lbp & 0.5 & 
    $\mathcal{U}[0.01,5]$ &  \rlambdap[] & 
    $\mathcal{N}[0.59,0.10]$ &  \slambdap[] & 
    $\mathcal{N}[0.59,0.10]$ &  \rrlambdap[] & 
    $\mathcal{U}[0.01,5]$ &  \lambdap[] \\
    $P_{\rm GP}$ (d) & 5 & 
    $\mathcal{U}[4,6]$ & \rPGP[] & 
    $\mathcal{N}[5.03,0.03]$ & \sPGP[] & 
    $\mathcal{N}[5.03,0.03]$ & \rrPGP[] & 
    $\mathcal{U}[4,6]$ & \PGP[] \\
    $I_1$ offset $S_1$ & 0 & 
    $\mathcal{U}[-0.5,0.5]$ & \rRVone[] &
    $\mathcal{U}[-0.5,0.5]$ & \sRVone[] & 
    $\mathcal{U}[-0.5,0.5]$ & \rrRVone[] & 
    $\mathcal{U}[-0.5,0.5]$ & \RVone[] \\
    $I_2$ offset $S_1$ & 0 & 
    $\mathcal{U}[-0.5,0.5]$  & \rRVtwo[] &
    $\mathcal{U}[-0.5,0.5]$  & \sRVtwo[] &
    $\mathcal{U}[-0.5,0.5]$  & \rrRVtwo[] &
    $\mathcal{U}[-0.5,0.5]$  &\RVtwo[] \\
    $I_1$ offset $S_2$ & 0 &
    $\cdots$  & $\cdots$ &
    $\cdots$  & $\cdots$ &
    $\cdots$  & $\cdots$ &
    $\mathcal{U}[-0.5,0.5]$  & \Actone[] \\
    $I_2$ offset $S_2$ & 0 &
    $\cdots$  & $\cdots$ &
    $\cdots$  & $\cdots$ &
    $\cdots$  & $\cdots$ &
    $\mathcal{U}[-0.5,0.5]$  & \Actone[] \\
    \hline
   \noalign{\smallskip}
  \end{tabular}
  \begin{tablenotes}\footnotesize
  \item \emph{Note} -- $^{(a)}$ and $^{(b)}$ defined as in Table~\ref{tab:parameterstoy}. 
\end{tablenotes}
\end{table*}

\subsubsection{Comparison between the different Runs}

\begin{figure}
    \centering
    \includegraphics[width=0.47\textwidth]{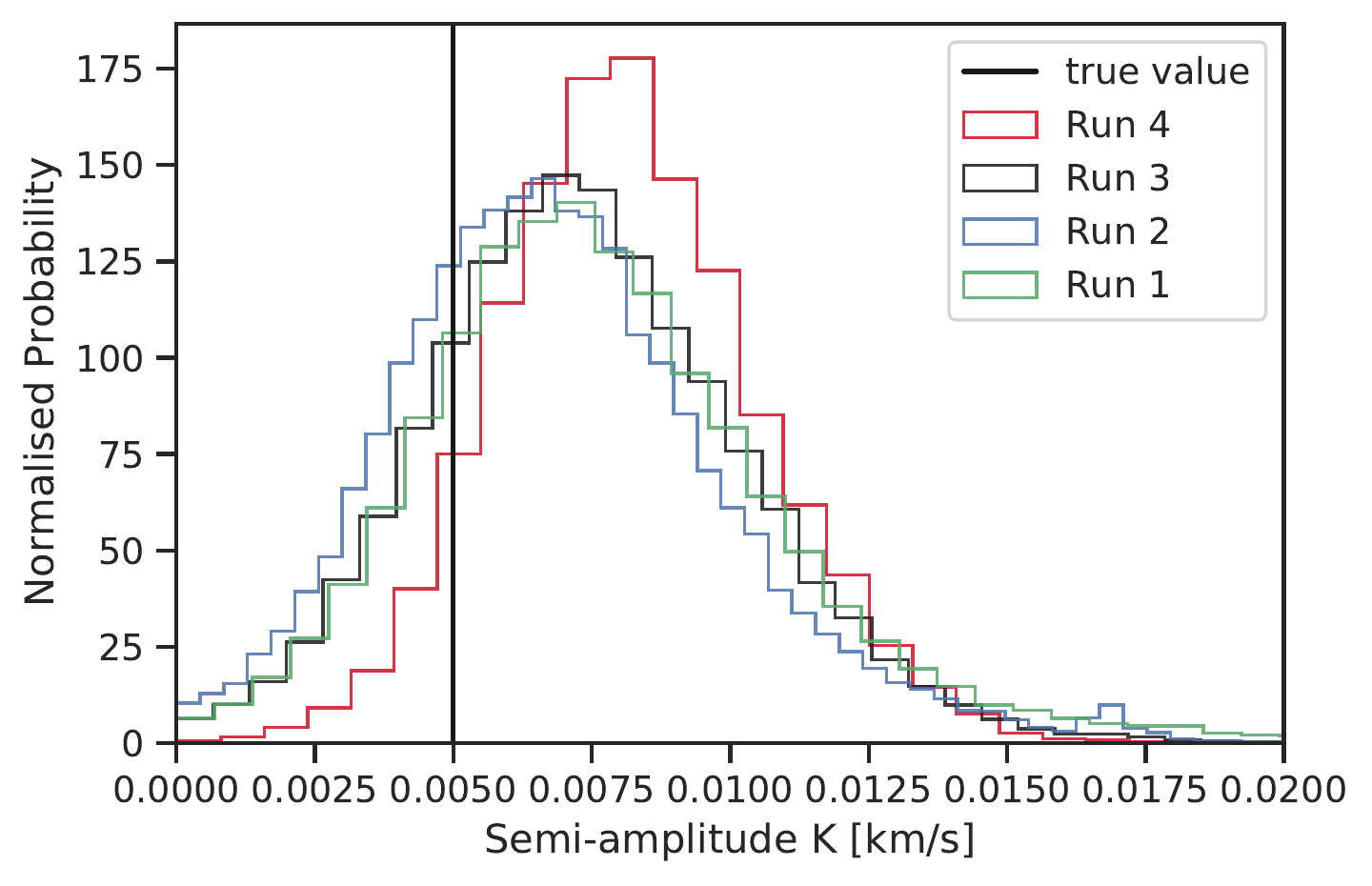}\\
    \includegraphics[width=0.47\textwidth]{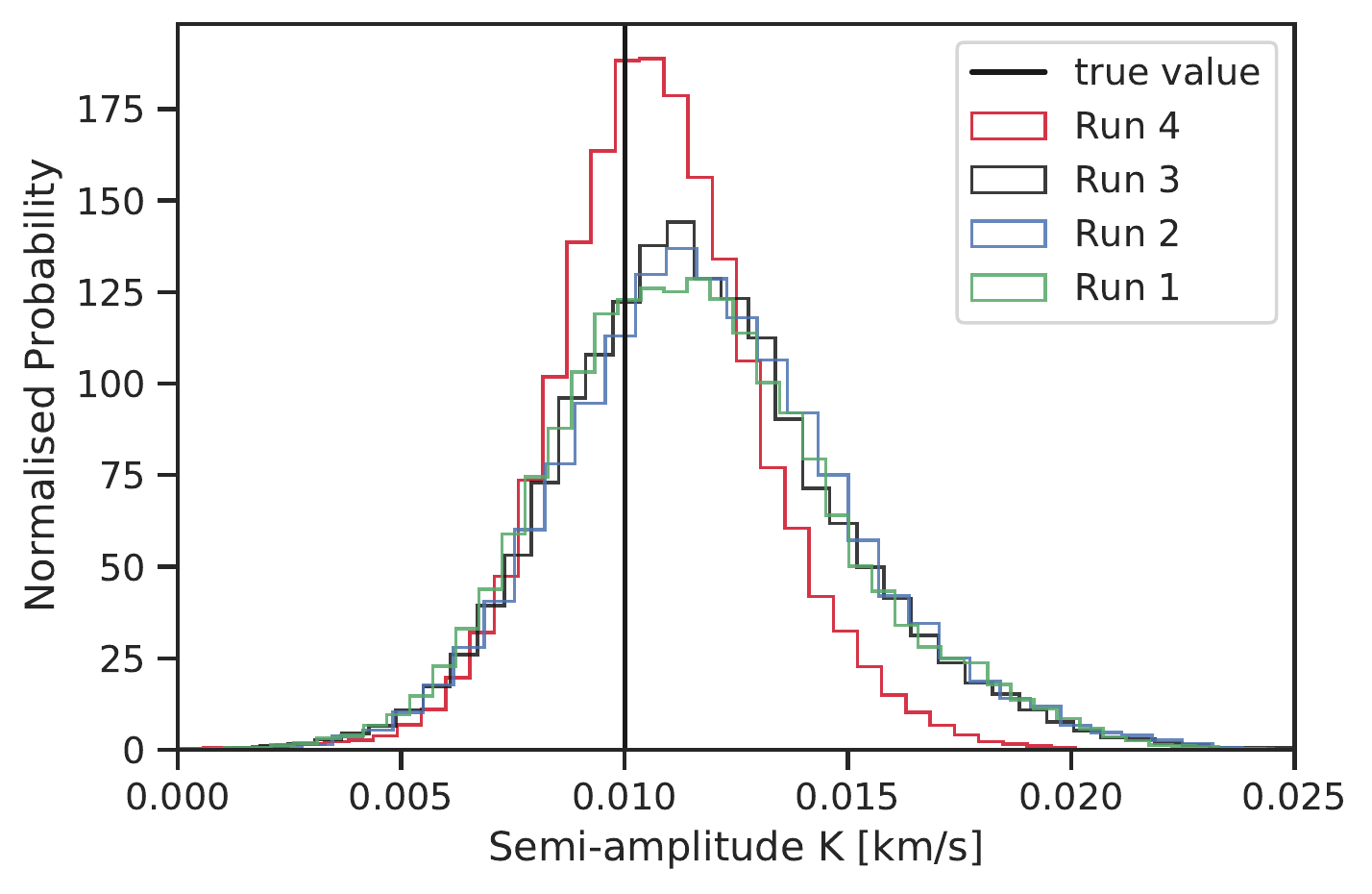}
    \caption{Probability distributions for the Doppler semi-amplitudes for the signals b (upper panel) and c (lower panel) as described in Sect.~\ref{sec:toy2}. Each colour corresponds to the different Runs described in the text. The true value of the parameter is shown with a vertical line in each case.}
    \label{fig:kas}
\end{figure}

Table~\ref{tab:parameterstoy2} shows the parameter value for all the four Runs described in this Section, and Figure~\ref{fig:kas} shows the recovered posterior distribution for the Keplerian signals b and c for each case.
From Table~\ref{tab:parameterstoy2} we can see that all Runs are able to recover the planetary-like signals within the error bars. This is relevant since we created this data set with activity-like amplitudes significantly larger than the Keplerian ones, with the intention to show that the code is able to recover coherent signals in the RV-like time-series.
This provides some tentative evidence that if RV observations are planned with suitable observing campaigns and with the right instruments, we may be able to find planetary signals even in cases with extreme activity.

From Table~\ref{tab:parameterstoy2} and Figure~\ref{fig:kas} we can also see that run that provides better precision on the detected Doppler semi-amplitudes is \emph{Run 4}. We can argue that \emph{Runs} 1 and 2 are not optimal to analyse this problem, because of the way we create the synthetic RV time-series. 
But we may think that Run 3 and 4 are equivalent: the model of the $S_2$ data is created with a function draw using a QP kernel, while the $S_1$ time-series with a function created with a QP kernel and its first time derivative. And from Table~\ref{tab:parameterstoy2}, we see that the QP kernel hyper-parameters are fully consistent within the error bars for both runs.
However, we obtain better detection of the Keplerian signals on \emph{Run 4} where we use the multidimensional GP approach.
This is explained by the discussion in Sect.~\ref{sec:gpcomparisons}, where we mentioned that the multidimensional GP approach ensures that the underlying function $G(t)$ is the same for all time-series.
This result shows the advantage of using the activity indicators to model stellar signals within a multidimensional GP framework.

These tests also illustrate \pyaneti's ability to handle different instruments. However, we caution that this capability should be used with care. In particular, multiple instruments do not only have different offsets between them, but they may also observe in different wavelength ranges. This means that the stellar signal that each instrument observes may be different. Therefore, they cannot necessarily be treated as the same underlying signal that the multi-GP approach described in this manuscript assumes.

\subsection{Multi-band transit modelling}

\begin{figure*}
    \centering
    \includegraphics[width=0.95\textwidth]{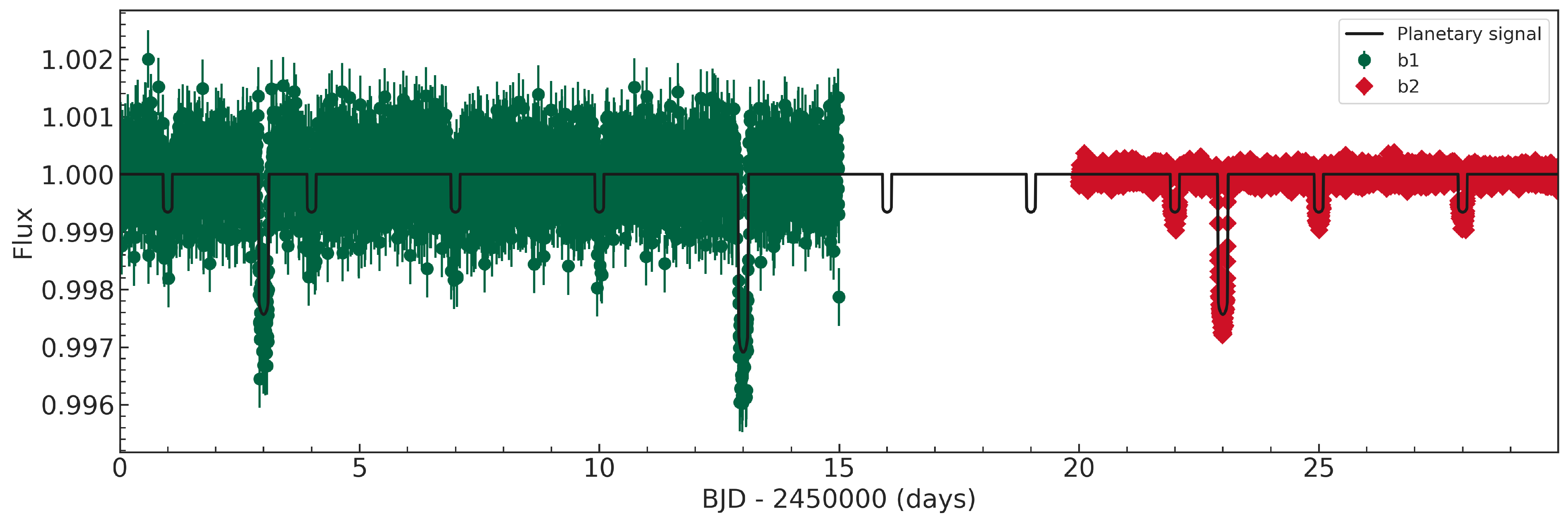} \\
    \includegraphics[width=0.45\textwidth]{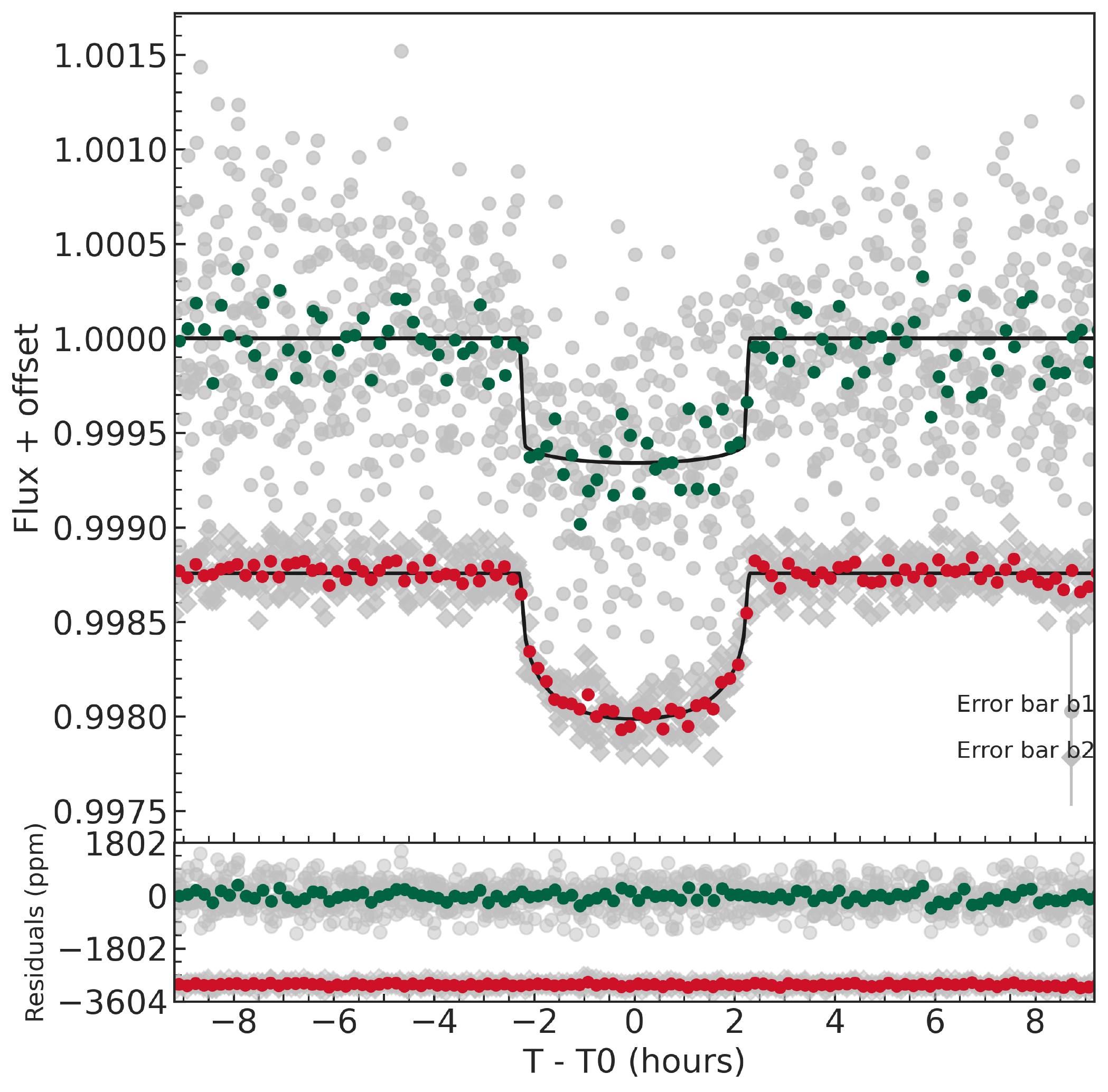}
    \includegraphics[width=0.45\textwidth]{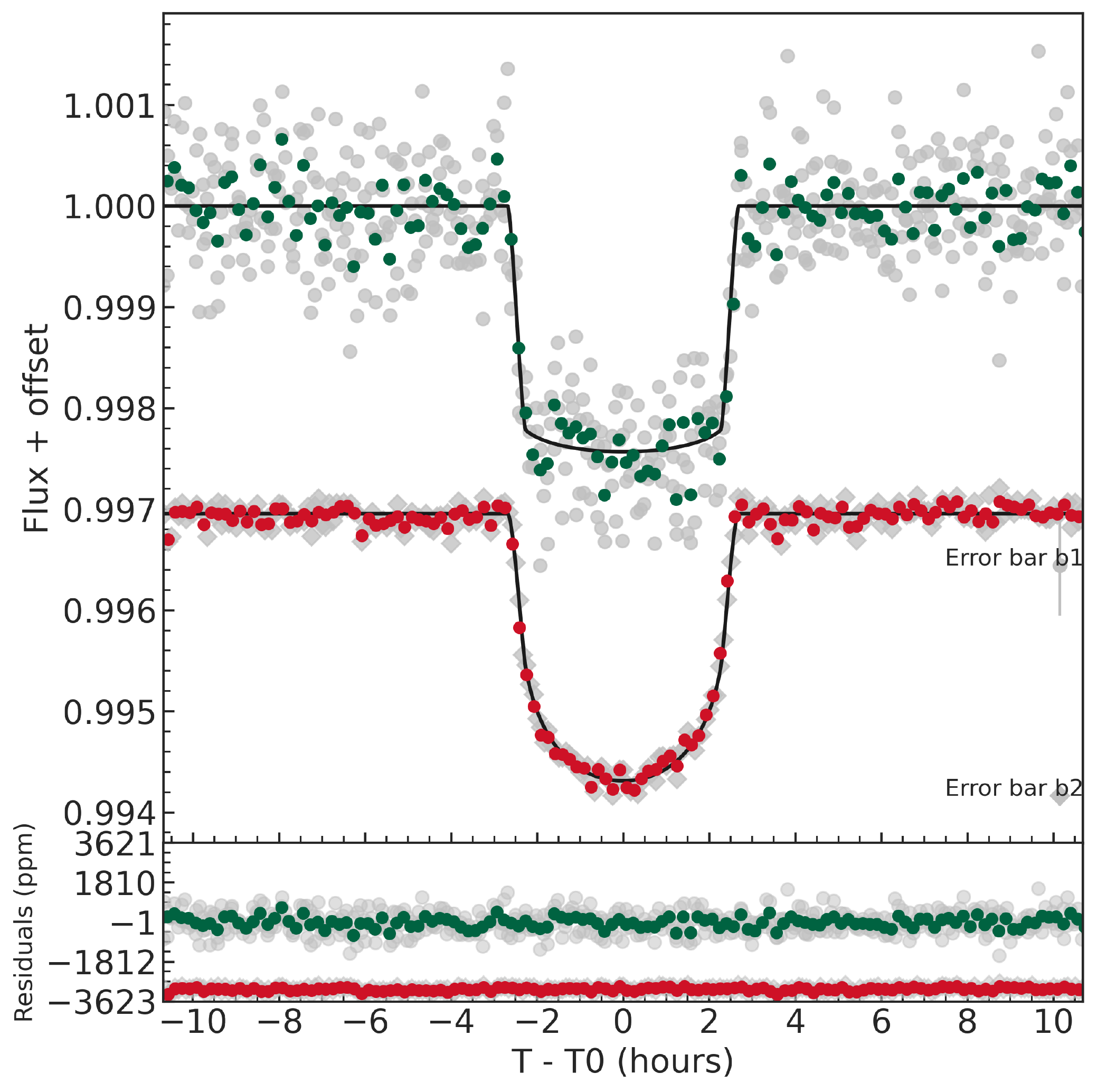}
    \caption{Top panel: Synthetic light curve created with the fictitious B1 (green circles) and B2 (red squares) instruments. The light curve model with the two transiting signals is shown with a black thick line.
    Bottom panel: Phase-folded light curves for injected planet signal $b$ (left) and $c$ (right). Each plot shows the data for instrument B1 (gray circles) and B2 (gray squares) separated by an offset. The plots also show the data in 10-min bins for each instrument (B1 as green and B2 as red circles) together with the inferred model (black line) for each case.
    These plots are generated automatically by \pyaneti.}
    \label{fig:toymodel3}
\end{figure*}

We create a synthetic light curve using \citlalicue\ (see Sect.~\ref{ap:citla}) in order to test the ability of \pyaneti\ to model multi-band transit modelling. 
We created data assuming we have a flattened light curve of a system with two transiting planets observed with two different instruments. 
For the first instrument, named ``B1'', the data ranges from 0 to 15 days, with one data point every 5 min, with a precision of 500 ppm per datum. 
The limb darkening coefficients for this fictitious observed star with instrument B1 are $u_1=0.25$ and $u_2=0$ \citep[$q1=0.06$ and $q2 =0.50$, following][]{Kipping2013}, following the quadratic law of \citet[][]{Mandel2002}. 
For the second instrument, named ``B2'', data go from 20 to 30 days, with one observations every 5 min with a precision of 100 ppm. 
The limb darkening coefficients for this fictitious star with instrument B2 are $u_1=0.50$ and $u_2=0.25$ \citep[$q1=0.56$ and $q2 =0.33$, following][]{Kipping2013}. 
We injected two transiting planets with circular orbits into the light curves. We also assume that both planets are transiting a star with a density of $1.4\,{\rm g\,cm^{-3}}$. The time of transit $T_0$, orbital period $P$, impact parameter $b$, and scaled planet radius $r_{\rm p}$ for each planet are given in Table~\ref{tab:parameterstoytransits}. 
The \texttt{Jupyter} notebook used to create this synthetic data set is provided \href{https://github.com/oscaribv/pyaneti/blob/master/inpy/example_multiband/multiband_transits.ipynb}{here \faGithub}, and
Figure \ref{fig:toymodel3} shows the synthetic light curves for both bands.

We perform a multi-band and multi-planet modelling with \pyaneti. Table~\ref{tab:parameterstoytransits} shows the sampled parameters and priors we use. We note that we sample for a different limb darkening coefficients for each band. We also sample for the stellar density, and recover the scaled semi-major axis for each planet using Kepler's third law.
We sample the parameter space with 100 independent chains. 
We create the posterior distributions for each sampled parameter with the last 5000 iterations of converged steps using a thin factor of 10.
This example can be reproduced in \pyaneti\ by running \texttt{./pyaneti.py example\_multiband} in the main \pyaneti\ directory.

Figure~\ref{fig:toymodel3} shows the phase-folded light curves for each transiting planet.
We show the inferred parameters in Table~\ref{tab:parameterstoytransits}.
We can see that the code is able to recover the orbital and planet parameters for both transiting signals. We also note that \pyaneti\ can recover the band-dependent parameters for this example (limb darkening coefficients). In Sect.~\ref{sec:realdata} we describe some real stellar systems where the multi-band capabilities of the code have been used.

\renewcommand{\Tzerob}[1][days]   {$3.9998 _{ - 0.0014 } ^ { + 0.0012 }$~#1} 
\renewcommand{\Pb}[1][days]   {$3.0001 _{ - 0.00018 } ^ { + 0.0002 }$~#1} 
\renewcommand{\eb}[1][ ]   {$0.0 _{ - 0.0 } ^ { + 0.0 }$~#1} 
\renewcommand{\wb}[1][deg]   {$90.0 _{ - 0.0 } ^ { + 0.0 }$~#1} 
\newcommand{\bb}[1][ ]   {$0.26 _{ - 0.17 } ^ { + 0.17 }$~#1} 
\newcommand{\arb}[1][ ]   {$4.98 _{ - 0.31 } ^ { + 0.14 }$~#1} 
\newcommand{\rrb}[1][ ]   {$0.025 _{ - 0.00027 } ^ { + 0.00039 }$~#1} 
\newcommand{\rpb}[1][$R_\mathrm{J}$]   {$0.243 _{ - 0.024 } ^ { + 0.024 }$~#1} 
\newcommand{\ib}[1][deg]   {$87.06 _{ - 2.32 } ^ { + 1.98 }$~#1} 
\newcommand{\ab}[1][AU]   {$0.0228 _{ - 0.0025 } ^ { + 0.0025 }$~#1} 
\newcommand{\insolationb}[1][${\rm F_{\oplus}}$]   {$1892.0 _{ - 178.0 } ^ { + 265.0 }$~#1} 
\newcommand{\denstrb}[1][${\rm g\,cm^{-3}}$]   {$0.26 _{ - 0.046 } ^ { + 0.023 }$~#1} 
\newcommand{\densspb}[1][${\rm g\,cm^{-3}}$]   {$1.41 _{ - 0.37 } ^ { + 0.55 }$~#1} 
\newcommand{\Teqb}[1][K]   {$1835.8 _{ - 44.7 } ^ { + 61.1 }$~#1} 
\newcommand{\ttotb}[1][hours]   {$4.608 _{ - 0.026 } ^ { + 0.028 }$~#1} 
\newcommand{\tfulb}[1][hours]   {$4.358 _{ - 0.027 } ^ { + 0.025 }$~#1} 
\newcommand{\tegb}[1][hours]   {$0.1212 _{ - 0.0078 } ^ { + 0.0198 }$~#1} 
\renewcommand{\Tzeroc}[1][days]   {$3.0008 _{ - 0.0011 } ^ { + 0.0011 }$~#1} 
\renewcommand{\Pc}[1][days]   {$9.9995 _{ - 0.00056 } ^ { + 0.00061 }$~#1} 
\renewcommand{\ec}[1][ ]   {$0.0 _{ - 0.0 } ^ { + 0.0 }$~#1} 
\renewcommand{\wc}[1][deg]   {$90.0 _{ - 0.0 } ^ { + 0.0 }$~#1} 
\newcommand{\bc}[1][ ]   {$0.652 _{ - 0.04 } ^ { + 0.027 }$~#1} 
\newcommand{\arc}[1][ ]   {$11.71 _{ - 0.36 } ^ { + 0.47 }$~#1} 
\newcommand{\rrc}[1][ ]   {$0.04898 _{ - 0.00059 } ^ { + 0.00051 }$~#1} 
\newcommand{\rpc}[1][$R_\mathrm{J}$]   {$0.476 _{ - 0.047 } ^ { + 0.048 }$~#1} 
\newcommand{\ic}[1][deg]   {$86.81 _{ - 0.24 } ^ { + 0.31 }$~#1} 
\newcommand{\ac}[1][AU]   {$0.0545 _{ - 0.0057 } ^ { + 0.0059 }$~#1} 
\newcommand{\Teqc}[1][K]   {$1191.1 _{ - 29.8 } ^ { + 29.2 }$~#1} 
\newcommand{\ttotc}[1][hours]   {$5.371 _{ - 0.034 } ^ { + 0.038 }$~#1} 
\newcommand{\tfulc}[1][hours]   {$4.528 _{ - 0.052 } ^ { + 0.056 }$~#1} 
\newcommand{\tegc}[1][hours]   {$0.424 _{ - 0.038 } ^ { + 0.032 }$~#1} 
\newcommand{\qonebone}[1][]   {$0.04 _{ - 0.025 } ^ { + 0.044 }$~#1} 
\newcommand{\qtwobone}[1][]   {$0.49 _{ - 0.34 } ^ { + 0.33 }$~#1} 
\newcommand{\uonebone}[1][]   {$0.17 _{ - 0.12 } ^ { + 0.14 }$~#1} 
\newcommand{\utwobone}[1][]   {$0.0 _{ - 0.11 } ^ { + 0.15 }$~#1} 
\newcommand{\qonebtwo}[1][]   {$0.666 _{ - 0.091 } ^ { + 0.111 }$~#1} 
\newcommand{\qtwobtwo}[1][]   {$0.231 _{ - 0.087 } ^ { + 0.093 }$~#1} 
\newcommand{\uonebtwo}[1][]   {$0.38 _{ - 0.13 } ^ { + 0.12 }$~#1} 
\newcommand{\utwobtwo}[1][]   {$0.44 _{ - 0.17 } ^ { + 0.18 }$~#1} 

\begin{table}
  \caption{System parameters for multi-band toy model. \label{tab:parameterstoytransits}}
  \begin{tabular}{lccc}
  \hline
  Parameter & Real value & Prior$^{(\mathrm{a})}$ & Inferred value$^{(\mathrm{b})}$ \\
  \hline
  \noalign{\smallskip}
    $T_{0,b}$ (days) & $4$ & $\mathcal{U}[3.95,4.05]$ & \Tzerob[]  \\
    $P_{b}$ (days) & $3$ & $\mathcal{U}[2.95,3.05]$ & \Pb[] \\
    $b_{b}$ & $0.25$ & $\mathcal{U}[0,1]$ & \bb[] \\
    $r_{p,b}/R_\star$ & $0.025$ & $\mathcal{U}[0,0.1]$ & \rrb[] \\
    $T_{0,c}$ (days) & $3$ & $\mathcal{U}[2.95,3.05]$ & \Tzeroc[]  \\
    $P_{c}$ (days) & $10$ & $\mathcal{U}[9.95,10.05]$ & \Pc[] \\
    $b_{c}$ & $0.7$ & $\mathcal{U}[0,1]$ & \bc[] \\
    $r_{p,c}/R_\star$ & $0.05$ & $\mathcal{U}[0,0.1]$ & \rrc[] \\
    $\rho_\star$ (${\rm g\,cm^{-3}}$)  & $1.4$ & $\mathcal{U}[0.01,5]$ & \densspb[] \\
    $q_{1,b1}$ &  0.06 &  $\mathcal{U}[0,1]$ & \qonebone[]  \\
    $q_{2,b1}$ &  0.50 &  $\mathcal{U}[0,1]$ &  \qtwobone[] \\
    $q_{1,b2}$ &  0.56 &  $\mathcal{U}[0,1]$ & \qonebtwo[] \\
    $q_{2,b2}$ &  0.33 &  $\mathcal{U}[0,1]$ &  \qtwobtwo[] \\
    \hline
   \noalign{\smallskip}
  \end{tabular}
  \begin{tablenotes}\footnotesize
  \item \emph{Note} -- Same Note as Table~\ref{tab:parameterstoy}. 
\end{tablenotes}
\end{table}

\subsection{Single transit event}

We create a single transit event light curves using \citlalicue\ (see Appendix~\ref{ap:citla}) in order to test the ability of \pyaneti\ to model single transits. 
We create the data assuming we have a flattened light curve of a system with one planetary transit. 
The data range from 9 to 11 d, with one data point every 5 min, with a precision of 100 ppm per datum. 
The limb darkening coefficients for this fictitious star are $u_1=0.25$ and $u_2=0$ \citep[$q1=0.06$ and $q2 =0.50$, following][]{Kipping2013}, following the quadratic law of \citet[][]{Mandel2002}. 
We injected a transiting planet assuming a circular orbit around a star with a Sun-like density.
The time of transit is $T_0 = 10$ d, orbital period $P=30$ d, scaled semi-major axis $a/R_\star = 40.6$, impact parameter $b=0.5$ and scaled planet radius $r_{\rm p}=0.25$.
The code needed to create this synthetic data set is provided in this \href{https://github.com/oscaribv/pyaneti/blob/master/inpy/example_single/example_single.ipynb}{link \faGithub}.

We perform a single transit modelling with \pyaneti\ by indicating \texttt{is\_single\_transit = True} in the input file for this system. Table~\ref{tab:parameterssingle} shows the sampled parameters and priors we use.
We sample the parameter space with 100 independent chains and create the posterior distributions with the last 5000 iterations of converged chains with  a thin factor of 10.
This example can be reproduced in \pyaneti\ by running \texttt{./pyaneti.py example\_single} within the main \pyaneti\ directory.

\begin{figure}
    \centering
    \includegraphics[width=0.45\textwidth]{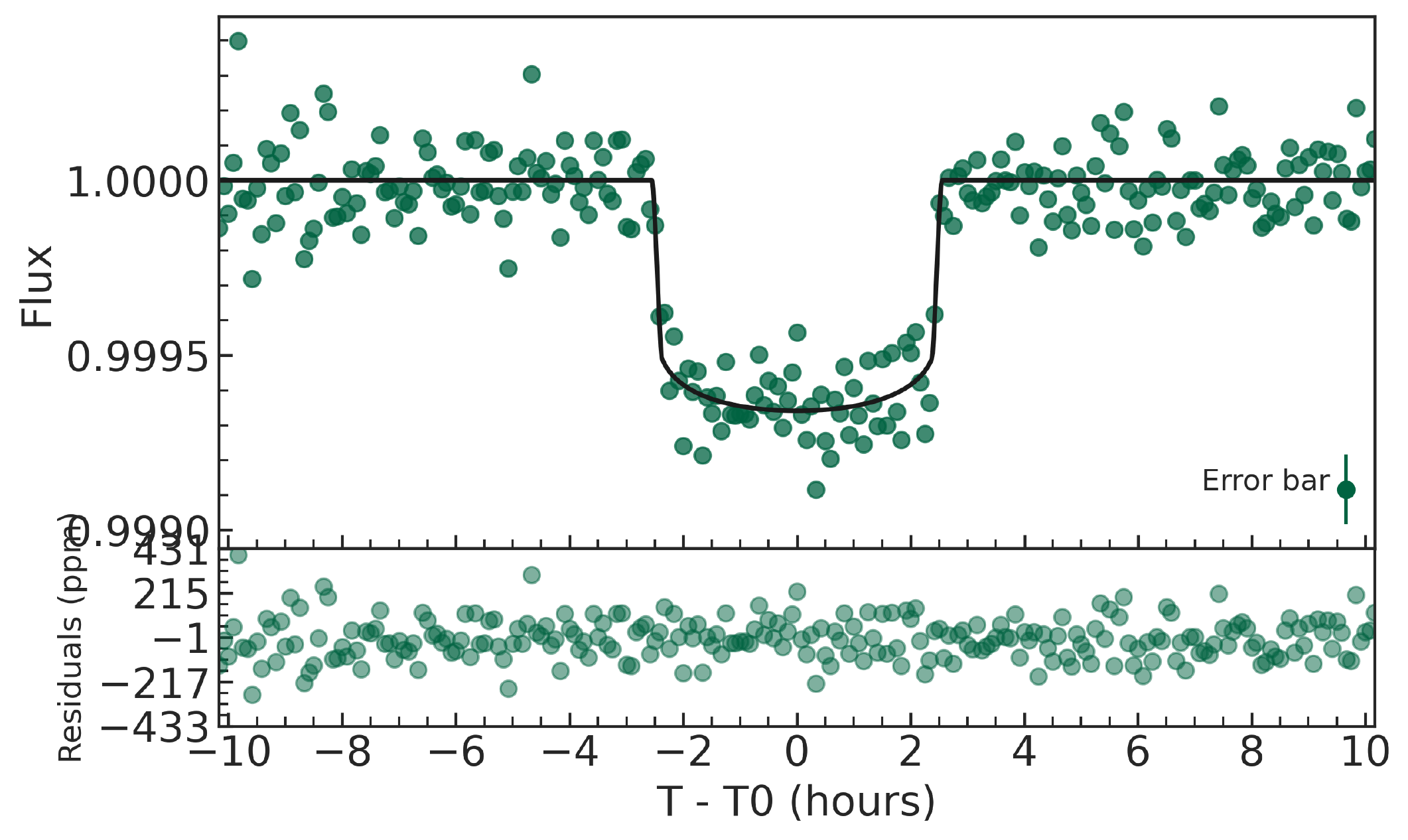} \\
    \caption{Model for the single transit modelling test. Synthetic data are shown as green circles. The inferred model is shown with a black line.
    This plot is shown as is provided automatically by \pyaneti.}
    \label{fig:toysingle}
\end{figure}

Figure \ref{fig:toysingle} shows the inferred model for the single transit and Table~\ref{tab:parameterssingle} shows the inferred parameters. \pyaneti\ is able to recover the injected values of $T_0$, $b$, $r_{\rm p}$, $q_1$, and $q_2$ within 1-sigma error bars. 
We note that for single transit fits, the scaled semi-major axis and periods given by the code do not have the same meaning as for normal runs fitting multiple transits. 
The scaled semi-major axis is a dummy value sampled to take into account the transit shape, while the orbital period is a derived parameter, i.e., we do not sample for it directly, we compute it with the other sampled parameters assuming the orbit is circular.
This capability of the code has been used before to estimate periods for transit signals detected by the \emph{TESS} mission \citep[e.g.,][]{Eisner2020b}.

\renewcommand{\Tzerob}[1][days]   {$9.99999 _{ - 0.00081 } ^ { + 0.00092 }$~#1} 
\renewcommand{\Pb}[1][days]   {$20.0 _{ - 0.0 } ^ { + 0.0 }$~#1} 
\renewcommand{\eb}[1][ ]   {$0.0 _{ - 0.0 } ^ { + 0.0 }$~#1} 
\renewcommand{\wb}[1][deg]   {$90.0 _{ - 0.0 } ^ { + 0.0 }$~#1} 
\renewcommand{\bb}[1][ ]   {$0.52 _{ - 0.37 } ^ { + 0.25 }$~#1} 
\renewcommand{\arb}[1][ ]   {$26.41 _{ - 6.8 } ^ { + 4.06 }$~#1} 
\renewcommand{\rrb}[1][ ]   {$0.02485 _{ - 0.00061 } ^ { + 0.0009 }$~#1} 
\renewcommand{\rpb}[1][$R_{\oplus}$]   {$2.72 _{ - 0.28 } ^ { + 0.31 }$~#1} 
\renewcommand{\Tperib}[1][days]   {$9.99999 _{ - 0.00081 } ^ { + 0.00092 }$~#1} 
\renewcommand{\ib}[1][deg]   {$88.87 _{ - 1.13 } ^ { + 0.84 }$~#1} 
\renewcommand{\ab}[1][AU]   {$0.12 _{ - 0.03 } ^ { + 0.024 }$~#1} 
\newcommand{\pcirb}[1][days]   {$34.1 _{ - 14.6 } ^ { + 47.6 }$~#1} 
\newcommand{\qone}[1][]   {$0.146 _{ - 0.083 } ^ { + 0.142 }$~#1} 
\newcommand{\qtwo}[1][]   {$0.3 _{ - 0.22 } ^ { + 0.39 }$~#1} 
\newcommand{\uone}[1][]   {$0.22 _{ - 0.16 } ^ { + 0.23 }$~#1} 
\newcommand{\utwo}[1][]   {$0.14 _{ - 0.25 } ^ { + 0.24 }$~#1} 
\newcommand{\jtr}[1][]   {$1.18e-05 _{ - 7.9e-06 } ^ { + 1.09e-05 }$~#1} 

\begin{table}
\caption{System parameters for single transit toy model. \label{tab:parameterssingle}}
\begin{tabular}{lccc}
  \hline
  Parameter & Real value & Prior$^{(\mathrm{a})}$ & Inferred value$^{(\mathrm{b})}$ \\
  \hline
  \noalign{\smallskip}
    $T_{0,b}$ (days) & $10$ & $\mathcal{U}[9.5,10.5]$ & \Tzerob[]  \\
    $b_{b}$ & $0.50$ & $\mathcal{U}[0,1]$ & \bb[] \\
    $r_{p,b}/R_\star$ & $0.025$ & $\mathcal{U}[0,0.1]$ & \rrb[] \\
    $q_{1}$ &  0.15 &  $\mathcal{U}[0,1]$ & \qone[]  \\
    $q_{2}$ &  0.50 &  $\mathcal{U}[0,1]$ &  \qtwo[] \\
    $a_{\rm dummy}$\,$^{(\mathrm{c})}$ & $\cdots$ & $\mathcal{U}[1.1,1000]$ & \arb[] \\
    \multicolumn{4}{l}{\emph{Derived parameters}} \\
    $P_{b,{\rm circ}}$\, (days)$^{(\mathrm{d})}$ & $30$ & $\cdots$ & \pcirb[] \\
    \hline
   \noalign{\smallskip}
  \end{tabular}
  \begin{tablenotes}\footnotesize
  \item \emph{Note} -- $(a)$ and $(b)$ same as Table~\ref{tab:parameterstoy}. $^{(\mathrm{c})}$ This is a dummy scaled semi-major axis that \pyaneti\ needs to sample to deal with the transit shape, it does not have a physical sense. $^{(\mathrm{d})}$ Note that the period is a derived parameter.
\end{tablenotes}
\end{table}

\subsection{Real planetary systems}
\label{sec:realdata}

We have shown how \pyaneti\ is able to recover the injected planetary and orbital parameters in specific examples of synthetic spectroscopic-like and photometry-like time-series. 
This demonstrates that if we believe that our RV and transit data behave as the models described in this paper, \pyaneti\ will be able to provide reliable parameter estimates. 
Fortunately, the new implementations of the code have already been applied to real data in peer-reviewed literature. 
In this section we describe how \pyaneti\ has been used in these analyses. We describe scenarios in which all the new additions of the code have been used combined.  
It is not our intention to reproduce the analyses or plots published in the aforesaid manuscripts. We only describe how \pyaneti\ was used in the relevant paper and we also provide examples to reproduce the analyses in such papers.

\subsubsection{K2-100}

\citet[][]{Barragan2019} published the RV detection of K2-100\,b, a transiting exoplanet orbiting a young active star in the Praesepe cluster. 
In that manuscript we used the ability of the code to fit multi-band transit photometry simultaneously with a 3-dimensional GP approach for spectroscopic time-series including a Keplerian component.

\begin{figure*}
    \centering
    \includegraphics[width=0.99\textwidth]{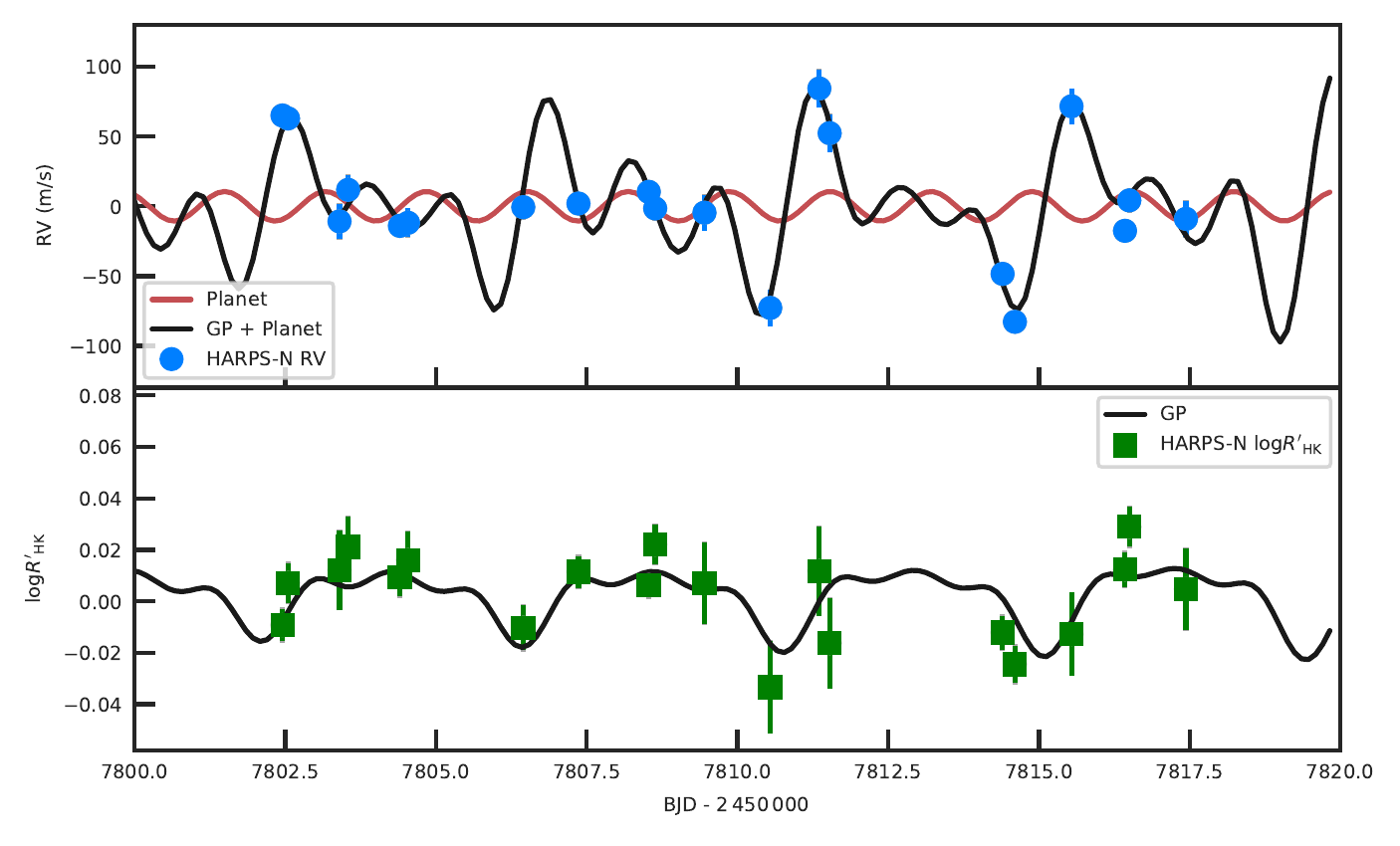}
    \caption{A 20 day subset of RV (upper panel) and \logr\ (lower panel) time-series for K2-100. Both time-series have had the inferred offsets in \citet[][]{Barragan2019} subtracted. 
    Measurements are shown as filled symbols with error bars. For the RV time-series, we show the planet-induced signal alone (red line) and also the subtracted activity signal (black line). For the \logr\ time-series, we show the activity model as a black line.}
    \label{fig:k2100}
\end{figure*}

The data were modelled using a multidimensional GP approach with a QP kernel, for which the harmonic complexity detected in the spectroscopic time-series was relatively high (\lbp$= 0.6$). 
Therefore, the use of the GP derivatives for the detection of the planetary signal was crucial in this case (see discussion in Sect.~\ref{sec:gpderivatives}).
Figure~\ref{fig:k2100} shows a 20 days subset of the RV and \logr\ time-series of Figure 2 in \citet[][]{Barragan2019}.
This figure shows that the \logr\ time-series behaves as the $S_1$ curve in the example in Sect~\ref{sec:gpderivatives}, and the RV data as the $S_2$ curve, i.e., the RV curve behaves as the time derivative of the \logr\ curve. 
We note that K2-100 is a fast rotating and spot-dominated young active star, and we expect that for young stars with similar characteristics, the GP derivative will be crucial to model the activity induced signals in the RV data.

We made available different setups to reproduce the analysis of the K2-100 system from the \pyaneti\ main directory. 
To reproduce the multi-band analysis of the system, the relevant command is \texttt{./pyaneti.py example\_multiband\_k2100}. To model the RV together with activity indicators, the command is \texttt{./pyaneti.py example\_timeseries\_k2100}. 
And to reproduce the full modelling including multi-band transits and RV together with activity indicators, type \texttt{./pyaneti.py example\_full\_k2100}. 

\subsubsection{TOI-1260}

\citet[][]{Georgieva2021} report the detection and characterisation of two mini-Neptunes transiting TOI-1260. They use \pyaneti\ to perform joint transit and RV modelling using a multi-planet, multi-band, and multidimensional GP configuration.

As for K2-100, the time-series were modelled using a multidimensional GP approach with a QP kernel. However, for this case, the harmonic complexity was moderate (\lbp$= 1.4_{-0.5}^{+1.0}$). Even in this case of moderate harmonic complexity, the multidimensional GP approach improved the planetary detection when compared with other common approaches \citep[see][for more details]{Georgieva2021}.

\subsection{Execution time}

All examples presented in Sect.~\ref{sec:tests} were run on a personal laptop with 8$\times$1.90 GHz Intel\textregistered\ Core\texttrademark\ i7-8650U CPUs with Ubuntu 18.04.
For comparison, we also ran all the examples with the same setup in a cluster with 12$\times$3.40 GHz Intel\textregistered\ Xeon\textregistered\ CPUs. 
We compiled the code with \texttt{GFortran} and ran all the examples in parallel.
Table~\ref{tab:times} show the execution time for the different setups presented in Section~\ref{sec:tests}.

We note that \pyaneti\ is able to produce ready-to-publish results in a relative short time even on a personal laptop. However, the execution time can be longer, depending on several parameters, such as the MCMC configuration, number of planets being modelled, number of data, number of CPUs used, etc.  
For example, the setup \texttt{example\_full\_k2100} is a complex problem that combines multi-band analysis and multidimensional GP regression. The total model samples 38 parameters simultaneously. Therefore, to reproduce the results by \citet[][]{Barragan2019} requires a relatively high computational power.

\begin{table}
\centering
\caption{Execution time for test runs in Sect.~\ref{sec:tests}. \label{tab:times}}
\begin{tabular}{lcc}
  \hline
  Setup & Time in laptop & Time in cluster \\
  \hline
  \noalign{\smallskip}
    \texttt{example\_toyp1} & 29m47s & 11m21s \\
    \texttt{example\_toyp2} & 23m10s & 8m35s \\
    \texttt{example\_multiband} & 17m59s & 6m1s \\
    \texttt{example\_single} & 1m03s & 35s \\
    \texttt{example\_timeseries\_k2100} & 38m34s & 18m51s \\
    \texttt{example\_multiband\_k2100} & 31m43s & 12m20s \\
    \texttt{example\_full\_k2100} & 8h40m  & 3h16m \\
    \hline
   \noalign{\smallskip}
  \end{tabular}
\end{table}

\section{Conclusions}
\label{sec:conlusions}

In this manuscript we present a major update to the \pyaneti\ code. 
This new version of \pyaneti\ allows one to perform GP regression, as well as multi-band and single transit fits. 
The biggest advantage of this new version of \pyaneti\ with respect to other codes is the multidimensional GP regression that is included within the RV modelling routines. 
We present some tests to show how \pyaneti\ can recover the model parameters in different data set configurations.
We will continue updating \pyaneti\ in the future according to the needs of the exoplanet {community}.

We described how we expanded the likelihood to account for data correlation using a multidimensional GP. In this manuscript we use the bulit-in MCMC described in \citet[][]{pyaneti} to sample the parameter space. However, we note that this new likelihood can be used with other sampling techniques. In the near future we plan to add different sampling techniques in \pyaneti\ with other MCMC \citep[e.g.,][]{emcee,zeus} sampling algorithms, as well as incorporate nested sampling \citep[e.g.][]{multinest,polychord,dynesty} methods.


Together with the code, we also provided some discussion on the use of GPs to model RV time-series. 
We pointed out how training GPs with activity indicators or light curves may not be the best approach for all cases of stellar activity.
Even if the discussion is open, these aspects should be taken into account when modelling RV time-series using GPs. 

We also presented \citlalicue\ and \citlalatonac. These numerical tools allow one to create synthetic photometric and spectroscopic time-series that may be useful to plan observing campaigns, especially in the cases in which the stellar signal is significantly larger than the planetary ones.

\section*{Acknowledgements}
{We thank the anonymous referee for their helpful suggestions that improved the quality of this paper.}
This publication is part of a project that has received funding from the European Research Council (ERC) under the European Union’s Horizon 2020 research and innovation programme (Grant agreement No. 865624).
NZ and SA acknowledge support from the UK Science and Technology Facilities Council (STFC) under Grant Code ST/N504233/1, studentship no. 1947725.

\section*{Data availability}

The data and software underlying this article are available as online supplementary material accessible following the links provided in the online version.



\bibliographystyle{mnras}
\bibliography{refs} 

\begin{thebibliography}{}
\makeatletter
\relax
\def\mn@urlcharsother{\let\do\@makeother \do\$\do\&\do\#\do\^\do\_\do\%\do\~}
\def\mn@doi{\begingroup\mn@urlcharsother \@ifnextchar [ {\mn@doi@}
  {\mn@doi@[]}}
\def\mn@doi@[#1]#2{\def\@tempa{#1}\ifx\@tempa\@empty \href
  {http://dx.doi.org/#2} {doi:#2}\else \href {http://dx.doi.org/#2} {#1}\fi
  \endgroup}
\def\mn@eprint#1#2{\mn@eprint@#1:#2::\@nil}
\def\mn@eprint@arXiv#1{\href {http://arxiv.org/abs/#1} {{\tt arXiv:#1}}}
\def\mn@eprint@dblp#1{\href {http://dblp.uni-trier.de/rec/bibtex/#1.xml}
  {dblp:#1}}
\def\mn@eprint@#1:#2:#3:#4\@nil{\def\@tempa {#1}\def\@tempb {#2}\def\@tempc
  {#3}\ifx \@tempc \@empty \let \@tempc \@tempb \let \@tempb \@tempa \fi \ifx
  \@tempb \@empty \def\@tempb {arXiv}\fi \@ifundefined
  {mn@eprint@\@tempb}{\@tempb:\@tempc}{\expandafter \expandafter \csname
  mn@eprint@\@tempb\endcsname \expandafter{\@tempc}}}

\bibitem[\protect\citeauthoryear{{Ahrer} et~al.,}{{Ahrer}
  et~al.}{2021}]{Ahrer2021}
{Ahrer} E.,  et~al., 2021, \mn@doi [\mnras] {10.1093/mnras/stab373}, \href
  {https://ui.adsabs.harvard.edu/abs/2021MNRAS.503.1248A} {503, 1248}

\bibitem[\protect\citeauthoryear{{Aigrain}, {Pont}  \& {Zucker}}{{Aigrain}
  et~al.}{2012}]{Aigrain2012}
{Aigrain} S.,  {Pont} F.,   {Zucker} S.,  2012, \mn@doi [\mnras]
  {10.1111/j.1365-2966.2011.19960.x}, \href
  {https://ui.adsabs.harvard.edu/abs/2012MNRAS.419.3147A} {419, 3147}

\bibitem[\protect\citeauthoryear{{Aigrain} et~al.,}{{Aigrain}
  et~al.}{2015}]{Aigrain2015}
{Aigrain} S.,  et~al., 2015, \mn@doi [\mnras] {10.1093/mnras/stv853}, \href
  {https://ui.adsabs.harvard.edu/abs/2015MNRAS.450.3211A} {450, 3211}

\bibitem[\protect\citeauthoryear{{Alvarez}, {Rosasco}  \& {Lawrence}}{{Alvarez}
  et~al.}{2011}]{Alvarez2011}
{Alvarez} M.~A.,  {Rosasco} L.,   {Lawrence} N.~D.,  2011, arXiv e-prints,
  \href {https://ui.adsabs.harvard.edu/abs/2011arXiv1106.6251A} {p.
  arXiv:1106.6251}

\bibitem[\protect\citeauthoryear{{Ambikasaran}, {Foreman-Mackey}, {Greengard},
  {Hogg}  \& {O'Neil}}{{Ambikasaran} et~al.}{2015}]{george}
{Ambikasaran} S.,  {Foreman-Mackey} D.,  {Greengard} L.,  {Hogg} D.~W.,
  {O'Neil} M.,  2015, \mn@doi [IEEE Transactions on Pattern Analysis and
  Machine Intelligence] {10.1109/TPAMI.2015.2448083}, \href
  {https://ui.adsabs.harvard.edu/abs/2015ITPAM..38..252A} {38, 252}

\bibitem[\protect\citeauthoryear{{Anderson} et~al.,}{{Anderson}
  et~al.}{2011}]{Anderson2011}
{Anderson} D.~R.,  et~al., 2011, \mn@doi [\apjl] {10.1088/2041-8205/726/2/L19},
  \href {http://adsabs.harvard.edu/abs/2011ApJ...726L..19A} {726, L19}

\bibitem[\protect\citeauthoryear{{Astropy Collaboration} et~al.,}{{Astropy
  Collaboration} et~al.}{2013}]{astropy1}
{Astropy Collaboration} et~al., 2013, \mn@doi [\aap]
  {10.1051/0004-6361/201322068}, \href
  {https://ui.adsabs.harvard.edu/abs/2013A&A...558A..33A} {558, A33}

\bibitem[\protect\citeauthoryear{{Astropy Collaboration} et~al.,}{{Astropy
  Collaboration} et~al.}{2018}]{astropy2}
{Astropy Collaboration} et~al., 2018, \mn@doi [\aj] {10.3847/1538-3881/aabc4f},
  \href {https://ui.adsabs.harvard.edu/abs/2018AJ....156..123A} {156, 123}

\bibitem[\protect\citeauthoryear{{Baluev}}{{Baluev}}{2013}]{planetpack}
{Baluev} R.~V.,  2013, \mn@doi [Astronomy and Computing]
  {10.1016/j.ascom.2013.07.001}, \href
  {https://ui.adsabs.harvard.edu/abs/2013A&C.....2...18B} {2, 18}

\bibitem[\protect\citeauthoryear{{Barrag{\'a}n} et~al.,}{{Barrag{\'a}n}
  et~al.}{2018}]{Barragan2018b}
{Barrag{\'a}n} O.,  et~al., 2018, \mn@doi [\aap] {10.1051/0004-6361/201732217},
  \href {https://ui.adsabs.harvard.edu/abs/2018A&A...612A..95B} {612, A95}

\bibitem[\protect\citeauthoryear{Barrag\'an, Gandolfi  \&
  Antoniciello}{Barrag\'an et~al.}{2019a}]{pyaneti}
Barrag\'an O.,  Gandolfi D.,   Antoniciello G.,  2019a, \mn@doi [\mnras]
  {10.1093/mnras/sty2472}, \href
  {https://ui.adsabs.harvard.edu/#abs/2019MNRAS.482.1017B} {482, 1017}

\bibitem[\protect\citeauthoryear{{Barrag{\'a}n} et~al.,}{{Barrag{\'a}n}
  et~al.}{2019b}]{Barragan2019}
{Barrag{\'a}n} O.,  et~al., 2019b, \mn@doi [\mnras] {10.1093/mnras/stz2569},
  \href {https://ui.adsabs.harvard.edu/abs/2019MNRAS.490..698B} {490, 698}

\bibitem[\protect\citeauthoryear{{Barrag{\'a}n} et~al.,}{{Barrag{\'a}n}
  et~al.}{2021a}]{Barragan2021b}
{Barrag{\'a}n} O.,  et~al., 2021a, arXiv e-prints, \href
  {https://ui.adsabs.harvard.edu/abs/2021arXiv211013069B} {p. arXiv:2110.13069}

\bibitem[\protect\citeauthoryear{{Barrag{\'a}n}, {Aigrain}, {Gillen}  \&
  {Guti{\'e}rrez-Canales}}{{Barrag{\'a}n} et~al.}{2021b}]{Barragan2021}
{Barrag{\'a}n} O.,  {Aigrain} S.,  {Gillen} E.,   {Guti{\'e}rrez-Canales} F.,
  2021b, \mn@doi [Research Notes of the American Astronomical Society]
  {10.3847/2515-5172/abef70}, \href
  {https://ui.adsabs.harvard.edu/abs/2021RNAAS...5...51B} {5, 51}

\bibitem[\protect\citeauthoryear{{Batalha}}{{Batalha}}{2014}]{Batalha2014}
{Batalha} N.~M.,  2014, \mn@doi [Proceedings of the National Academy of
  Science] {10.1073/pnas.1304196111}, \href
  {https://ui.adsabs.harvard.edu/abs/2014PNAS..11112647B} {111, 12647}

\bibitem[\protect\citeauthoryear{{Boisse} et~al.,}{{Boisse}
  et~al.}{2009}]{Boisse2009}
{Boisse} I.,  et~al., 2009, \mn@doi [\aap] {10.1051/0004-6361:200810648}, \href
  {https://ui.adsabs.harvard.edu/abs/2009A&A...495..959B} {495, 959}

\bibitem[\protect\citeauthoryear{{Bonfanti} \& {Gillon}}{{Bonfanti} \&
  {Gillon}}{2020}]{mcmci}
{Bonfanti} A.,  {Gillon} M.,  2020, \mn@doi [\aap]
  {10.1051/0004-6361/201936326}, \href
  {https://ui.adsabs.harvard.edu/abs/2020A&A...635A...6B} {635, A6}

\bibitem[\protect\citeauthoryear{{Bozza}, {Mancini}  \& {Sozzetti}}{{Bozza}
  et~al.}{2016}]{exoplanets2016}
{Bozza} V.,  {Mancini} L.,   {Sozzetti} A.,  2016, {Methods of Detecting
  Exoplanets}.
~ Vol. 428, \mn@doi{10.1007/978-3-319-27458-4, }

\bibitem[\protect\citeauthoryear{{Broeg} et~al.,}{{Broeg}
  et~al.}{2013}]{Broeg2013}
{Broeg} C.,  et~al., 2013, in European Physical Journal Web of Conferences. p.
  03005 (\mn@eprint {arXiv} {1305.2270}), \mn@doi{10.1051/epjconf/20134703005}

\bibitem[\protect\citeauthoryear{Bruno}{Bruno}{1584}]{Bruno1584}
Bruno G.,  1584, De l'infinito, universo e mondi

\bibitem[\protect\citeauthoryear{{Carleo} et~al.,}{{Carleo}
  et~al.}{2020}]{Carleo2020}
{Carleo} I.,  et~al., 2020, \mn@doi [\aj] {10.3847/1538-3881/aba124}, \href
  {https://ui.adsabs.harvard.edu/abs/2020AJ....160..114C} {160, 114}

\bibitem[\protect\citeauthoryear{{Charbonneau}, {Brown}, {Latham}  \&
  {Mayor}}{{Charbonneau} et~al.}{2000}]{Charbonneau2000}
{Charbonneau} D.,  {Brown} T.~M.,  {Latham} D.~W.,   {Mayor} M.,  2000, \mn@doi
  [\apjl] {10.1086/312457}, \href
  {http://adsabs.harvard.edu/abs/2000ApJ...529L..45C} {529, L45}

\bibitem[\protect\citeauthoryear{{Charbonneau}, {Brown}, {Noyes}  \&
  {Gilliland}}{{Charbonneau} et~al.}{2002}]{Charbonneau2002}
{Charbonneau} D.,  {Brown} T.~M.,  {Noyes} R.~W.,   {Gilliland} R.~L.,  2002,
  \mn@doi [\apj] {10.1086/338770}, \href
  {https://ui.adsabs.harvard.edu/abs/2002ApJ...568..377C} {568, 377}

\bibitem[\protect\citeauthoryear{{Chen}, {Fan}  \& {Wang}}{{Chen}
  et~al.}{2020}]{Chen2020}
{Chen} Z.,  {Fan} J.,   {Wang} K.,  2020, arXiv e-prints, \href
  {https://ui.adsabs.harvard.edu/abs/2020arXiv201009830C} {p. arXiv:2010.09830}

\bibitem[\protect\citeauthoryear{Coleman}{Coleman}{1974}]{Coleman1974}
Coleman R.,  1974, What is a Stochastic Process?.
Springer Netherlands, Dordrecht, pp~1--5, \mn@doi{10.1007/978-94-010-9796-3_1},
  \url {https://doi.org/10.1007/978-94-010-9796-3_1}

\bibitem[\protect\citeauthoryear{{Collier Cameron} et~al.,}{{Collier Cameron}
  et~al.}{2019}]{CollierCameron2019}
{Collier Cameron} A.,  et~al., 2019, \mn@doi [\mnras] {10.1093/mnras/stz1215},
  \href {https://ui.adsabs.harvard.edu/abs/2019MNRAS.487.1082C} {487, 1082}

\bibitem[\protect\citeauthoryear{{Collier Cameron} et~al.,}{{Collier Cameron}
  et~al.}{2020}]{CollierCameron2020}
{Collier Cameron} A.,  et~al., 2020, arXiv e-prints, \href
  {https://ui.adsabs.harvard.edu/abs/2020arXiv201100018C} {p. arXiv:2011.00018}

\bibitem[\protect\citeauthoryear{{Cretignier}, {Dumusque}, {Allart}, {Pepe}  \&
  {Lovis}}{{Cretignier} et~al.}{2020}]{Cretignier2020}
{Cretignier} M.,  {Dumusque} X.,  {Allart} R.,  {Pepe} F.,   {Lovis} C.,  2020,
  \mn@doi [\aap] {10.1051/0004-6361/201936548}, \href
  {https://ui.adsabs.harvard.edu/abs/2020A&A...633A..76C} {633, A76}

\bibitem[\protect\citeauthoryear{{Csizmadia}}{{Csizmadia}}{2020}]{tlcm}
{Csizmadia} S.,  2020, \mn@doi [\mnras] {10.1093/mnras/staa349}, \href
  {https://ui.adsabs.harvard.edu/abs/2020MNRAS.496.4442C} {496, 4442}

\bibitem[\protect\citeauthoryear{{D{\'\i}az}, {Almenara}, {Santerne}, {Moutou},
  {Lethuillier}  \& {Deleuil}}{{D{\'\i}az} et~al.}{2014}]{pastis}
{D{\'\i}az} R.~F.,  {Almenara} J.~M.,  {Santerne} A.,  {Moutou} C.,
  {Lethuillier} A.,   {Deleuil} M.,  2014, \mn@doi [\mnras]
  {10.1093/mnras/stu601}, \href
  {https://ui.adsabs.harvard.edu/abs/2014MNRAS.441..983D} {441, 983}

\bibitem[\protect\citeauthoryear{{Donati} et~al.,}{{Donati}
  et~al.}{2018}]{Donati2017}
{Donati} J.-F.,  et~al., 2018, {SPIRou: A NIR Spectropolarimeter/High-Precision
  Velocimeter for the CFHT}.
p.~107, \mn@doi{10.1007/978-3-319-55333-7_107}

\bibitem[\protect\citeauthoryear{{Dumusque}, {Boisse}  \& {Santos}}{{Dumusque}
  et~al.}{2014}]{Dumusque2014}
{Dumusque} X.,  {Boisse} I.,   {Santos} N.~C.,  2014, \mn@doi [\apj]
  {10.1088/0004-637X/796/2/132}, \href
  {https://ui.adsabs.harvard.edu/abs/2014ApJ...796..132D} {796, 132}

\bibitem[\protect\citeauthoryear{{Eastman}, {Gaudi}  \& {Agol}}{{Eastman}
  et~al.}{2013}]{exofast}
{Eastman} J.,  {Gaudi} B.~S.,   {Agol} E.,  2013, \mn@doi [\pasp]
  {10.1086/669497}, \href
  {https://ui.adsabs.harvard.edu/abs/2013PASP..125...83E} {125, 83}

\bibitem[\protect\citeauthoryear{{Eastman} et~al.,}{{Eastman}
  et~al.}{2019}]{exofast2}
{Eastman} J.~D.,  et~al., 2019, arXiv e-prints, \href
  {https://ui.adsabs.harvard.edu/abs/2019arXiv190709480E} {p. arXiv:1907.09480}

\bibitem[\protect\citeauthoryear{{Eisner} et~al.,}{{Eisner}
  et~al.}{2020}]{Eisner2020a}
{Eisner} N.~L.,  et~al., 2020, \mn@doi [\mnras] {10.1093/mnras/staa138}, \href
  {https://ui.adsabs.harvard.edu/abs/2020MNRAS.494..750E} {494, 750}

\bibitem[\protect\citeauthoryear{{Eisner} et~al.,}{{Eisner}
  et~al.}{2021}]{Eisner2020b}
{Eisner} N.~L.,  et~al., 2021, \mn@doi [\mnras] {10.1093/mnras/staa3739}, \href
  {https://ui.adsabs.harvard.edu/abs/2021MNRAS.501.4669E} {501, 4669}

\bibitem[\protect\citeauthoryear{{Espinoza}, {Kossakowski}  \&
  {Brahm}}{{Espinoza} et~al.}{2019}]{juliet}
{Espinoza} N.,  {Kossakowski} D.,   {Brahm} R.,  2019, \mn@doi [\mnras]
  {10.1093/mnras/stz2688}, \href
  {https://ui.adsabs.harvard.edu/abs/2019MNRAS.490.2262E} {490, 2262}

\bibitem[\protect\citeauthoryear{{Feroz}, {Hobson}  \& {Bridges}}{{Feroz}
  et~al.}{2009}]{multinest}
{Feroz} F.,  {Hobson} M.~P.,   {Bridges} M.,  2009, \mn@doi [\mnras]
  {10.1111/j.1365-2966.2009.14548.x}, \href
  {https://ui.adsabs.harvard.edu/abs/2009MNRAS.398.1601F} {398, 1601}

\bibitem[\protect\citeauthoryear{{Foreman-Mackey}}{{Foreman-Mackey}}{2018}]{celerite}
{Foreman-Mackey} D.,  2018, \mn@doi [Research Notes of the American
  Astronomical Society] {10.3847/2515-5172/aaaf6c}, \href
  {https://ui.adsabs.harvard.edu/abs/2018RNAAS...2...31F} {2, 31}

\bibitem[\protect\citeauthoryear{{Foreman-Mackey}, {Hogg}, {Lang}  \&
  {Goodman}}{{Foreman-Mackey} et~al.}{2013}]{emcee}
{Foreman-Mackey} D.,  {Hogg} D.~W.,  {Lang} D.,   {Goodman} J.,  2013, \mn@doi
  [\pasp] {10.1086/670067}, \href
  {https://ui.adsabs.harvard.edu/abs/2013PASP..125..306F} {125, 306}

\bibitem[\protect\citeauthoryear{{Foreman-Mackey} et~al.,}{{Foreman-Mackey}
  et~al.}{2021}]{exoplanet}
{Foreman-Mackey} D.,  et~al., 2021, arXiv e-prints, \href
  {https://ui.adsabs.harvard.edu/abs/2021arXiv210501994F} {p. arXiv:2105.01994}

\bibitem[\protect\citeauthoryear{{Fulton}, {Petigura}, {Blunt}  \&
  {Sinukoff}}{{Fulton} et~al.}{2018}]{radvel}
{Fulton} B.~J.,  {Petigura} E.~A.,  {Blunt} S.,   {Sinukoff} E.,  2018, \mn@doi
  [\pasp] {10.1088/1538-3873/aaaaa8}, \href
  {https://ui.adsabs.harvard.edu/abs/2018PASP..130d4504F} {130, 044504}

\bibitem[\protect\citeauthoryear{Gelman, Carlin, Stern  \& Rubin}{Gelman
  et~al.}{2004}]{Gelman2004}
Gelman A.,  Carlin J.~B.,  Stern H.~S.,   Rubin D.~B.,  2004, Bayesian Data
  Analysis, 2nd ed. edn.
Chapman and Hall/CRC

\bibitem[\protect\citeauthoryear{{Georgieva} et~al.,}{{Georgieva}
  et~al.}{2021}]{Georgieva2021}
{Georgieva} I.~Y.,  et~al., 2021, \mn@doi [\mnras] {10.1093/mnras/stab1464},
  \href {https://ui.adsabs.harvard.edu/abs/2021MNRAS.505.4684G} {505, 4684}

\bibitem[\protect\citeauthoryear{{Gilbertson}, {Ford}, {Jones}  \&
  {Stenning}}{{Gilbertson} et~al.}{2020}]{Gilbertson2020}
{Gilbertson} C.,  {Ford} E.~B.,  {Jones} D.~E.,   {Stenning} D.~C.,  2020,
  arXiv e-prints, \href {https://ui.adsabs.harvard.edu/abs/2020arXiv200901085G}
  {p. arXiv:2009.01085}

\bibitem[\protect\citeauthoryear{{Grunblatt}, {Howard}  \&
  {Haywood}}{{Grunblatt} et~al.}{2015}]{Grunblatt2015}
{Grunblatt} S.~K.,  {Howard} A.~W.,   {Haywood} R.~D.,  2015, \mn@doi [\apj]
  {10.1088/0004-637X/808/2/127}, \href
  {https://ui.adsabs.harvard.edu/abs/2015ApJ...808..127G} {808, 127}

\bibitem[\protect\citeauthoryear{{G{\"u}nther} \& {Daylan}}{{G{\"u}nther} \&
  {Daylan}}{2020}]{allesfitter}
{G{\"u}nther} M.~N.,  {Daylan} T.,  2020, arXiv e-prints, \href
  {https://ui.adsabs.harvard.edu/abs/2020arXiv200314371G} {p. arXiv:2003.14371}

\bibitem[\protect\citeauthoryear{{Handley}, {Hobson}  \& {Lasenby}}{{Handley}
  et~al.}{2015}]{polychord}
{Handley} W.~J.,  {Hobson} M.~P.,   {Lasenby} A.~N.,  2015, \mn@doi [\mnras]
  {10.1093/mnrasl/slv047}, \href
  {https://ui.adsabs.harvard.edu/abs/2015MNRAS.450L..61H} {450, L61}

\bibitem[\protect\citeauthoryear{Harris et~al.,}{Harris et~al.}{2020}]{numpy}
Harris C.~R.,  et~al., 2020, \mn@doi [Nature] {10.1038/s41586-020-2649-2}, 585,
  357

\bibitem[\protect\citeauthoryear{{Hatzes} et~al.,}{{Hatzes}
  et~al.}{2010}]{Hatzes2010}
{Hatzes} A.~P.,  et~al., 2010, \mn@doi [\aap] {10.1051/0004-6361/201014795},
  \href {https://ui.adsabs.harvard.edu/abs/2010A&A...520A..93H} {520, A93}

\bibitem[\protect\citeauthoryear{{Hatzes} et~al.,}{{Hatzes}
  et~al.}{2011}]{Hatzes2011}
{Hatzes} A.~P.,  et~al., 2011, \mn@doi [\apj] {10.1088/0004-637X/743/1/75},
  \href {https://ui.adsabs.harvard.edu/abs/2011ApJ...743...75H} {743, 75}

\bibitem[\protect\citeauthoryear{{Haywood} et~al.,}{{Haywood}
  et~al.}{2014}]{Haywood2014}
{Haywood} R.~D.,  et~al., 2014, \mn@doi [\mnras] {10.1093/mnras/stu1320}, \href
  {https://ui.adsabs.harvard.edu/abs/2014MNRAS.443.2517H} {443, 2517}

\bibitem[\protect\citeauthoryear{{Henry}, {Marcy}, {Butler}  \& {Vogt}}{{Henry}
  et~al.}{1999}]{Henry1999}
{Henry} G.~W.,  {Marcy} G.,  {Butler} R.~P.,   {Vogt} S.~S.,  1999, \iaucirc,
  \href {http://adsabs.harvard.edu/abs/1999IAUC.7307....1H} {7307}

\bibitem[\protect\citeauthoryear{{Hippke}, {David}, {Mulders}  \&
  {Heller}}{{Hippke} et~al.}{2019}]{Hippke2019}
{Hippke} M.,  {David} T.~J.,  {Mulders} G.~D.,   {Heller} R.,  2019, \mn@doi
  [\aj] {10.3847/1538-3881/ab3984}, \href
  {https://ui.adsabs.harvard.edu/abs/2019AJ....158..143H} {158, 143}

\bibitem[\protect\citeauthoryear{{Isaacson} \& {Fischer}}{{Isaacson} \&
  {Fischer}}{2010}]{Isaacson2010}
{Isaacson} H.,  {Fischer} D.,  2010, \mn@doi [\apj]
  {10.1088/0004-637X/725/1/875}, \href
  {https://ui.adsabs.harvard.edu/abs/2010ApJ...725..875I} {725, 875}

\bibitem[\protect\citeauthoryear{{Jones}, {Stenning}, {Ford}, {Wolpert},
  {Loredo}, {Gilbertson}  \& {Dumusque}}{{Jones} et~al.}{2017}]{Jones2017}
{Jones} D.~E.,  {Stenning} D.~C.,  {Ford} E.~B.,  {Wolpert} R.~L.,  {Loredo}
  T.~J.,  {Gilbertson} C.,   {Dumusque} X.,  2017, arXiv e-prints, \href
  {https://ui.adsabs.harvard.edu/abs/2017arXiv171101318J} {p. arXiv:1711.01318}

\bibitem[\protect\citeauthoryear{{Karamanis}, {Beutler}  \&
  {Peacock}}{{Karamanis} et~al.}{2021}]{zeus}
{Karamanis} M.,  {Beutler} F.,   {Peacock} J.~A.,  2021, arXiv e-prints, \href
  {https://ui.adsabs.harvard.edu/abs/2021arXiv210503468K} {p. arXiv:2105.03468}

\bibitem[\protect\citeauthoryear{{Kipping}}{{Kipping}}{2010}]{Kipping2010}
{Kipping} D.~M.,  2010, \mn@doi [\mnras] {10.1111/j.1365-2966.2010.17242.x},
  \href {https://ui.adsabs.harvard.edu/abs/2010MNRAS.408.1758K} {408, 1758}

\bibitem[\protect\citeauthoryear{{Kipping}}{{Kipping}}{2013}]{Kipping2013}
{Kipping} D.~M.,  2013, \mn@doi [\mnras] {10.1093/mnras/stt1435}, \href
  {https://ui.adsabs.harvard.edu/abs/2013MNRAS.435.2152K} {435, 2152}

\bibitem[\protect\citeauthoryear{{Mandel} \& {Agol}}{{Mandel} \&
  {Agol}}{2002}]{Mandel2002}
{Mandel} K.,  {Agol} E.,  2002, \mn@doi [\apjl] {10.1086/345520}, \href
  {http://adsabs.harvard.edu/abs/2002ApJ...580L.171M} {580, L171}

\bibitem[\protect\citeauthoryear{{Mayo} et~al.,}{{Mayo}
  et~al.}{2019}]{Mayo2019}
{Mayo} A.~W.,  et~al., 2019, \mn@doi [\aj] {10.3847/1538-3881/ab3e2f}, \href
  {https://ui.adsabs.harvard.edu/abs/2019AJ....158..165M} {158, 165}

\bibitem[\protect\citeauthoryear{{Mayor} \& {Queloz}}{{Mayor} \&
  {Queloz}}{1995}]{Mayor1995}
{Mayor} M.,  {Queloz} D.,  1995, \mn@doi [\nat] {10.1038/378355a0}, \href
  {http://adsabs.harvard.edu/abs/1995Natur.378..355M} {378, 355}

\bibitem[\protect\citeauthoryear{{Osborn} et~al.,}{{Osborn}
  et~al.}{2016}]{Osborn2016}
{Osborn} H.~P.,  et~al., 2016, \mn@doi [\mnras] {10.1093/mnras/stw137}, \href
  {https://ui.adsabs.harvard.edu/abs/2016MNRAS.457.2273O} {457, 2273}

\bibitem[\protect\citeauthoryear{{Osborn} et~al.,}{{Osborn}
  et~al.}{2021}]{Osborn2021}
{Osborn} H.~P.,  et~al., 2021, \mn@doi [\mnras] {10.1093/mnras/stab182}, \href
  {https://ui.adsabs.harvard.edu/abs/2021MNRAS.502.4842O} {502, 4842}

\bibitem[\protect\citeauthoryear{{Parviainen}}{{Parviainen}}{2015}]{pytransit}
{Parviainen} H.,  2015, \mn@doi [\mnras] {10.1093/mnras/stv894}, \href
  {https://ui.adsabs.harvard.edu/abs/2015MNRAS.450.3233P} {450, 3233}

\bibitem[\protect\citeauthoryear{{Parviainen} et~al.,}{{Parviainen}
  et~al.}{2019}]{Parviainen2019}
{Parviainen} H.,  et~al., 2019, \mn@doi [\aap] {10.1051/0004-6361/201935709},
  \href {https://ui.adsabs.harvard.edu/abs/2019A&A...630A..89P} {630, A89}

\bibitem[\protect\citeauthoryear{{Pepe} et~al.,}{{Pepe}
  et~al.}{2010}]{Pepe2010}
{Pepe} F.~A.,  et~al., 2010, in Ground-based and Airborne Instrumentation for
  Astronomy III. p. 77350F, \mn@doi{10.1117/12.857122}

\bibitem[\protect\citeauthoryear{{Pepe} et~al.,}{{Pepe}
  et~al.}{2013}]{Pepe2013}
{Pepe} F.,  et~al., 2013, \mn@doi [\nat] {10.1038/nature12768}, \href
  {http://adsabs.harvard.edu/abs/2013Natur.503..377P} {503, 377}

\bibitem[\protect\citeauthoryear{{Petigura}, {Howard}  \& {Marcy}}{{Petigura}
  et~al.}{2013}]{Petigura2013}
{Petigura} E.~A.,  {Howard} A.~W.,   {Marcy} G.~W.,  2013, \mn@doi [Proceedings
  of the National Academy of Science] {10.1073/pnas.1319909110}, \href
  {https://ui.adsabs.harvard.edu/abs/2013PNAS..11019273P} {110, 19273}

\bibitem[\protect\citeauthoryear{{Queloz} et~al.,}{{Queloz}
  et~al.}{2001}]{Queloz2001}
{Queloz} D.,  et~al., 2001, \mn@doi [\aap] {10.1051/0004-6361:20011308}, \href
  {https://ui.adsabs.harvard.edu/abs/2001A&A...379..279Q} {379, 279}

\bibitem[\protect\citeauthoryear{{Rajpaul}, {Aigrain}, {Osborne}, {Reece}  \&
  {Roberts}}{{Rajpaul} et~al.}{2015}]{Rajpaul2015}
{Rajpaul} V.,  {Aigrain} S.,  {Osborne} M.~A.,  {Reece} S.,   {Roberts} S.,
  2015, \mn@doi [\mnras] {10.1093/mnras/stv1428}, \href
  {https://ui.adsabs.harvard.edu/abs/2015MNRAS.452.2269R} {452, 2269}

\bibitem[\protect\citeauthoryear{{Rajpaul}, {Aigrain}  \& {Roberts}}{{Rajpaul}
  et~al.}{2016}]{Rajpaul2016}
{Rajpaul} V.,  {Aigrain} S.,   {Roberts} S.,  2016, \mn@doi [\mnras]
  {10.1093/mnrasl/slv164}, \href
  {https://ui.adsabs.harvard.edu/abs/2016MNRAS.456L...6R} {456, L6}

\bibitem[\protect\citeauthoryear{{Rajpaul}, {Aigrain}  \& {Buchhave}}{{Rajpaul}
  et~al.}{2020}]{Rajpaul2020}
{Rajpaul} V.~M.,  {Aigrain} S.,   {Buchhave} L.~A.,  2020, \mn@doi [\mnras]
  {10.1093/mnras/stz3599}, \href
  {https://ui.adsabs.harvard.edu/abs/2020MNRAS.492.3960R} {492, 3960}

\bibitem[\protect\citeauthoryear{{Rasmussen} \& {Williams}}{{Rasmussen} \&
  {Williams}}{2006}]{Rasmussen2006}
{Rasmussen} C.~E.,  {Williams} C. K.~I.,  2006, {Gaussian Processes for Machine
  Learning}

\bibitem[\protect\citeauthoryear{{Rauer} et~al.,}{{Rauer}
  et~al.}{2014}]{Rauer2014}
{Rauer} H.,  et~al., 2014, \mn@doi [Experimental Astronomy]
  {10.1007/s10686-014-9383-4}, \href
  {http://adsabs.harvard.edu/abs/2014ExA....38..249R} {38, 249}

\bibitem[\protect\citeauthoryear{{Ricker} et~al.,}{{Ricker}
  et~al.}{2015}]{Ricker2015}
{Ricker} G.~R.,  et~al., 2015, \mn@doi [Journal of Astronomical Telescopes,
  Instruments, and Systems] {10.1117/1.JATIS.1.1.014003}, \href
  {https://ui.adsabs.harvard.edu/#abs/2015JATIS...1a4003R} {1, 014003}

\bibitem[\protect\citeauthoryear{Roberts, Osborne, Ebden, Reece, Gibson  \&
  Aigrain}{Roberts et~al.}{2013}]{Roberts2013}
Roberts S.,  Osborne M.,  Ebden M.,  Reece S.,  Gibson N.,   Aigrain S.,  2013,
  \mn@doi [Philosophical Transactions of the Royal Society A: Mathematical,
  Physical and Engineering Sciences] {10.1098/rsta.2011.0550}, 371, 20110550

\bibitem[\protect\citeauthoryear{{Santerne} et~al.,}{{Santerne}
  et~al.}{2015}]{pastis2}
{Santerne} A.,  et~al., 2015, \mn@doi [\mnras] {10.1093/mnras/stv1080}, \href
  {https://ui.adsabs.harvard.edu/abs/2015MNRAS.451.2337S} {451, 2337}

\bibitem[\protect\citeauthoryear{{Speagle}}{{Speagle}}{2020}]{dynesty}
{Speagle} J.~S.,  2020, \mn@doi [\mnras] {10.1093/mnras/staa278}, \href
  {https://ui.adsabs.harvard.edu/abs/2020MNRAS.493.3132S} {493, 3132}

\bibitem[\protect\citeauthoryear{{Struve}}{{Struve}}{1952}]{Struve1952}
{Struve} O.,  1952, The Observatory, \href
  {https://ui.adsabs.harvard.edu/abs/1952Obs....72..199S} {72, 199}

\bibitem[\protect\citeauthoryear{{Su{\'a}rez Mascare{\~n}o}
  et~al.,}{{Su{\'a}rez Mascare{\~n}o} et~al.}{2020}]{Suarez2020}
{Su{\'a}rez Mascare{\~n}o} A.,  et~al., 2020, \mn@doi [\aap]
  {10.1051/0004-6361/202037745}, \href
  {https://ui.adsabs.harvard.edu/abs/2020A&A...639A..77S} {639, A77}

\bibitem[\protect\citeauthoryear{{Thompson}, {Watson}, {de Mooij}  \&
  {Jess}}{{Thompson} et~al.}{2017}]{Thompson2017}
{Thompson} A.~P.~G.,  {Watson} C.~A.,  {de Mooij} E.~J.~W.,   {Jess} D.~B.,
  2017, \mn@doi [\mnras] {10.1093/mnrasl/slx018}, \href
  {https://ui.adsabs.harvard.edu/abs/2017MNRAS.468L..16T} {468, L16}

\bibitem[\protect\citeauthoryear{{Tracey} \& {Wolpert}}{{Tracey} \&
  {Wolpert}}{2018}]{Tracey2018}
{Tracey} B.~D.,  {Wolpert} D.~H.,  2018, arXiv e-prints, \href
  {https://ui.adsabs.harvard.edu/abs/2018arXiv180106147T} {p. arXiv:1801.06147}

\bibitem[\protect\citeauthoryear{{Trifonov}}{{Trifonov}}{2019}]{exostricker}
{Trifonov} T.,  2019, {The Exo-Striker: Transit and radial velocity interactive
  fitting tool for orbital analysis and N-body simulations} (\mn@eprint {ascl}
  {1906.004})

\makeatother
\end{thebibliography}




\appendix


\section{Covariance functions and their derivatives}
\label{ap:derivatives}

In this appendix we show the form of the $\gamma^{G,G}_{i,j} = \gamma^{G,G}(t_i,t_j)$, 

\begin{equation}
    \gamma_{i,j}^{G,dG} =  \gamma^{G,dG}(t_i,t_j) = \frac{\partial}{\partial t} \gamma^{G,G}(t_i,t) \Bigr|_{t=t_j}, 
    \label{eq:gdg}
\end{equation}

\begin{equation}
    \gamma_{i,j}^{dG,G} =  \gamma^{dG,G}(t_i,t_j) = \frac{\partial}{\partial t} \gamma^{G,G}(t,t_j) \Bigr|_{t=t_i}, 
\end{equation}

\noindent and

\begin{equation}
    \gamma_{i,j}^{dG,dG} =  \gamma^{dG,dG}(t_i,t_j) = 
    \frac{\partial}{\partial t'}
    \left(
    \frac{\partial}{\partial t} \gamma^{G,G}(t',t) \Bigr|_{t=t_i}
    \right)
    \Bigr|_{t'=t_j}, 
\end{equation}

\noindent 
for the squared exponential, Mat\'ern 5/2, and Quasi-periodic kernels. These quantities can be used to compute the matrices given in eq.~\eqref{eq:smallks} needed to compute the big covariance matrix given in eq.~\eqref{eq:bigk} for the multidimensional GP regression.

\subsection{Square exponential kernel}

The squared exponential covariance function is written as
\begin{equation}
    \gamma_{{\rm SE},i,j}^{G,G}  = 
    \exp \left[
    - \frac{ \left( t_i - t_j \right)^2 }{2 \lambda^2}
    \right]
    ,
    \label{eq:segg}
\end{equation}
\noindent
where we have omitted the amplitude term shown in eq~\eqref{eq:se}. The covariance between the derivative observation $i$ and the non-derivative observation $j$ is 
\begin{equation}
    \gamma_{{\rm SE},i,j}^{G,dG}  = - \gamma_{{\rm SE},i,j}^{dG,G} = 
    \frac{t_i - t_j}{\lambda^2}
    \gamma_{{\rm SE},i,j}^{G,G}
    ,
    \label{eq:segdg}
\end{equation}
\noindent
and the covariance between the derivative observation $i$ and the derivative observation $j$ is 
\begin{equation}
    \gamma_{{\rm SE},i,j}^{dG,dG}  =
    \left[
    \frac{1}{\lambda^2} -
    \frac{\left(t_i - t_j \right)^2}{\lambda^4}
    \right]
    \gamma_{{\rm SE},i,j}^{G,G}
    .
    \label{eq:sedgdg}
\end{equation}

\subsection{Mat\'ern 5/2 Kernel}

If we define
\begin{equation}
    t_{5/2} \equiv \frac{\sqrt{5} \left\lvert t_i - t_j \right\rvert }{\lambda},
\end{equation}
\noindent
we can write the Mat\'ern 3/2 Kernel as
\begin{equation}
    \gamma_{{\rm M52},i,j}^{G,G} = \left( 
    1 + t_{5/2} + \frac{t_{5/2}^2}{3}
    \right)
    \exp \left( - t_{5/2} \right),
\end{equation}
\noindent
where we have omitted the amplitude term shown in eq~\eqref{eq:m32}.
The covariance between the derivative of the observations $i$ and the non-derivative observations of $j$ is then
\begin{equation}
    \gamma_{{\rm M52},i,j}^{G,dG} = - \gamma_{{\rm M52},i,j}^{dG,G} = \frac{\sqrt{5}}{3 \lambda}
   t_{5/2}
    \left(  1 + t_{5/2}  \right)
    \exp \left( - t_{5/2} \right)  \mathrm{sgn}(t_i - t_j),
\end{equation}
\noindent
where $\mathrm{sgn}$ is the sign function.
 The covariance between the derivative of the observations $i$ and the derivative observations of $j$ is then
\begin{equation}
    \gamma_{{\rm M52},i,j}^{dG,dG} =  - \frac{5}{3 \lambda^2} 
    \left(  t_{5/2}^2 - t_{5/2} - 1  \right)
    \exp \left( - t_{5/2} t \right).
\end{equation}

\subsection{Quasi-Periodic Kernel}

In order to use the QP kernel in the multidimensional GP approach, we need to use eq.~\eqref{eq:qp} without amplitude term as 
\begin{equation}
    \gamma_{{\rm QP},i,j}^{G,G} =  \exp 
    \left\{
    - \frac{\sin^2\left[\pi \left(t_i - t_j \right)/P_{\rm GP}\right]}{2 \lambda_{\rm p}^2}
    - \frac{\left(t_i - t_j\right)^2}{2\lambda_{\rm e}^2}
    \right\}
    \label{eq:apqp}
    ,
\end{equation}

If we define
\begin{equation}
    \tau \equiv \frac{2 \pi (t_i - t_j)}{P_{\rm GP}}, \\
\end{equation}
\noindent
the covariance between the derivative observation $i$ and the non-derivative observation $j$ for the QP kernel given by eq.~\eqref{eq:apqp} is
 \begin{equation}
     \gamma^{G,dG}_{{\rm QP},i,j} = - \gamma^{dG,G}_{{\rm QP},i,j} = - \gamma^{G,G}_{{\rm QP},i,j} 
     \left(
      \frac{\pi \sin \tau}{2 P_{\rm GP} \lambda_{\rm p}^2}
      + \frac{t_i - t_j}{\lambda_{\rm e}^2}
     \right).
     \label{eq:qpgdg}
 \end{equation}
\noindent
Finally, the covariance between the derivative observation $i$ and the derivative observation $j$ is then
 \begin{equation}
 \begin{aligned}
      \gamma^{dG,dG}_{{\rm QP},i,j} =  \gamma^{G,G}_{{\rm QP},i,j} 
      &
     \left[
      - \left(\frac{\pi \sin \tau}{2 P_{\rm GP} \lambda_{\rm p}^2} \right)^2
     - \frac{\tau \sin \tau}{2 \lambda_{\rm p}^2 \lambda_{\rm e}^2}
      + \frac{\pi^2 \cos \tau}{P_{\rm GP}^2 \lambda_{\rm p}^2}
      \right.
      \\
      &
      \left.
     - \left(\frac{t_i - t_j}{\lambda_{\rm e}^2}\right)^2
     + \frac{1}{\lambda_{\rm e}^2}
     \right].
 \end{aligned}
     \label{eq:qpdgdg}
 \end{equation}

\bsp	
\label{lastpage}
\end{document}